\documentclass[prb,twocolumn,amsmath,amssymb,showpacs,floatfix]{revtex4}

\usepackage{graphicx,bm,epsfig,setspace}
\usepackage{color}

\makeatletter
\def\graphicscale{\twocolumn@sw{0.33}{0.4}}
\makeatother

\def\spose#1{\hbox to 0pt{#1\hss}}
\def\lesssim{\mathrel{\spose{\lower 3pt\hbox{$\mathchar"218$}}
 \raise 2.0pt\hbox{$\mathchar"13C$}}}
\def\gtrsim{\mathrel{\spose{\lower 3pt\hbox{$\mathchar"218$}}
 \raise 2.0pt\hbox{$\mathchar"13E$}}}

\def\<{\langle}
\def\>{\rangle}

\newcommand*{\beq}{\begin{eqnarray}}
\newcommand*{\eeq}{\end{eqnarray}}
\newcommand*{\bea}{\begin{eqnarray}}
\newcommand*{\eea}{\end{eqnarray}}

\def\simge{\mathrel{%
       \rlap{\raise 0.511ex \hbox{$>$}}{\lower 0.511ex \hbox{$\sim$}}}}
\def\simle{\mathrel{
       \rlap{\raise 0.511ex \hbox{$<$}}{\lower 0.511ex \hbox{$\sim$}}}}

\begin{document}

\title{Phase diagram of mixtures of colloids and polymers\\
   in the thermal crossover from good to $\theta$ solvent.}

\author{Giuseppe D'Adamo}
\email{giuseppe.dadamo@sissa.it}
\affiliation{SISSA, V. Bonomea 265, I-34136 Trieste, Italy}
\author{Andrea Pelissetto}
\email{andrea.pelissetto@roma1.infn.it}
\affiliation{Dipartimento di Fisica, Sapienza Universit\`a di Roma and
INFN, Sezione di Roma I, P.le Aldo Moro 2, I-00185 Roma, Italy}
\author{Carlo Pierleoni}
\email{carlo.pierleoni@aquila.infn.it}
\affiliation{Dipartimento di Scienze Fisiche e Chimiche, Universit\`a dell'Aquila and
CNISM, UdR dell'Aquila, V. Vetoio 10, Loc.~Coppito, I-67100  L'Aquila, Italy}

\date{\today}

\begin{abstract}
We determine the phase diagram of mixtures of spherical colloids and 
neutral nonadsorbing 
polymers in the thermal crossover region between the $\theta$ point and 
the good-solvent regime. We use the generalized free-volume theory (GFVT),
which turns out to be quite accurate as long as $q = R_g/R_c\lesssim 1$ 
($R_g$ is the radius of gyration of the polymer and $R_c$ is 
the colloid radius).
Close to the $\theta$ point the phase diagram is not very sensitive to 
solvent quality, while, close to the good-solvent region, changes of the 
solvent quality modify significantly the position of the critical point and of
the binodals. We also analyze the phase behavior of aqueous solutions 
of charged colloids and polymers, using the extension of GFVT proposed by
Fortini {\em et al.}, J. Chem. Phys. {\bf 128}, 024904 (2008).
\end{abstract}

% 61.25.he Polymer solutions
% 65.20.De 
%     General theory of thermodynamic properties of liquids, including computer
%     simulation
% 82.35.Lr Physical properties of polymers

\maketitle

\section{Introduction}

Mixtures of colloids
and polymers are physical systems of great interest for their many
technological applications. In particular, they
are used, e.g., in the production of inks, paints,
and personal-care products. There is also an extensive theoretical interest, 
since the addition of polymers to a colloid suspension can induce aggregation
and allows one to modify the rheological properties in 
a controlled fashion. 
Here, we will consider dispersions of spherical 
colloids and nonadsorbing neutral polymers in an organic solvent. 
These systems show 
\cite{Poon-02,FS-02,TRK-03,MvDE-07,FT-08,ME-09}
a very interesting phenomenology, which only depends 
to a large extent on the nature of the solvent and 
on the ratio $q \equiv R_g/R_c$, where $R_g$ is the zero-density radius of 
gyration of the polymer and $R_c$ is the radius of the colloid.
Experiments and numerical simulations indicate that such 
polymer-colloid mixtures
have a solid colloidal phase for large enough colloidal concentrations
and a corresponding fluid-solid coexistence. 
The presence of a fluid-fluid coexistence of a 
colloid-rich, polymer-poor phase (colloid liquid) with a 
colloid-poor, polymer-rich phase (colloid gas) is much less obvious. 
Extensive theoretical and experimental work
indicates that such a transition occurs only if the size of the polymers is 
sufficiently large, i.e., for $q > q_{CEP}$, where
\cite{Poon-02} $q_{CEP} \approx 0.3$-0.4. 

The phase behavior of colloid-polymer mixtures depends on solvent quality. 
If the polymer solution is close to the $\theta$ point, polymers behave
approximately as ideal chains---the second virial coefficient is approximately
equal to zero. In this regime, 
the Asakura-Oosawa-Vrij (AOV) model,
\cite{AO-54,Vrij-76} which gives a coarse-grained description of the 
mixture, provides quantitatively accurate results as long as $q\lesssim 1$. 
The polymers are treated as an ideal gas 
of point particles, which interact with the colloids by means 
of a simple hard-core potential. 
This model is extremely crude since it ignores the polymeric structure. 
Nonetheless, it correctly predicts
polymer-colloid demixing as a result of the entropy-driven 
effective attraction (depletion interaction) 
between colloidal pairs due to the presence 
of the polymers.
\cite{GHR-83,LPPSW-92,MF-94,DBE-99,SLBE-00,BLH-02,DvR-02,SD-02,LLPR-03,VH-04,%
VH-04bis,VHB-05,ZTHB-10}  
Polymer-polymer interactions, which are necessary for a correct description
of the good-solvent regime, have also been included.
Refs.~\onlinecite{SDB-03,VS-05,ZVBHV-09,AP-12} discuss phenomenological 
generalizations of the 
AOV model, while 
Refs.~\onlinecite{BLH-02,FBD-07,Dzubiella-etal-01,DLL-02,VJDL-05}  
discuss coarse-grained models
in which the polymer-colloid and polymer-polymer potentials are 
derived either numerically \cite{BLH-02,FBD-07} from full-monomer
simulations or by means of general theoretical
considerations.\cite{Dzubiella-etal-01,DLL-02,VJDL-05}   
Another successfull approach is free-volume
theory\cite{LPPSW-92} which has been originally developed 
for mixtures of colloids and ideal polymers and later generalized
to include polymer-polymer and polymer-colloid
interactions.\cite{FT-08,ATL-02,FT-07,TSPEALF-08,LT-11} 
Also the PRISM approach, \cite{FS-00,FS-01,RFSZ-02} density functional theory,
\cite{SDB-03,Bryk-05} and thermodynamic perturbation theory
\cite{PVJ-03,PH-06} have been used. 
Full-monomer simulations have also been performed,
\cite{BML-03,CVPR-06,MLP-12,MIP-13} 
providing the phase diagram in the protein regime $q\gtrsim 1$, in which 
coarse-grained models, which identify each polymer with a 
monoatomic molecule,  are not expected to be accurate.
For a comparison with experimental data, polydispersity effects should also
be included as they affect quite significantly the thermodynamics
of polymer solutions.\cite{Schaefer-99,Pelissetto-Polydisp} 
They are discussed in Refs.~\onlinecite{SF-97,FS-05},
where it is shown that gas-liquid phase separation is favored and fluid-solid
segregation is retarded by increased polydispersity at fixed distribution of
the polymer sizes.

While significant work has been devoted to polymer systems under good-solvent
conditions or at the $\theta$ point, no systematic investigation has been 
performed of colloid-polymer segregation in the thermal crossover region. 
Such a regime can be parametrized  by using the 
Zimm-Stockmayer-Fixmann variable\cite{ZSF-53} 
\begin{equation}
   z = a^{(1)} {(T - T_\theta)\over T_\theta} M_w^{1/2},
\label{ZSF-variable}
\end{equation}
where $T$ is the temperature (in K), 
$T_\theta$ its value at the $\theta$ point, 
$M_w$ is the weight average molar mass\cite{Rubinstein-Colby-03}
(equivalently, one could 
use the degree of polymerization $L$),
and $a^{(1)}$ is a constant which is fixed 
once the normalization of $z$ is specified. The $\theta$ point corresponds to
$z=0$, while the good-solvent regime corresponds to $z=\infty$. 
Choice (\ref{ZSF-variable}) is not unique.
Another slightly different parametrization which is often used in the 
experimental analyses is 
\begin{equation}
   z = a^{(2)} {(T - T_\theta)\over T} M_w^{1/2},
\label{ZSF-variable2}
\end{equation}
which differs by the presence of $T$, instead of $T_\theta$, in the
denominator. A third possibility, which is the one preferred by theorists, is 
\begin{equation}
z = a^{(3)} (T - T_\theta) L^{1/2},
\label{ZSF-variable3}
\end{equation}
where $L$ is the degree of polymerization, i.e. the number of monomers
present in each chain.

Note that Eqs.~(\ref{ZSF-variable}), (\ref{ZSF-variable2}), and 
(\ref{ZSF-variable3}) are approximations that
neglect\cite{Schaefer-99} logarithmic corrections in $\ln M_w$ or $\ln L$
and that, moreover, are only valid close to $\theta$ point. Indeed, in 
general\cite{Schaefer-99} $z$ is related to the temperature $T$ by
$z = \alpha_1 f_T(T) L^{1/2} (\ln L)^{-4/11}$ (or with $M_w$ replacing $L$), 
where $f_T(T)$ is an analytic function 
vanishing at the $\theta$ point and $\alpha_1$ is 
a system-dependent constant. However, as we show in the supplementary
material\cite{suppl} for three typical polymer solutions, the 
previous simple parametrizations without the logarithmic factor
are fully adequate to describe experimental
data in a large temperature interval, given the typical experimental
accuracy. 

In this work 
we aim at determining the phase behavior of colloid-polymer mixtures in the 
intermediate crossover region in which $z$ is finite and positive. We will
use the generalized free-volume theory 
\cite{FT-08,ATL-02,FT-07,TSPEALF-08,LT-11} (GFVT), which has been shown to be 
quite accurate both at the $\theta$ point and under good-solvent conditions.
To implement the GFVT approach, one needs explicit expressions for the 
polymer equation of state and for the depletion thickness \cite{LT-11} as 
a function of the polymer density and of $z$. We will use here the accurate 
results reported in Refs.~\onlinecite{Pelissetto-08,DPP-thermal,DPP-depletion}. 
The GFVT results will be compared with numerical 
full-monomer data (a meaningful comparison
requires an extrapolation of the numerical results to the scaling, $L\to
\infty$, limit) and experimental results. It turns out that GFVT is 
only predictive for $q\lesssim 1$. For larger values of $q$, 
significant discrepancies are observed. This is not surprising given that 
GFVT treats polymers as soft colloids, an approximation that only makes 
sense for $q$ small.
Finally, we will consider the generalized model of Fortini {\em et al.},
\cite{FDT-05} which allows one to determine the main features of the 
phase diagram of 
aqueous solutions of charged colloids.

The paper is organized as follows. In Section \ref{sec2} we introduce 
the general ideas that allow us to describe the thermal crossover 
in terms of the two-parameter model variable $z$ (a more extensive 
discussion can be found in Ref.~\onlinecite{DPP-thermal}).
In Section \ref{sec3} we define GFVT and validate its use
against full-monomer results in the homogeneous phase for two values of 
$q$, $q = 1$ and $q = 2$. In Section~\ref{sec4} we determine the phase diagram 
of the system as a function of $q$ and $z$ and carefully compare the
results with previous work. In Section~\ref{sec5} we extend the discussion
to dispersions of charged colloids. Finally, in Section~\ref{sec6}
we present our conclusions. 
In the supplementary material\cite{suppl} we collect the 
relevant formulae for the polymer equation of state and the polymer-colloid
depletion thickness that are used throughout the paper, and provide  extensive 
tables of results. Moreover, we reanalize the experimental data 
presented in Refs.~\onlinecite{Berry-66,NKTF-68,MNF-72,FFKH-74} and 
discuss carefully the effects of polydispersity. In particular, we determine
the nonuniversal constants that allow us to map the experimental data 
onto the two-parameter model expressions we use to parametrize thermodynamic
experimental quantities in the  thermal crossover region.

\section{Thermal crossover for polymer solutions} \label{sec2}

In this paper we consider polymer solutions in 
the thermal crossover regime, which is observed in a
relatively large temperature interval above the $\theta$ temperature $T_\theta$.
\cite{DaoudJannink} For $T\gg T_\theta$ in which interactions are dominated by
the pairwise repulsion (good-solvent regime), polymers are swollen and 
the radius of gyration $R_g$ scales as \cite{Flory,deGennes-72} $L^\nu$, 
where \cite{Clisby-10} $\nu = 0.587597(7)\simeq 3/5$ is 
the Flory exponent. On the other hand, for $T=T_\theta$, polymers behave 
approximately as noninteracting random chains and $R_g\sim L^{1/2}$. 
In the intermediate region
experiments and computer simulations show that,
for sufficiently large values of the degree of polymerization $L$ (i.e.,
for large molar masses $M_w$), thermodynamic and large-scale structural 
properties of the solution obey general relations of the form 
\begin{equation}
\mathcal{O}(T, L,\rho_p) = \alpha_2 \mathcal{O}_G(L,\rho_p) 
f_{\mathcal{O}}(z, \phi_p),
\label{scaling-gen}
\end{equation}
where $\mathcal{O}_G(L,\rho_p)$ is the expression of $\mathcal{O}$ for the 
Gaussian-chain model, the function $f_{\mathcal{O}}$ is a universal---hence
independent of chemical details---{\em crossover} 
function, $\rho_p = N_p/V$ is the number concentration of the polymers, 
$\phi_p = 4 \pi R^3_g \rho_p/3$, $R_g$ is the {\em zero-density} 
radius of gyration, and $\alpha_2$ is a nonuniversal constant that 
embodies all chemical details. The quantity $z$ is the 
Zimm-Stockmayer-Fixmann variable defined in Eqs.~(\ref{ZSF-variable}),
(\ref{ZSF-variable2}), or (\ref{ZSF-variable3}).
By varying $z$ one obtains 
the full crossover behavior, from the $\theta$ region, corresponding to 
small values of $z$, to the good-solvent regime, which is obtained for 
$z\to\infty$.

Theory\cite{Duplantier,dCJ-book,Schaefer-99}
supports the scaling behavior (\ref{scaling-gen}), albeit with a 
slightly different scaling variable. Indeed, the crossover 
limit should be taken by keeping $\alpha_1 (T-T_{\theta})L^{1/2} 
(\ln L)^{-4/11}$ fixed, which differs by a power of $\ln L$ 
from the scaling variable $z \sim (T-T_{\theta})L^{1/2}$.
Such a logarithmic dependence
is irrelevant in most practical applications, since the observation of this 
slowly varying term would require accurate data in a very large interval of 
polymer lengths/molecular weights, both in experiments and in numerical
simulations. Moreover, additive logarithmic corrections, vanishing as 
inverse powers of $1/\ln L$, should also be considered. They are 
expected to be relevant only very close to the $\theta$ point. For instance,
they are responsible for the nonideal density corrections to the (osmotic) 
pressure $P$
at $T_\theta$. Indeed, if the $\theta$ point is defined as the temperature 
at which the second virial coefficient vanishes for each value of $L$,
then we have $P = k_B T \rho_p (1 + a \rho_p^2)$, 
with\cite{dCJ-book,Schaefer-99,PH-05} $a \propto 1/\ln L$.
In this work, we do not consider such terms, assuming that $L$ is so large that 
such corrections are negligible.

It is important to note that the nontrivial universal behavior is 
obtained by taking simultaneously the limits $L\to\infty$,
$T\to T_{\theta}$, and $\rho_p\to 0$ (dilute and semidilute regime) 
in such a way that the arguments $z$ and $\phi_p$ 
of the crossover function $f_{\mathcal{O}}$
remain constant. If the limits are taken differently, one would obtain 
a different result. For instance, in the zero-density case,
if one takes the limit $L\to \infty$ 
at fixed $T>T_\theta$, one would obtain good-solvent behavior in all
cases, while, if one decreases $T$ towards $T_\theta$ at fixed large $L$, 
only the $\theta$ behavior would be observed. Analogously, if one takes 
the limit $L\to \infty$ at fixed $\rho_p$,  one ends up with a melt. 

\begin{figure}[t!]
\centering
\begin{tabular}{c}
\epsfig{file=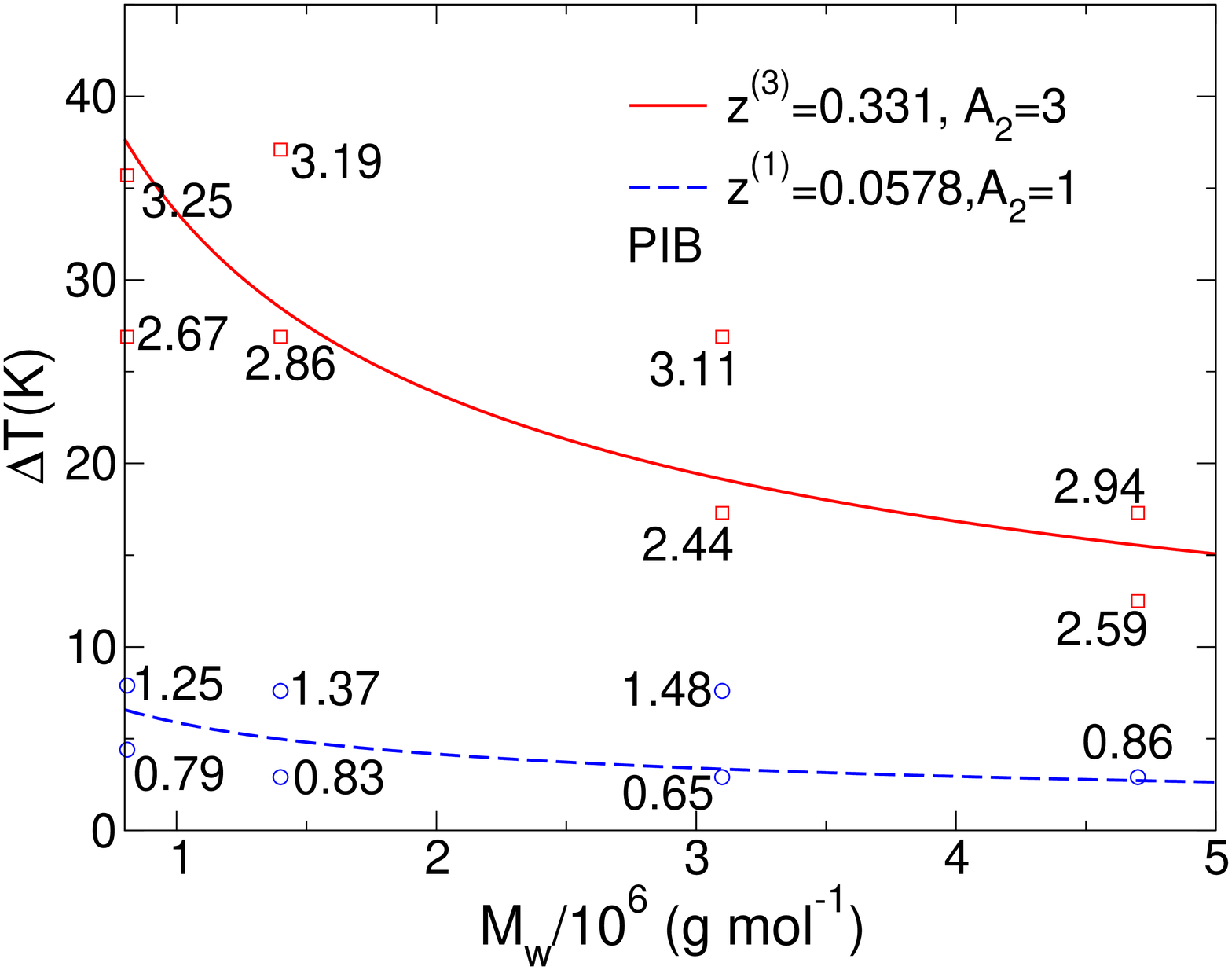,width=6.5cm,angle=0} \\
\epsfig{file=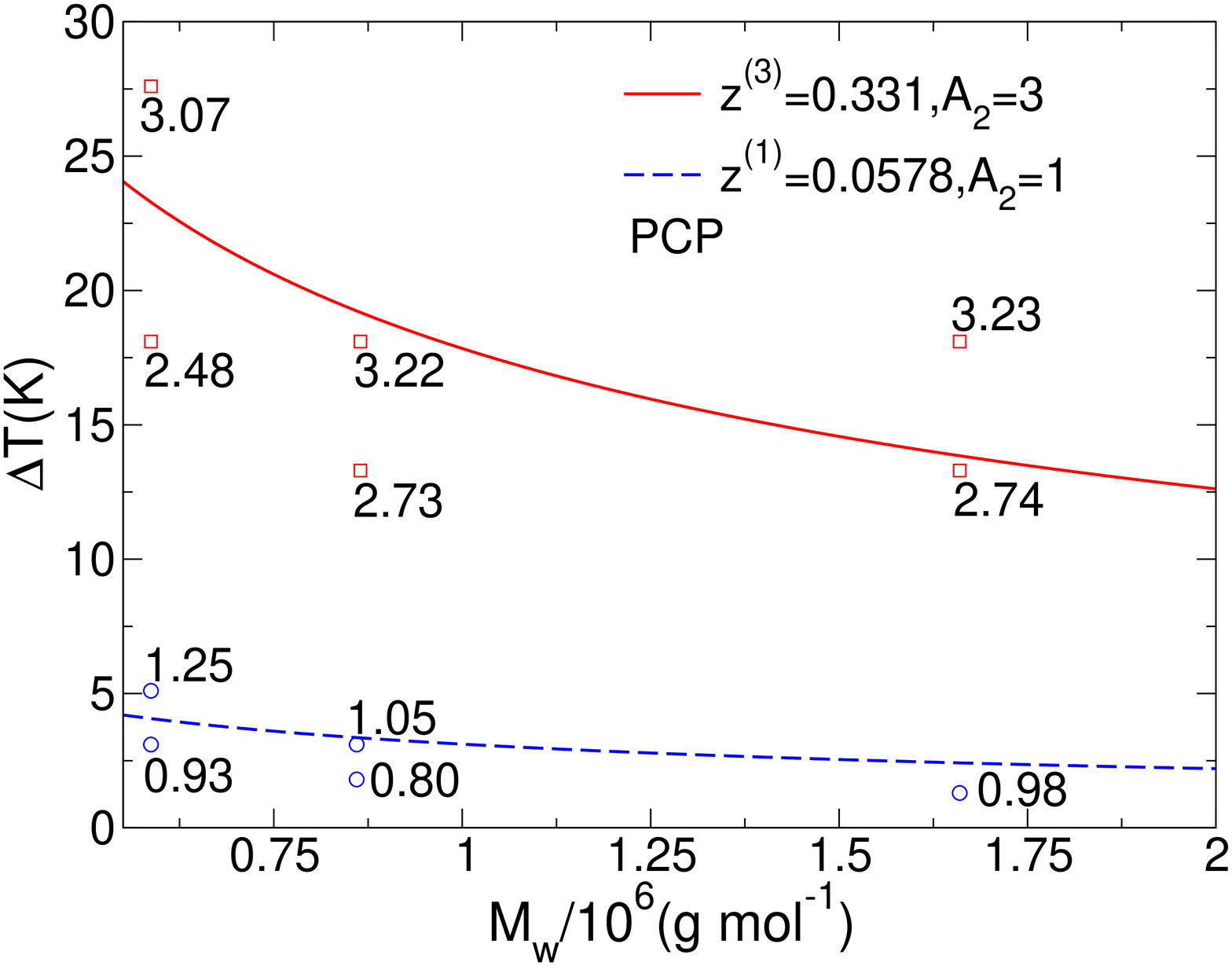,width=6.5cm,angle=0} \\
\epsfig{file=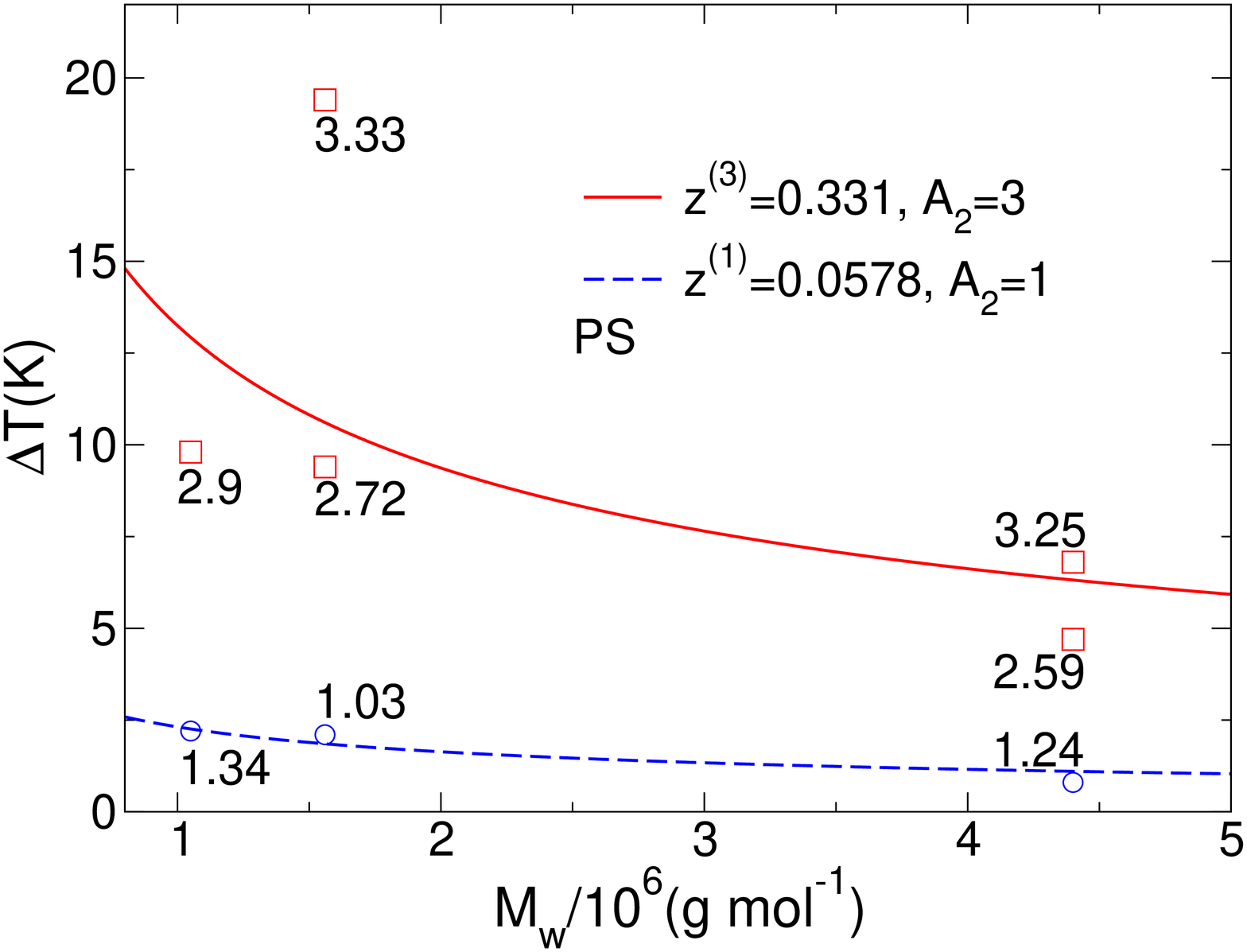,width=6.5cm,angle=0}
\end{tabular}
\caption{Thermal crossover for polyisobutylene (PIB) in
isoamyl-isovalerate (top), polychloroprene (PCP) in transdecalin (middle), and
polystyrene (PS) in decalin (bottom):
Lines $z = z^{(1)}$ (dashed line) and $z = z^{(3)}$ (continuous line) 
in the plane $(\Delta T = T - T_\theta$, $M_w)$. The variable $z$ is defined 
in Eq.~(\ref{ZSF-variable}) and the constant $a^{(1)}$ is estimated in the 
supplementary material.\cite{suppl} 
Empty points correspond to the experimental results,
taken from  Refs.~\onlinecite{Berry-66,NKTF-68,MNF-72}.
The number which is reported close to each point gives the 
corresponding experimental estimate of $A_{2,pp}$.
}
\label{fig:z-expt}
\end{figure}

Experimentally, the quality of the solution is usually determined by measuring 
the swelling ratio $\alpha$ in the zero-concentration limit or 
the second virial combination $A_{2,pp}$ (in the experimental literature,
one often quotes the interpenetration ratio $\Psi = 2 (4\pi)^{-3/2} A_{2,pp}$). 
The first quantity,
\begin{equation}
\alpha(T,L) = {R_g(T,L)\over R_g(T_\theta,L)}
\end{equation}
measures how the radius of gyration in the zero-concentration limit 
changes as $T$ is increased away from
$T_\theta$. It varies from 1 to a number of order $L^{\nu-1/2}$ 
(it tends to infinity in the scaling limit $L\to\infty$) as $T$ increases. 
The combination
$A_{2,pp}$ is defined as 
\begin{equation}
A_{2,pp}(T,L) = {B_{2,pp}(T,L)\over R_g(T,L)^3}, 
\end{equation}
where $B_{2,pp}(T,L)$ is the second virial coefficient\cite{footPsi} 
defined by the expansion of the (osmotic) pressure $P$ of the pure polymer
solution,  
\begin{equation}
P = k_B T \rho_p (1 + B_{2,pp} \rho_p + \ldots). 
\end{equation}
For $L\to\infty$, 
both functions $\alpha(T,L)$ and $A_{2,pp}(T,L)$ converge to 
functions $\alpha(z)$ and $A_{2,pp}(z)$, which are universal 
provided one chooses appropriately the 
constant $a^{(1)}$ appearing in Eq.~(\ref{ZSF-variable}). If the 
other definitions
are used, $a^{(2)}$ or $a^{(3)}$ should be chosen appropriately. 
We will fix these constants 
by requiring that $A_{2,pp}(z) \approx 4 \pi^{3/2} z$,
or equivalently $\Psi \approx z$, for $z\to 0$. For large $z$, 
$A_{2,pp}$ converges to the good-solvent value,\cite{CMP-06}
$A_{2,pp}(z) = 5.50 + O(z^{-\Delta})$, where\cite{Clisby-10} 
$\Delta = 0.528(12)$, while $\alpha(z) \sim z^{2\nu-1}$. 
Because of universality, the $z$ dependence 
of these two quantities can be determined in any model which describes 
the thermal crossover. High-precision numerical simulations of lattice polymer 
models provided an accurate expression for these two functions:\cite{CMP-08}
\begin{eqnarray}
&& A_{2,pp}(z) = 
4 \pi^{3/2} z (1 + 19.1187 z 
\label{A2-to-z}
\\
&& \qquad\quad
    + 126.783 z^2 + 331.99 z^3 + 268.96 z^4)^{-1/4},
\nonumber 
\end{eqnarray}
and 
\begin{equation}
\alpha(z) = (1 + 10.9288 z + 35.1869 z^2 + 30.4463 z^3)^{0.0583867}.
\label{alpha-to-z}
\end{equation}
To determine the $z$ dependence of the phase diagram, we have considered in
detail the behavior for two values of $z$: $z = z^{(1)} = 0.056215$ and
$z = z^{(3)} = 0.321650$, which correspond to\cite{CMP-08}
$A_{2,pp}(z^{(1)}) = 0.9926(10)$ and 
$A_{2,pp}(z^{(3)}) = 2.9621(27)$. Since $A_{2,pp} \approx 5.50$ 
under good-solvent
conditions,\cite{CMP-06} we have $A_{2,pp}(z)/A_{2,pp}(z=\infty) = 0.18$
and 0.54 for $z = z^{(1)}$ and $z = z^{(3)}$, respectively. 
Hence, the solution is quite 
close to the $\theta$ point for $z=z^{(1)}$, while for $z = z^{(3)}$ the
behavior is intermediate between the good-solvent and the $\theta$ regimes. 
In Fig.~\ref{fig:z-expt} we report the lines $z = z^{(1)}$ 
and $z = z^{(3)}$ 
in the $(\Delta T,M_w)$ plane for three experimental systems, 
which are analyzed in detail in the supplementary material.\cite{suppl}
For the typical molar masses used in experiments, systems with 
$z = z^{(1)}$ correspond to temperatures that are only a few degrees above 
$T_\theta$. Significantly larger temperature 
differences occur instead for $z = z^{(3)}$. In this case
the solution is quite far
from $\theta$ conditions.

\section{Generalized free-volume theory} \label{sec3}
 
\subsection{Definitions}  \label{sec3.1}

\begin{figure}
\begin{tabular}{c}
\epsfig{file=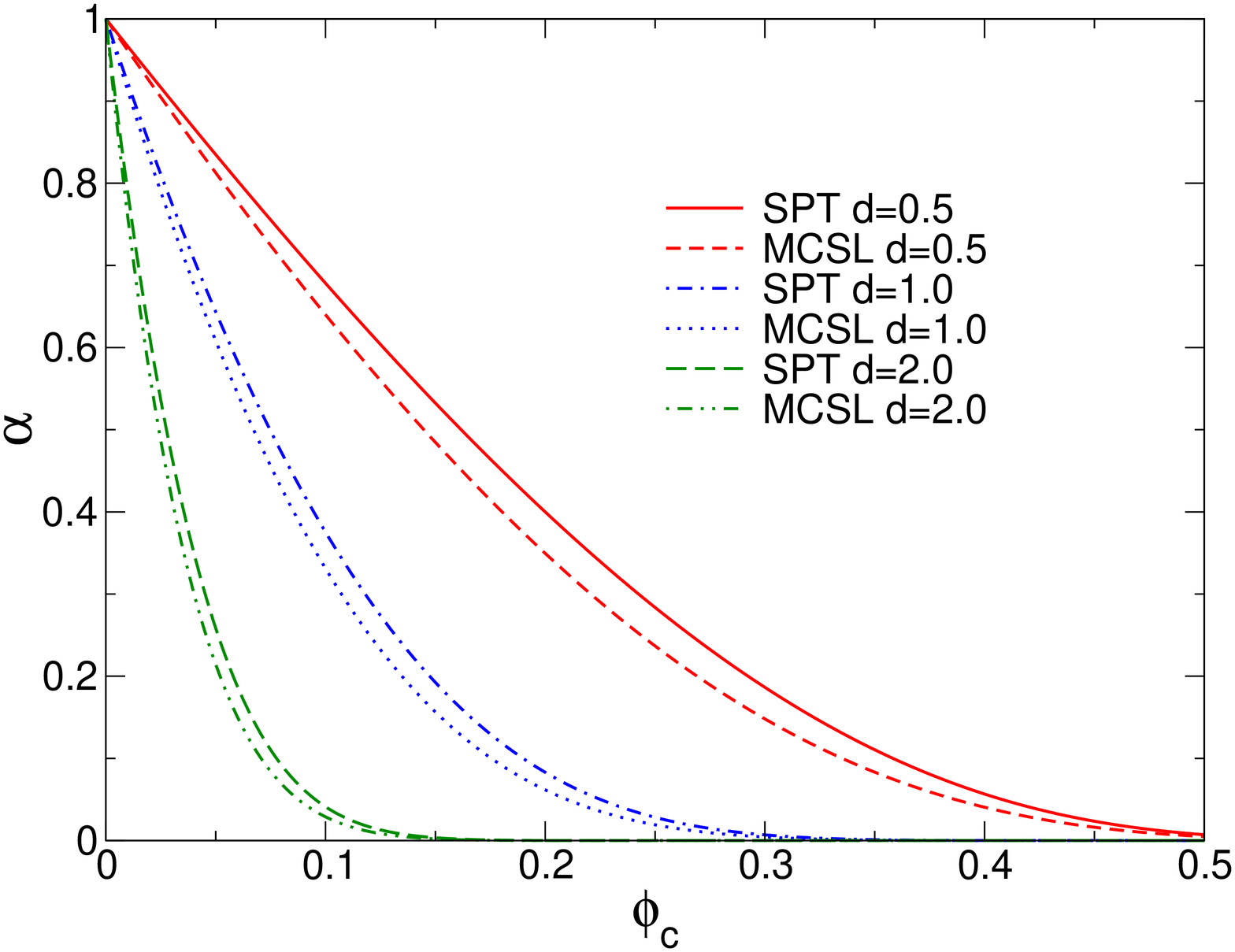,width=8cm,angle=0}  \\
\epsfig{file=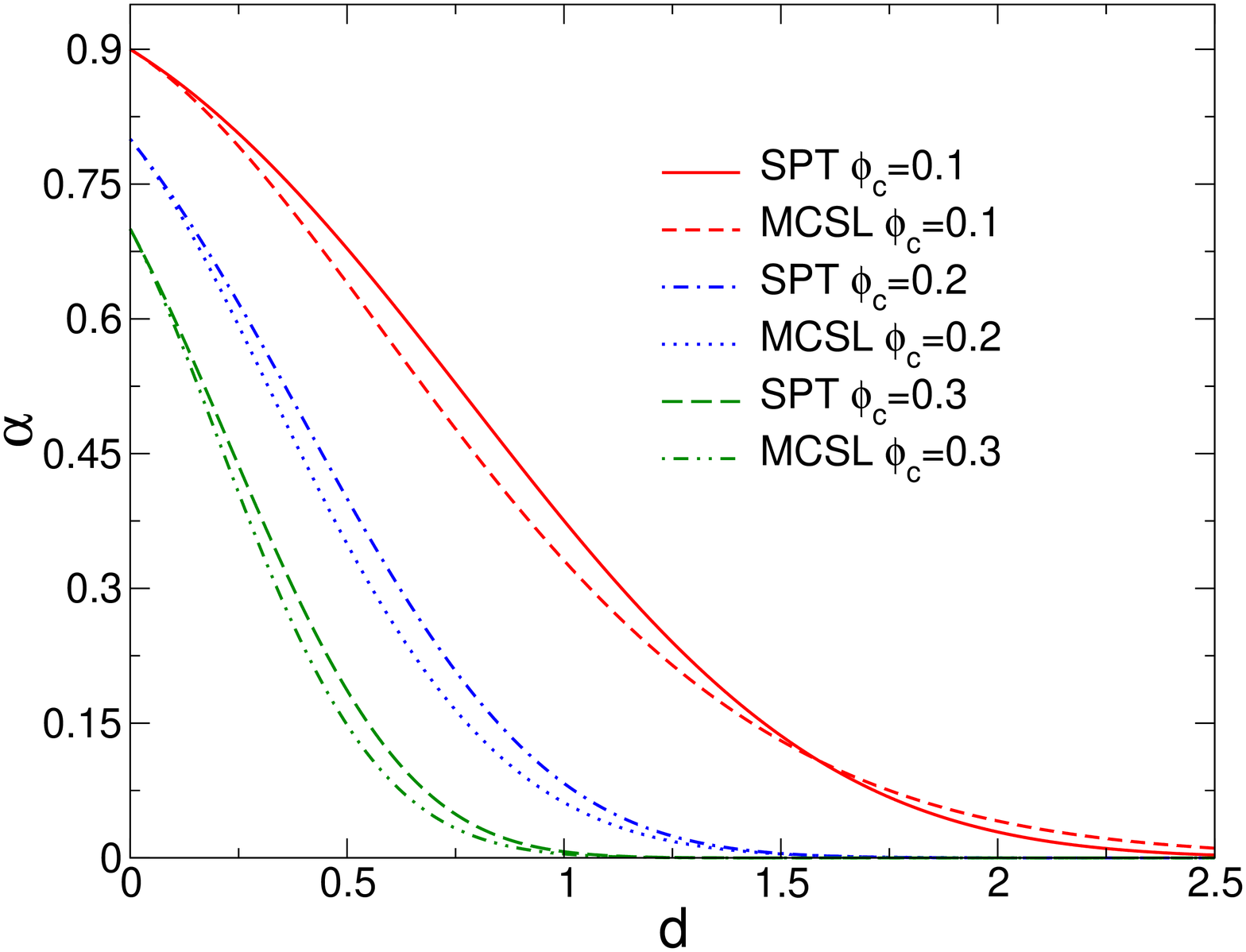,width=8cm,angle=0}  
\end{tabular}
\caption{Plot of $\alpha(\phi_c,d)$ obtained by using SPT and 
by using the Mansouri {\em et al.} equation of state (MCSL).
In the top panel we plot $\alpha(\phi_c,d)$ versus $\phi_c$ for three
different values of $d$, in the lower panel we plot it 
versus $d$ for three values of $\phi_c$. 
}
\label{comparison-alpha}
\end{figure}

We work in the semigrand canonical ensemble, in which the fundamental
thermodynamic variables are the volume $V$, the temperature $T$, 
the number of colloids $N_c$,
and the polymer chemical potential $\mu_p$. In practice, it is more convenient
to use as basic variables the crossover variable $z$ (it will be always 
implicit in the notation), 
the colloid volume fraction $\phi_c$ defined by 
\begin{equation}
\phi_c = V_c {N_c\over V}, 
\end{equation}
where $V_c = 4\pi R_c^3/3$ is the volume of the colloid,
and the reservoir polymer volume fraction $\phi_p^{(r)}$, which is related
to the polymer chemical potential $\mu_p$ by the polymer equation of state 
in the absence of colloids:
\begin{equation}
\beta \mu_p(\phi_p^{(r)}) = \beta \mu_p^{\rm id}(\phi_p^{(r)}) + 
   \int_0^{\phi_p^{(r)} } {K_p(\eta_p) - 1\over \eta_p}d\eta_p.
\end{equation}
Here $\mu_p^{\rm id}$ is the ideal-gas contribution, $K_p(\phi_p)$ is the 
inverse isothermal compressibility, 
\begin{equation}
K_p(\phi_p) = {\partial \beta P\over \partial \rho_p},
\end{equation}
$P$ is the (osmotic) pressure in the absence of colloids,
$\phi_p = V_p \rho_p$, $V_p = 4 \pi R_g^3/3$, $R_g$ is the zero-density polymer 
radius of gyration, and $\beta = 1/k_B T$.
In the GFVT one makes the following ansatz for the semigrand potential 
$\Omega$:\cite{ATL-02} 
\begin{eqnarray}
\omega(\phi_c,\phi_p^{(r)}) &=& {V_c\over V} \beta \Omega = {f}(\phi_c) 
\nonumber \\
&& - 
  {1\over q^3}
 \int_0^{\phi_p^{(r)}} \alpha[\phi_c,\Delta(\eta_p)] K_p(\eta_p) d\eta_p,
\label{GFVT-potential}
\end{eqnarray}
where $f(\phi_c) = V_c\beta F_{\rm coll}/V$,
$F_{\rm coll}$ is the colloid canonical free energy
in the absence of polymers, $\Delta(\eta_p)$ is a function
that will be specified below, and $\alpha(\phi_c,d)$ is the so-called 
free-volume factor for the insertion of a particle of radius 
$R_p = R_c d$ in a colloidal system at volume fraction $\phi_c$. 
For ${f}(\phi_c)$ we use the Carnahan-Starling approximation in the 
fluid phase,\cite{CS-69} 
\begin{equation}
{f}(\phi_c) = \phi_c \log \left({\phi_c \lambda_c^3\over V_c}\right) - 
\phi_c + {4 \phi_c^2 - 3 \phi_c^3\over (1-\phi_c)^2},
\label{CS-f}
\end{equation}
where $\lambda_c$ is the colloid thermal length. 
In the solid phase we use \cite{FL-84,LT-11}
\begin{equation}
{f}(\phi_c) = 2.1178 \phi_c - 
    3 \phi_c \log {1 - \phi_c/\phi_{cp}\over \phi_c} + 
    \phi_c \ln {\lambda_c^3\over V_c},
\label{solid-f}
\end{equation}
where $\phi_{cp} = \pi/(3 \sqrt{2})$. The pressure derivative $K_p(\phi_p)$
is obtained by using the accurate polymer equation of state reported in 
Ref.~\onlinecite{Pelissetto-08} (good-solvent case) and 
in Ref.~\onlinecite{DPP-thermal} (thermal crossover region). They are 
reported for completeness in the supplementary material.\cite{suppl}
For the free-volume factor we use the expression obtained in the 
framework of scaled particle theory (SPT), 
\cite{LPPSW-92,Reiss-92,LT-11} both in the fluid and solid phases:
\begin{equation}
\alpha(\phi_c,d) = (1 - \phi_c) e^{-Q(\phi_c,d)},
\label{alpha-SPT}
\end{equation}
with\cite{footnote-err} 
\begin{equation}
Q(\phi_c,d) = (3 d + 3 d^2 + d^3) y + (9 d^2/2 + 3 d^3) y^2 + 3 d^3 y^3,
\end{equation}
where $y = \phi_c/(1 - \phi_c)$. We have also computed the function 
$\alpha(\phi_c,d)$ from the equation of state for a bidisperse
system of hard spheres of Mansouri {\em et al.},\cite{MCSL-71} obtaining 
\cite{footnote-MCSL}
\begin{equation}
\alpha_{MCSL}(\phi_c,d) = (1 - \phi_c)^{(1-d)^2(1+2d)} e^{-Q_{MCSL}(\phi_c,d)},
\label{alpha-MCSL}
\end{equation}
where
\begin{equation}
Q_{MCSL}(\phi_c,d) = (3 d + 6 d^2 - d^3) y + (3 d^2 + 4 d^3) y^2 + 2 d^3 y^3.
\end{equation}
The two expressions (\ref{alpha-SPT}) and (\ref{alpha-MCSL}) 
are equivalent for $d\ll 1$ ($\phi_c$ arbitrary) 
(in this limit we have 
$\alpha \approx \alpha_{MCSL} \approx (1 - \phi_c) e^{-3 d y}$), 
for $1\ll d \ll \phi_c^{-1/3}$ 
($\alpha \approx \alpha_{MCSL} \approx e^{-d^3 \phi_c}$), and for 
$\phi_c$ small ($d \ll \phi_c^{-1/3}$)
($\alpha \approx \alpha_{MCSL} \approx 1 - (1 + d)^3 \phi_c$).
In general, see Fig.~\ref{comparison-alpha}, differences are tiny 
in the relevant region in which $\alpha(\phi_c,d)$ is not too small.
For $d \lesssim 2$ 
and $\phi_c\lesssim 0.6$, differences are small in all cases, making the two
expressions equivalent for our purposes. In the following we use
Eq.~(\ref{alpha-SPT}). For comparison, some calculations will be repeated using
the more accurate expression (\ref{alpha-MCSL}).

Finally, we should specify the function $\Delta(\phi_p)$, which gives 
the ratio of the effective size of the polymer and the radius of the colloid.
We will define it as 
$\Delta(\phi_p) = \delta_s(q,\phi_p)/R_c$, where $\delta_s(q,\phi_p)$ is 
the depletion thickness as defined in Refs.~\onlinecite{FT-08,LT-11}.
For such a quantity we use the 
accurate expressions determined in Ref.~\onlinecite{DPP-depletion} and 
reported for completeness in the supplementary material.\cite{suppl}

Once $\omega$ is known, the pressure and the colloid chemical potential
can be computed by using 
\begin{equation}
\beta P V_c = \phi_c \left({\partial {\omega} \over \partial \phi_c}
   \right)_{\phi_p^{(r)}} - {\omega} \qquad\qquad
\beta \mu_c = V_c \left({\partial {\omega} \over \partial \phi_c}
   \right)_{\phi_p^{(r)}}.
\end{equation}
The polymer volume fraction $\phi_p$ can be derived by using 
\begin{eqnarray}
\phi_p = V_p \rho_p &=& 
     - q^3 {\phi^{(r)}_p \over K_p(\phi^{(r)}_p)}
       {\partial \omega \over \partial \phi^{(r)}_p} 
\nonumber \\
    &=& \alpha[\phi_c,\Delta(\phi^{(r)}_p)] \phi^{(r)}_p.
\label{phip-GFVT}
\end{eqnarray}

\subsection{Comparison with full-monomer data} \label{sec3.2}

By construction GFVT  provides the exact semigrand potential on the polymer
($\phi_c = 0$) and colloid ($\phi_p^{(r)} = 0$) axis, provided one uses 
the exact expression for $F_{\rm coll}$ and $K_p(\phi_p)$. 
Moreover, if one uses an accurate 
estimate of $\delta_s(q,\phi_p)$ (we remind the reader that the knowledge of 
$\delta_s(q,\phi_p)$ 
is equivalent to the knowledge of the insertion free energy 
of a single colloid in the pure polymer solution), 
the potential $\omega(\phi_c,\phi_p^{(r)})$ 
is also accurate close to the polymer axis to first order in $\phi_c$. 
We wish now to study the accuracy of the approximation in the
$(\phi_c,\phi_p)$ plane, below the phase separation line.

Let us first consider the dilute limit, in which the 
polymer and colloid densities $\rho_p$ and $\rho_c$ 
are both small. In this regime the pressure can be expanded as 
\begin{eqnarray}
&& \beta P = \rho_c + \rho_p + B_{2,cc} \rho_c^2 + B_{2,pp} \rho_p^2 +
        B_{2,cp} \rho_c \rho_p
\label{virial-expansion} \\
        && \quad + B_{3,ccc} \rho_c^3 + B_{3,ppp} \rho_p^3 +
        B_{3,ccp} \rho_c^2 \rho_p + B_{3,cpp} \rho_c \rho_p^2 +
        \ldots
\nonumber 
\end{eqnarray}
The virial coefficients are not universal as they depend on the details of 
the system. On the other hand, the combinations
\begin{equation}
A_{2,\#} = B_{2,\#} R_g^{-3} \qquad 
A_{3,\#} = B_{3,\#} R_g^{-6} 
\end{equation}
are model independent in the limit of large degree of polymerization. 
These combinations have been determined quite accurately in
Refs.~\onlinecite{CMP-06,CMP-08,DPP-depletion}. 
If we start from the GFVT semigrand potential
(\ref{GFVT-potential}) we obtain a low-density expansion analogous to 
Eq.~(\ref{virial-expansion}), with corresponding coefficients 
$A_{2,\#}^{(GFVT)}$ and $A_{3,\#}^{(GFVT)}$. If we use accurate estimates 
for the colloid free energy density $f(\phi_c)$, for the derivative 
$K_p(\phi_p^{(r)})$, and for the depletion thickness, we have 
$A_{2,\#}^{(GFVT)} \approx A_{2,\#}$ and 
\begin{eqnarray}
&& A_{3,ppp}^{(GFVT)} \approx A_{3,ppp} \qquad
A_{3,ppc}^{(GFVT)} \approx A_{3,ppc} \qquad \nonumber \\
&& A_{3,ccc}^{(GFVT)} \approx A_{3,ccc}.
\end{eqnarray}
The only virial combination for which the approximate equality does not hold
is $A_{3,ccp}$, for which we obtain (see Appendix)
\begin{eqnarray}
A_{3,ccp}^{(GFVT)}&=& {32 \pi A_{2,cp}^{(GFVT)}\over 3 q^3} + 
     {32 \pi^2\over 9 q^6} \nonumber \\
&& - 9 \left[A_{2,cp}^{(GFVT)}\right]^{2/3} 
\left({4\pi\over 3 q^3}\right)^{4/3}.
\label{A3ccp-GFVT}
\end{eqnarray}
Since we use the accurate expression of the depletion thickness 
of Ref.~\onlinecite{DPP-depletion}, $A_{2,cp}^{(GFVT)} \approx A_{2,cp}$, 
hence Eq.~(\ref{A3ccp-GFVT}) gives us a prediction which can be compared with 
the full-monomer result. For $q\to 0$, we have 
$A_{2,cp} \approx 4\pi/(3 q^3)$ for any value of $z$, so that 
$A_{3,ccp}^{(GFVT)} \approx 16\pi^2/(9 q^6)$ in the same limit,
which is indeed the correct small-$q$ limiting behavior.\cite{DPP-depletion}
For large values of $q$, we have 
$A_{3,ccp}^{(GFVT)} q^3/A_{2,cp}^{(GFVT)}\approx 32 \pi/3  \approx 34$,
which should be compared with the numerical result
$A_{3,ccp} q^3/A_{2,cp} \approx 17$, 21, 15 (these estimates are obtained
by using the interpolations reported in Ref.~\onlinecite{DPP-depletion})
for $z=\infty$, $z^{(3)}$, and $z^{(1)}$.
These results indicate that $A_{3,ccp}$ is quite precisely reproduced for 
$q\to 0$, while discrepancies are expected for large $q$. 
We can also make a more detailed check for 
$q = 0.5, 1$ and $2$, comparing the Monte Carlo estimates\cite{DPP-depletion}
 of $A_{3,ccp}$ and the prediction $A_{3,ccp}^{(GFVT)}$ obtained by
using the Monte Carlo estimate\cite{DPP-depletion} of $A_{2,cp}$.
The results are reported in Table~\ref{A3ccp-comparison}. GFVT
overstimates $A_{3,ccp}$ by 5\%, 13\%, 23-30\% for $q=0.5$, 1, and 2.
Clearly, the approximation works very well for small values of 
$q$ and worsens somewhat as $q$ increases. Apparently, the accuracy is also
higher for good-solvent conditions than close to the $\theta$ point.

\begin{table}
\caption{Estimates of the third virial combination $A_{3,ccp}$ obtained 
by using GFVT and full-monomer (FM) simulations.}
\label{A3ccp-comparison}
\begin{tabular}{ccccccc}
\hline\hline
 & \multicolumn{2}{c}{$q = 0.5$} 
 & \multicolumn{2}{c}{$q = 1$} 
 & \multicolumn{2}{c}{$q = 2$} \\
$z$ & FM & GFVT & FM & GFVT & FM & GFVT \\
\hline
$\infty$ & 8630(45) & 9056 & 360(2) & 403 & 16.8(1) & 20.7 \\
$z^{(3)}$& 9300(30) & 9788 & 399(1) & 451 & 19.2(1) & 24.2 \\
$z^{(1)}$& 9500(30) &10025 & 409(1) & 466 & 19.6(1) & 25.4 \\
\hline \hline
\end{tabular}
\end{table}

\begin{table}
\caption{Comparison of the isothermal compressibility. We report 
$\beta R_c^3/\kappa_T$ for $L=600$ DJ walks in a volume 
$V=256^3$ ($L=600$,FM), 
the corresponding infinite-volume scaling ($L,V\to\infty$)
extrapolation (FM,scal), and the GFVT prediction.}
\label{kappaT-FM}
\begin{tabular}{clllccc}
\hline\hline
$z$ & $q$ & $\phi_c$ & $\phi_p$ & $L=600$,FM & FM,scal & GFVT \\
\hline
$\infty$ & 1   & 0.1   & 0.6    & 1.05(1) &       & 1.056 \\
    && 0.2   & 0.2    & 0.695(2)&  0.73(1) & 0.666 \\
    && 0.3   & 0.2    & 1.772(6)&       & 1.625 \\
& 2   & 0.1   & 1.0    & 0.481(5)&       & 0.469 \\
     && 0.2   & 0.8    & 0.946(9)&       & 0.892 \\
     && 0.3   & 0.2    & 1.063(3)&       & 0.954 \\
$z_3$ & 1 & 0.3 & 0.05 & 0.965(3)& 0.98(1) & 0.902 \\
      & 2 & 0.3 & 0.1  & 0.810(10)& 0.79(1) & 0.853 \\
\hline\hline
\end{tabular}
\end{table}

To check the accuracy of the GFVT predictions at finite densities,
we have performed finite-density simulations of a lattice polymer-colloid 
system in a finite box of volume $V$. 
Polymers are modelled as Domb-Joyce \cite{DJ-72} walks of length $L$ on 
a cubic lattice, 
while colloids are modelled as hard spheres of radius $R_c$, whose 
centers lie in the continuum space. This model, already 
discussed in D'Adamo {\em et al.},\cite{DPP-depletion} allows us 
to study both the good-solvent regime ($z=+\infty$) and the thermal crossover
region. 
We compare GFVT and full-monomer results for the adimensional combination 
$\beta R_c^3/\kappa_T$, where $\kappa_T$ is the isothermal compressibility
\begin{equation}
\kappa_T = - {1\over V} \left( {\partial V\over \partial p}\right)_{N_c,N_p}.
\end{equation}
Such a quantity can be expressed \cite{KB-51,BenNaim}
in terms of the zero-momentum partial structure 
factors, quantities that can be easily computed in simulations 
(see Ref.~\onlinecite{DPP-depletion} for a discussion). 
To obtain a universal, i.e., model independent, 
prediction, the numerical results must be carefully extrapolated. 
First, since simulations are performed in a finite box [typically we take 
$V\approx (15 R_g)^3$], we must determine the infinite-volume limit. 
Second, we must perform an extrapolation in the length $L$ of the chains,
taking into account that the leading corrections behave as $L^{-\nu}$,
where $\nu = 0.588$ in the good-solvent case and $\nu = 1/2$ in the crossover
region (see the supplementary material of Ref.~\onlinecite{DPP-depletion}
for a careful check of this behavior). 
To perform these two extrapolations\cite{foot-extrapola}
one must obtain results for several values of $L$ and $V$, which is 
very CPU-time consuming. We have therefore performed the full analysis only
in a few cases. We find that the results 
corresponding to $L=600$ and $V = 256^3$
differ at most by 10\% from the asymptotic ones, a not too large
systematic deviation (note that this conclusion is valid only for our model,
since scaling corrections are not universal, i.e., they are 
system dependent). 
Therefore, it is possible to perform a
meaningful comparison with the GFVT data even by using the $L=600$ data.
The results for a few selected points (they have been chosen to lie close
to the GFVT binodal) are reported in Table~\ref{kappaT-FM}. It is evident that 
the approximation works better for $\phi_c$ small. Quantitatively,
deviations are less than 10\% in all cases, confirming the relative accuracy of
the approximation in the homogeneous phase for $q\lesssim 2$.

\section{Polymer-colloid phase diagram} \label{sec4}

\subsection{GFVT results} \label{sec4.1}

\begin{figure*}[t]
\begin{center}
\begin{tabular}{c}
\epsfig{file=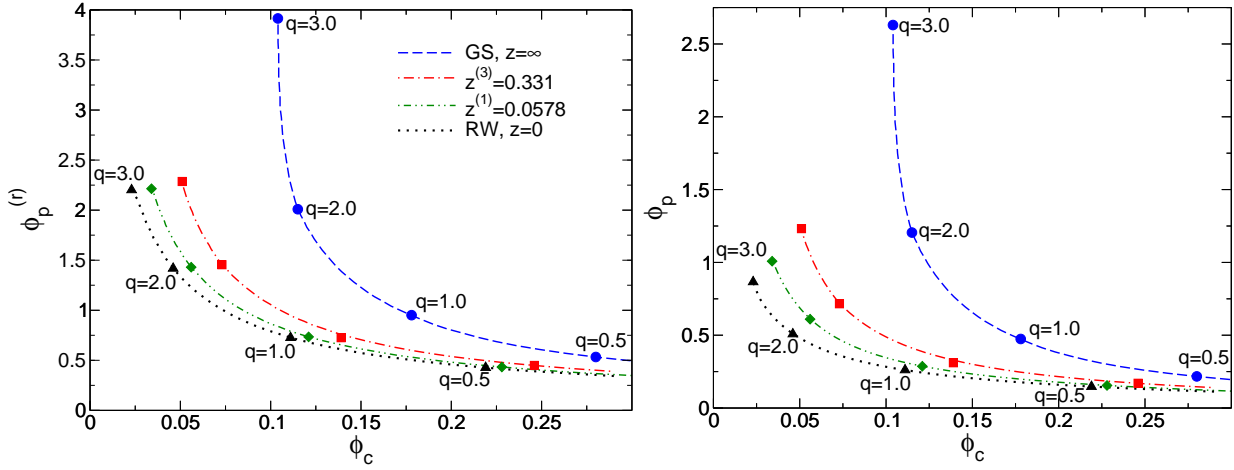,angle=0,width=16.5truecm} \hspace{0.5truecm} \\
\end{tabular}
\end{center}
\caption{Critical-point positions in the $\phi_c,\phi_p^{(r)}$ (left)
and $\phi_c,\phi_p$ (right) planes as $q$ varies between $0.4$ and 3. 
We explicitly report the positions of the critical points corresponding to 
$q=0.5$,1,2,3.
From top to bottom, the four curves correspond to
$z = \infty$ (good-solvent case), $z = z_3$, $z= z_1$, and $z = 0$ 
($\theta$ conditions).
}
\label{CP}
\end{figure*}

\begin{figure}[t]
\begin{center}
\begin{tabular}{c}
\epsfig{file=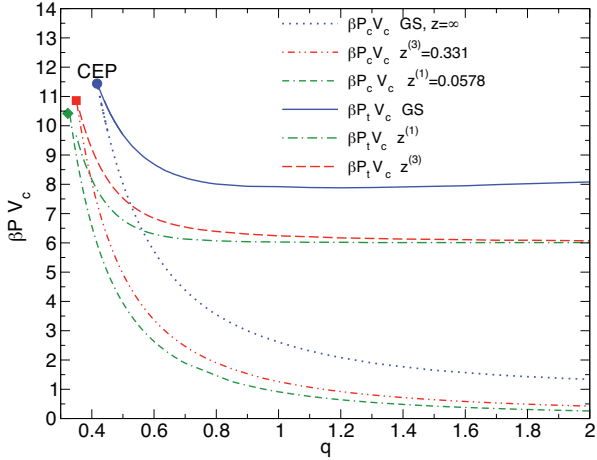,angle=0,width=8truecm} \hspace{0.5truecm} \\
\end{tabular}
\end{center}
\caption{Critical-point ($P_c$) and triple-point ($P_t$) pressure. We report 
$\widetilde{P} = \beta P V_c$ 
as a function of $q$ up to the critical endpoint (CEP) for 
good-solvent (GS) conditions, $z = z^{(3)}$, and $z = z^{(1)}$.
}
\label{CP-press}
\end{figure}

We wish now to use GFVT to determine the phase diagram for the colloid-polymer
system. The colloid volume fractions $\phi_{c1}$ and $\phi_{c2}$ of the 
coexisting phases are obtained by equating pressure and colloid chemical
potential at fixed $\phi_p^{(r)}$:
\begin{equation}
P(\phi_{c1},\phi_p^{(r)}) = 
P(\phi_{c2},\phi_p^{(r)}), \quad  
\mu_c(\phi_{c1},\phi_p^{(r)}) = 
\mu_c(\phi_{c2},\phi_p^{(r)}).
\end{equation}
As is well known,\cite{LT-11} for small values of $q$ only a solid-fluid
transition occurs between a polymer-rich phase in which the colloids 
are disordered and a colloid crystal phase with a very low density of polymers. 
As $q$ increases, the phase diagram becomes more complex. For $q$ 
larger than the critical endpoint value $q_{CEP}$,
beside the solid-liquid transition,
the system undergoes a 
fluid transition (analogous to the liquid-gas transition in
simple fluids) between a polymer-rich colloid-poor phase and a polymer-poor
colloid-rich phase. The fluid transition is characterized by a critical point,
which belongs to the Ising universality class. Moreover, the different phases
merge at a triple point, with three coexisting phases. The critical endpoint
value $q_{CEP}$ is characterized by the fact that the triple point merges 
with the critical one.

We will now use GFVT to identify the binodals, the triple and the critical 
points, as a function of $q$ and for different solvent quality. 
For the good-solvent case and for noninteracting polymers results have already
been presented in Ref.~\onlinecite{ATL-02,FT-08,LT-11}. We will repeat here the
same calculation using our more precise expressions for the polymer equation
of state and depletion thickness. We anticipate here that differences are
small for $q\lesssim 2$, but significantly increase for large values of $q$,
since the phenomenological expression for
$\delta_s(q,\phi_p)$ of Ref.~\onlinecite{FT-08} 
becomes inaccurate as $q$ increases beyond 4.\cite{DPP-depletion}
We anticipate that such a discrepancy is not very relevant, since GFVT
turns out to be not predictive in the protein regime $q > 1$.

We begin by computing the critical endpoint 
values $q_{CEP}$ for different values of 
$z$.  We obtain $q_{CEP} \approx 0.42, 0.35, 0.32, 0.31$ for 
$z = \infty$, $z_3$, $z_1$, and for $z = 0$, respectively. 
Clearly, $q_{CEP}$ depends somewhat on solvent quality---it decreases
as one approaches the $\theta$ point---but the change is relatively small. 
If one uses the expressions of $K$ and $\delta_s$ reported in 
Ref.~\onlinecite{FST-07,FT-08,LT-11} (see supplementary material\cite{suppl}),
one would obtain $q_{CEP} \approx 0.39$ for the 
good-solvent case. As anticipated the difference is small.

\begin{figure*}[t]
\begin{center}
\begin{tabular}{c}
\epsfig{file=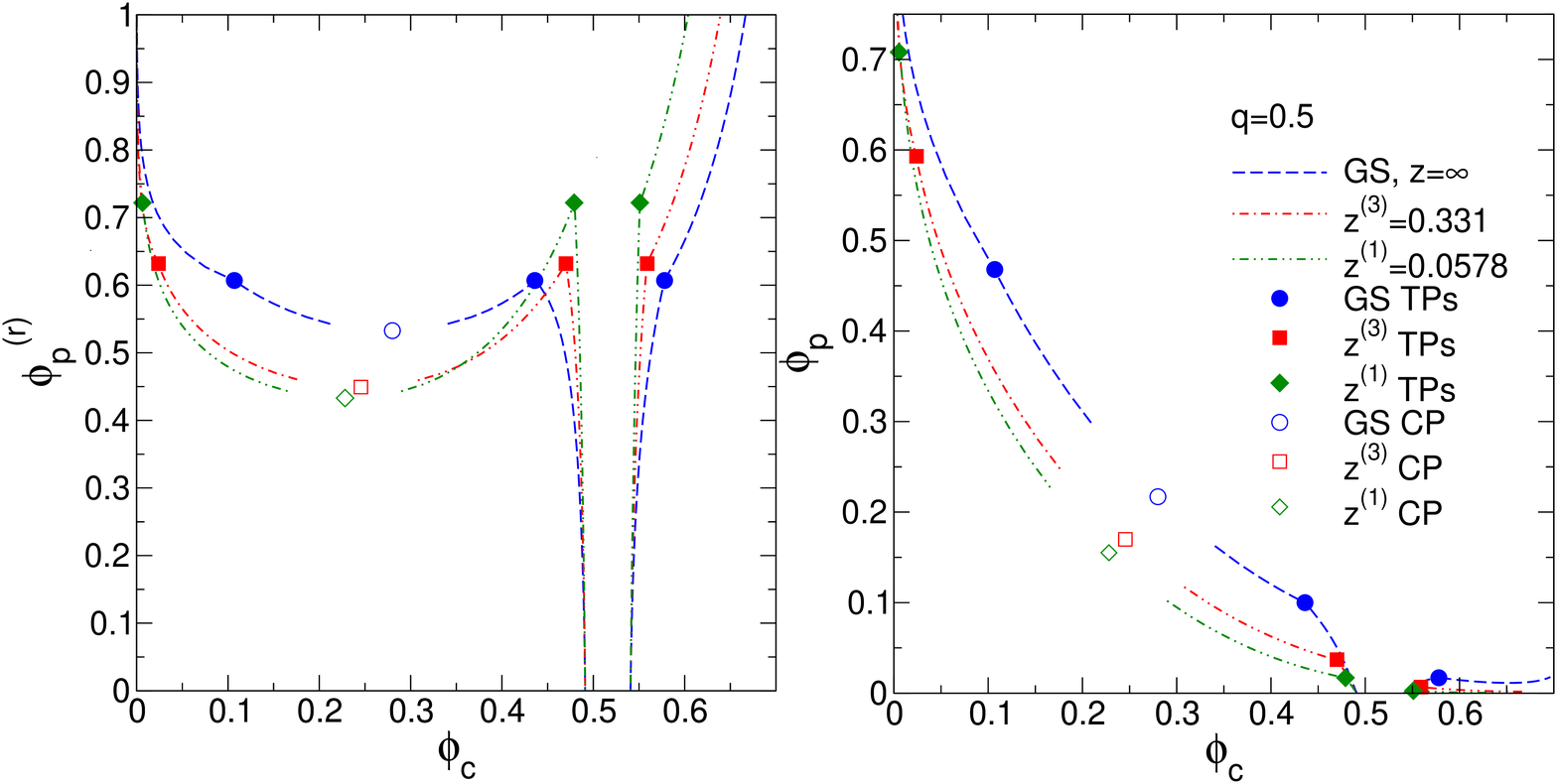,angle=0,width=12truecm} \hspace{0.5truecm} \\
\epsfig{file=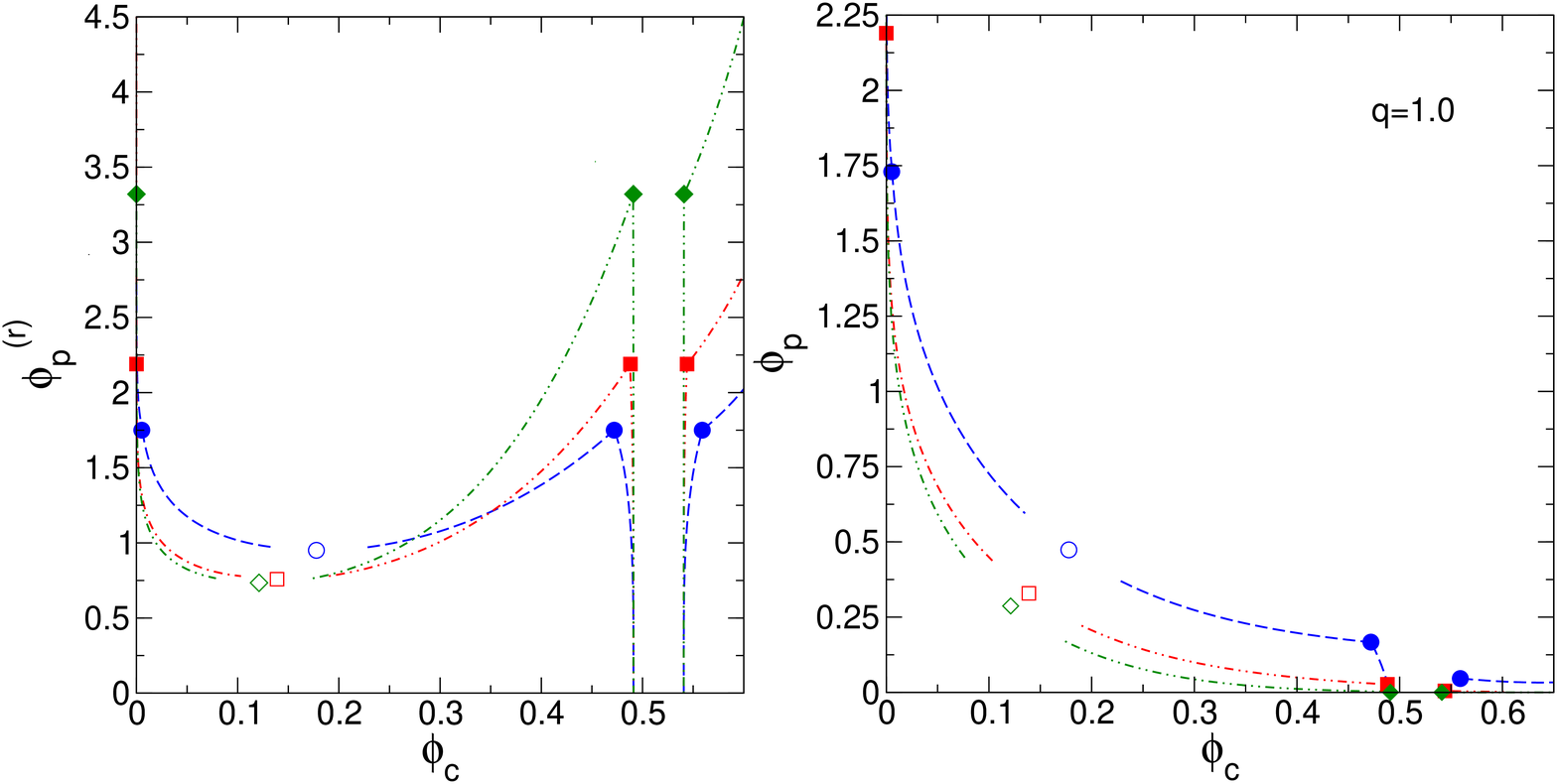,angle=0,width=12truecm} \hspace{0.5truecm} \\
\epsfig{file=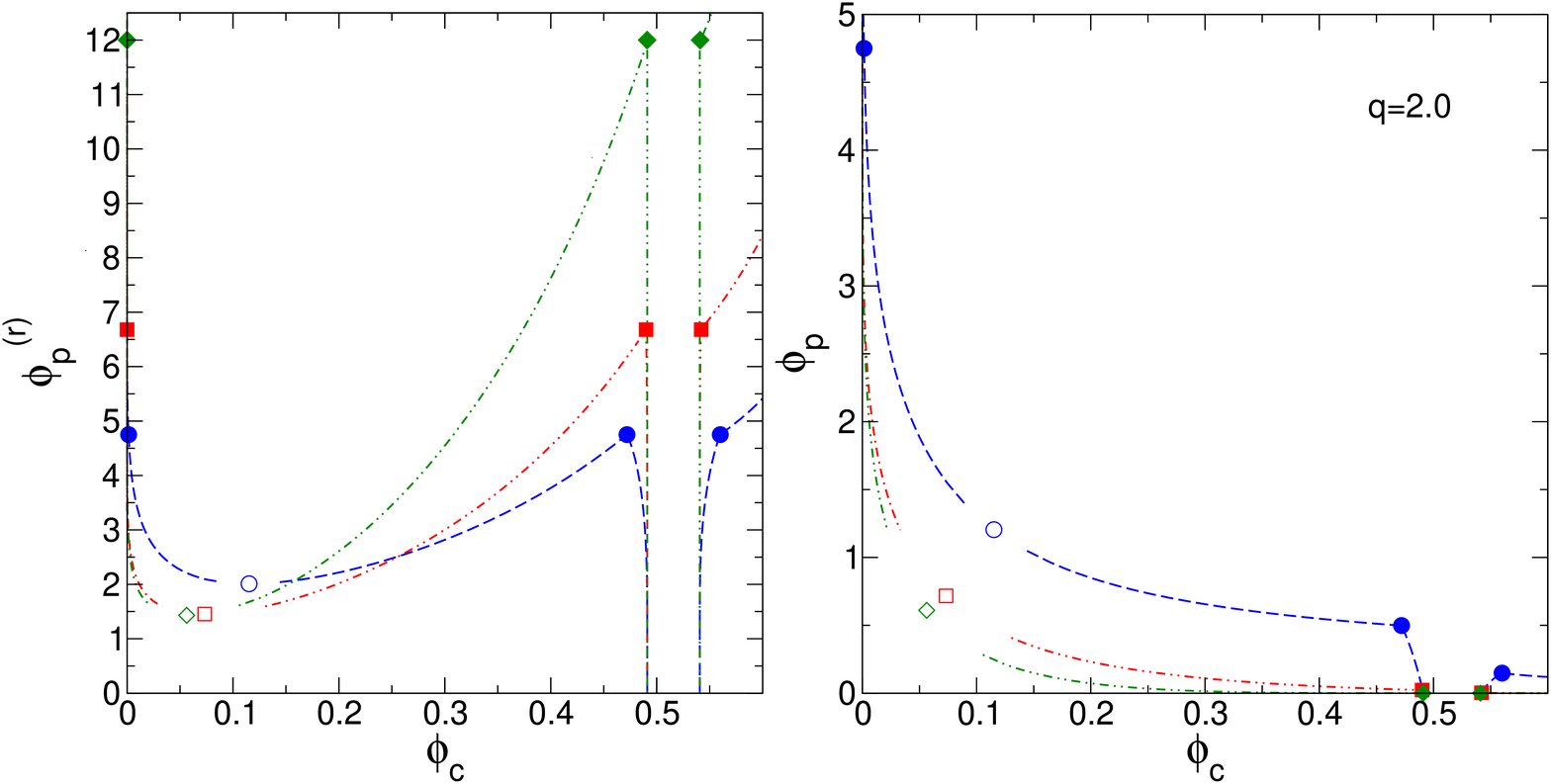,angle=0,width=12truecm} \hspace{0.5truecm} \\
\end{tabular}
\end{center}
\caption{Binodals as a function of $\phi_c$ and $\phi_p^{(r)}$ (left) and 
of $\phi_c$ and $\phi_p$ (right). We report data for $q=0.5$ (top),
$q = 1$ (middle), $q = 2$ (bottom). Empty symbols correspond to the critical 
point (CP), filled symbols label the different phases 
coexisting at the triple point (TPs).
}
\label{Binodals}
\end{figure*}

In Fig.~\ref{CP} we report the position of the critical points as $q$ varies
(an extensive table of results is given in the supplementary
material\cite{suppl}),
while in Fig.~\ref{CP-press} we report the reduced pressure $\widetilde{P} = 
\beta P V_c$, at the critical and at the triple point, as a function of 
$q$. From the results it is evident that solvent quality is an 
important variable. If $z$ increases, both critical volume fractions 
$\phi_{c,\rm crit}$ and $\phi_{p,\rm crit}$ increase significantly. For
$q\gtrsim 2$, $\phi_{c,\rm crit}$ and $\phi_{p,\rm crit}$ change approximately
by a factor of 2 as $z$ varies between 0 and $\infty$. Note that the change 
of the critical point from $z = z^{(3)}$ to the good-solvent regime (which
correspond to a change of the second-virial combination
$A_{2,pp}$ from 3.0 to 5.5) is larger than that between the 
$\theta$ point and $z = z^{(3)}$
(correspondingly, $A_{2,pp}$ varies between 0 and 3), 
an indication that phase coexistence is
more sensitive to small deviations from the good-solvent regime than to
deviations from $\theta$ behavior. This is also evident from the binodals
reported in Fig.~\ref{Binodals}. For the considered values of $q$, the 
binodals in the $(\phi_c,\phi_p)$ plane  
for $z = z^{(1)}$ and $z = z^{(3)}$ are very close and significantly
lower than the good-solvent ones. We also report the position of the triple
points (extensive tables of numerical data are reported in the supplementary
material \cite{suppl}), whose position varies significantly with $z$. 
In particular, the
polymer volume fractions $\phi_{pl}$ and $\phi_{ps}$ in the coexisting
liquid and solid phases are very small for $z = z^{(3)}$ and $z = z^{(1)}$
and typically an order of magnitude smaller than in the good-solvent
regime. Moreover, the $q$ dependence of these quantities is qualitatively
different in the two cases.
In the good-solvent case $\phi_{pl}$ and $\phi_{ps}$ 
increase with increasing $q$ (for instance, $\phi_{pl} = 0.167$,
0.498 for $q=1,2$ respectively), while in the thermal crossover region the 
opposite occurs.  For instance, for $z = z^{(3)}$ we have 
$\phi_{pl} = 0.027, 0.023$ for $q = 1,2$ respectively. 
We have also determined the pressure. Its value at 
the critical and triple points is reported in Fig.~\ref{CP-press}.
At the critical endpoint (CEP) the pressure
has a small dependence on solvent quality. As $q$ increases,
the relative difference between the results for the good-solvent case and 
for $z = z^{(1)}$, $z^{(3)}$, also increases. 
For instance, for $q = 2$ we have 
$\widetilde{P} = \beta P_{\rm crit} V_c = 1.34$
for good-solvent conditions, and $\widetilde{P} =0.26$ for $z = z^{(1)}$. 
On the other hand,
results for $z = z^{(3)}$ and $z = z^{(1)}$ are always very close.

The results we have discussed have been obtained by using the 
SPT expression (\ref{alpha-SPT}) for the free-volume factor. 
The same calculations can be repeated by using Eq.~(\ref{alpha-MCSL}). 
Differences are tiny in all cases. For instance, for $q = 1$ and good-solvent
polymers, we obtain 
$\phi_{p,\rm crit}^{(r)} = 0.951$, $\phi_{c,\rm crit} = 0.178$, 
and $\phi_{p,\rm crit} = 0.474$ using the SPT expression and 
$\phi_{p,\rm crit}^{(r)} = 0.942$, $\phi_{c,\rm crit} = 0.173$, 
and $\phi_{p,\rm crit} = 0.479$ using Eq.~(\ref{alpha-MCSL}).
Clearly, the two expressions are equivalent for our purposes.

\begin{figure}[t]
\begin{center}
\begin{tabular}{c}
\epsfig{file=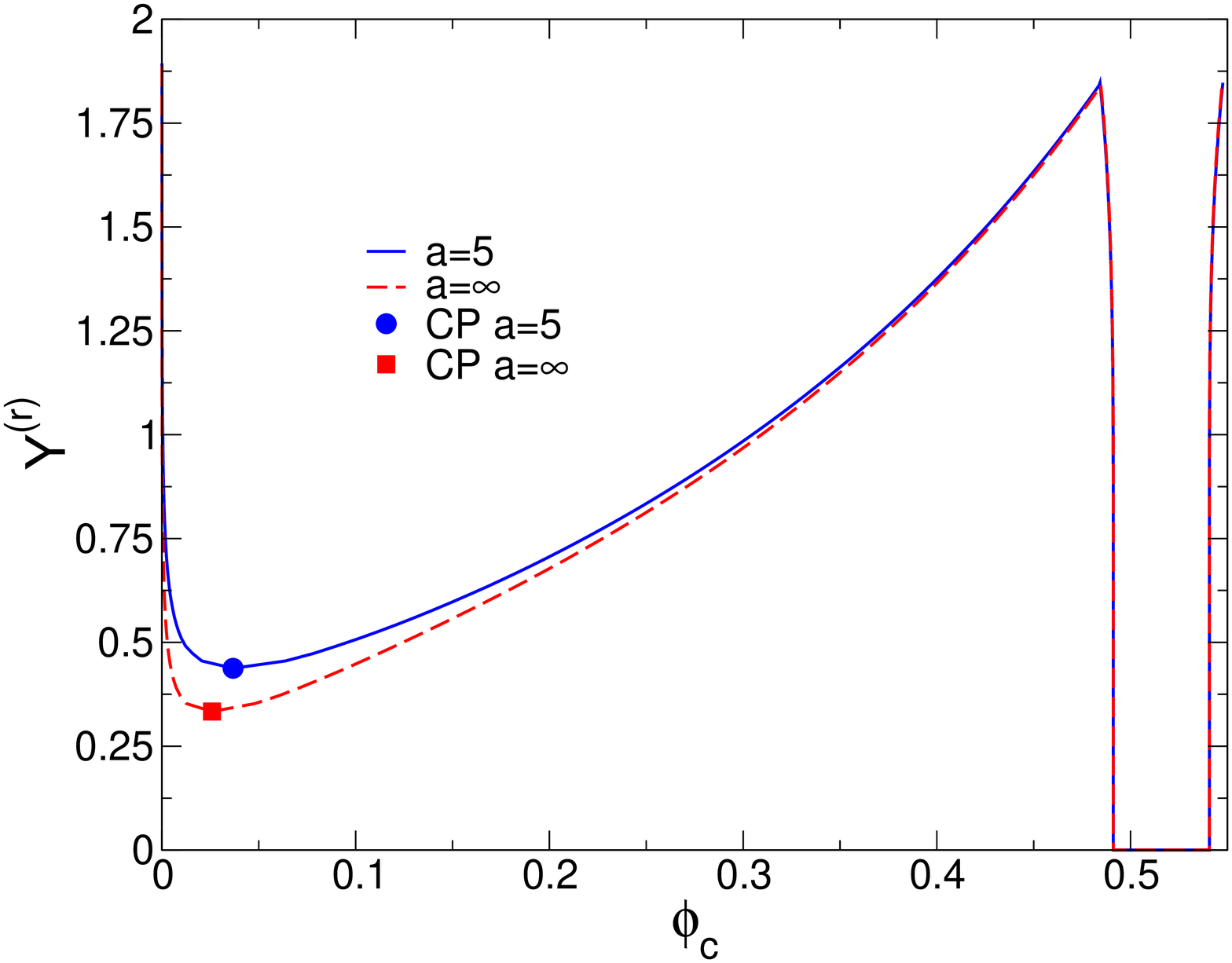,angle=0,width=7truecm} \hspace{0.5truecm} \\
\epsfig{file=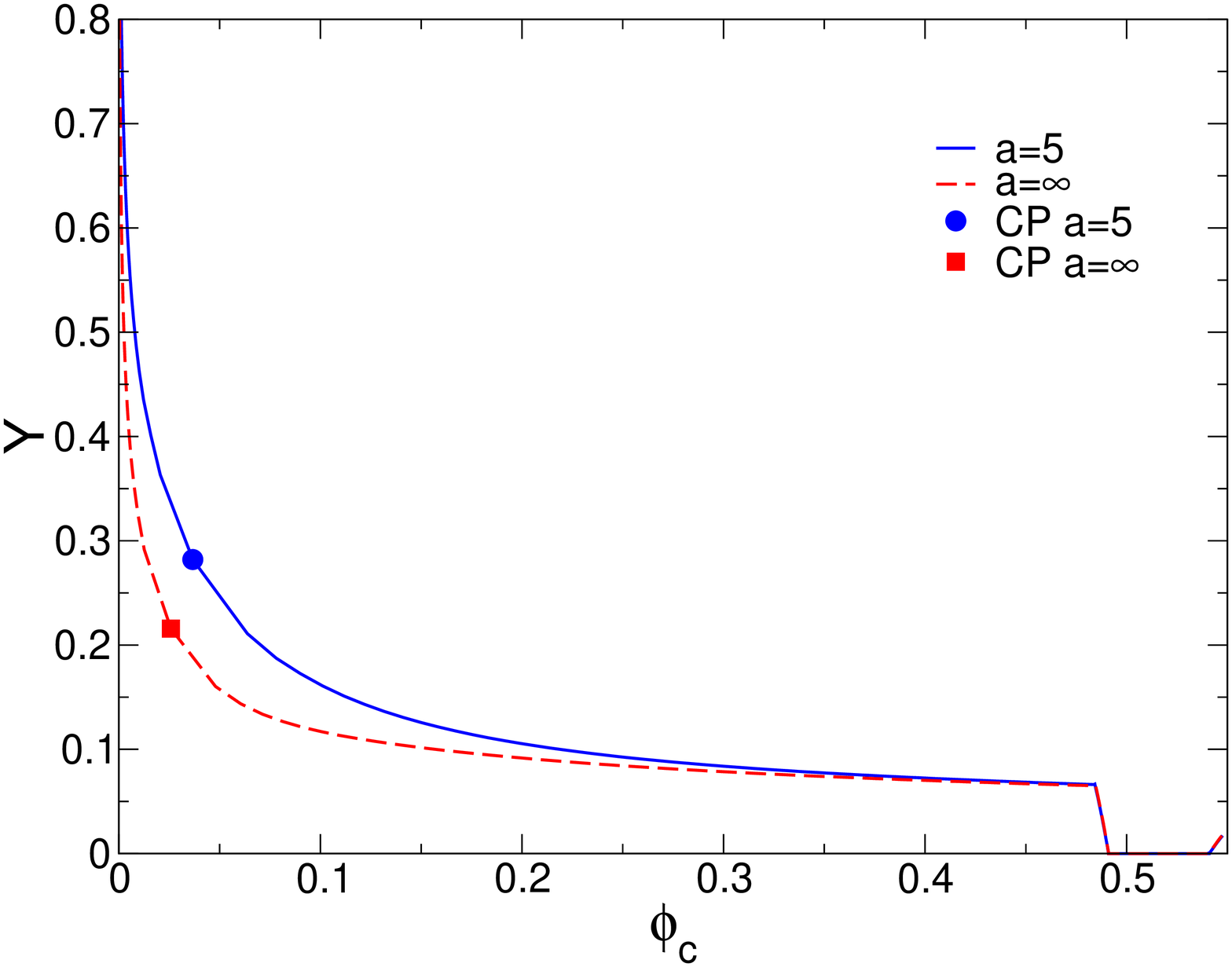,angle=0,width=7truecm} \hspace{0.5truecm} \\
\end{tabular}
\end{center}
\caption{Universal large-$q$ binodals as a function of $\phi_c$ and $Y^{(r)}$ 
(top) and 
of $\phi_c$ and $Y$ (bottom).  
}
\label{Binodals-Y}
\end{figure}

Let us now determine the GFVT binodals in the protein regime, in which $q$ is 
large. We will only consider the good-solvent case, in which our results 
significantly differ from those of Ref.~\onlinecite{FT-08}.
The results presented in Fig.~\ref{CP} apparently suggest
$\phi_{c,\rm crit}\approx 0.10$ as $q\to \infty$. 
As we now discuss, this is not 
correct, since, as $q$ increases beyond 2-3, $\delta_s(q,\phi_p)$ shows a 
crossover\cite{DPP-depletion} to a different, large-$q$ behavior. In
particular, general renormalization-group arguments
predict\cite{HED-99,DPP-depletion}
\begin{equation}
{\delta_s\over R_c} = \Delta_<(q,\phi_p) =  \left({3 A_{\gamma,\infty} \over 
   K_p(\phi_p)}\right)^{1/3} q^{1/3\nu},
\label{Delta_small}
\end{equation}
with\cite{HED-99,DPP-depletion} 
$A_{\gamma,\infty}\approx 1.41$ as long as $R_c\ll \xi\ll R_g$, where $\xi$ is 
the polymer correlation length. In the opposite case, 
$R_c\gg \xi$, we have instead \cite{MEB-01,DPP-depletion}
\begin{equation}
{\delta_s\over R_c} = \Delta_>(q,\phi_p) = 
 0.649 q \phi_p^{-\gamma},
\label{Delta_large}
\end{equation}
where
\begin{equation}
\gamma = {\nu\over 3\nu-1} \approx 0.7703.
\end{equation}
As discussed in Ref.~\onlinecite{BML-03}, in the protein limit $q\to\infty$ 
a universal scaling behavior is observed provided one uses 
\begin{equation}
Y_q = \phi_p q^{-1/\gamma}
\end{equation}
as basic scaling variable. Indeed, 
phase separation occurs deep in the semidilute regime for large values of
$q$. The relevant length scale is therefore the correlation length $\xi$
and the relevant adimensional volume fraction is proportional to 
$\rho_p \xi^3$, which, in turn, is proportional to $Y_q$,
since \cite{deGennes-79} $\xi \sim R_g(\phi_p=0) \phi_p^{-\gamma}$.
The GFVT binodals show this universal scaling behavior \cite{FT-08} 
for large $q$. Hence, for large $q$ one finds 
\begin{equation}
\phi_{p,\rm crit}^{(r)} q^{-1/\gamma} \approx Y_{c,\infty}^{(r)},  \qquad
\phi_{p,\rm crit} q^{-1/\gamma} \approx Y_{c,\infty}, 
\end{equation}
while $\phi_{c,\rm crit}$ and $\widetilde{P} = \beta P V_c$ 
converge to constants 
$\phi_{c,c,\infty}$ and $\widetilde{P}_{c,\infty}$. To
estimate these quantities,
we must have an expression for $\delta_s/R_c$ 
which is valid for large $q$ and all values of $\phi_p$ and which, therefore,
interpolates between the two expressions (\ref{Delta_small}) and 
(\ref{Delta_large}).
We have used two different interpolants, in order to be able to estimate 
roughly how important the interpolation is. 
We consider the family of interpolants
\begin{equation}
{\delta_s(q,\phi_p)\over R_c} = 
    \left[\Delta_<(q,\phi_p)^{-a} + \Delta_>(q,\phi_p)^{-a}\right]^{-1/a},
\end{equation}
which depend on the parameter $a>0$. In the following we quote results 
for $a = 5$ and $a = \infty$, the latter choice corresponding to 
\begin{equation}
{\delta_s(q,\phi_p)\over R_c} = 
    \min \left[\Delta_<(q,\phi_p),\Delta_>(q,\phi_p)\right].
\end{equation}
The results for the critical and the triple point are reported in 
Table~\ref{largeq-binodal}. 

\begin{table}
\caption{Large-$q$ critical and triple points. The subscript 
$c$ refers to the critical point, $t$ to the triple point,
$tg$, $tl$, and $ts$ refer to the three phases (gas, liquid, solid) 
coexisting at the triple point.
We report results corresponding to two different interpolations for 
$\delta_s/R_c$ ($a=5$ and $a=\infty$) and the results (FT) of 
Ref.~\onlinecite{FT-08}.} 
\label{largeq-binodal}
\begin{tabular}{cccc}
\hline\hline
& $a=5$ & $a=\infty$ & FT\\
\hline
\multicolumn{3}{c}{Critical point} \\
\hline
$\phi_{c,c,\infty}$    & 0.037 & 0.025 & 0.104\\
$Y_{c,\infty}^{(r)}$ & 0.440 & 0.337   & 0.97 \\
$Y_{c,\infty}$       & 0.296 & 0.218   & 0.665 \\
$\widetilde{P}_{c,\infty}$ & 0.25 & 0.13& 0.360 \\
\hline
\multicolumn{3}{c}{Triple point} \\
\hline
$\phi_{c,tg,\infty}$    & $2\cdot 10^{-4}$ & $2\cdot 10^{-4}$ & 0 \\
$\phi_{c,tl,\infty}$    & 0.484 & 0.485   & 0.469 \\
$\phi_{c,ts,\infty}$    & 0.548 & 0.547   & 0.565 \\
$Y_{t,\infty}^{(r)}$    & 1.85  & 1.84    & 2.08  \\
$Y_{tg,\infty}$         & 1.85  & 1.85    & 2.08  \\
$Y_{tl,\infty}$         & 0.068 & 0.065   & 0.25  \\
$Y_{ts,\infty}$         & 0.018 & 0.017   & 0.09 \\
$\widetilde{P}_{t,\infty}$& 6.61 &  6.58  & 8.82 \\
\hline\hline
\end{tabular}
\end{table}

The results for the critical point depend significantly on the chosen 
interpolation, indicating how crucial the expression for
$\delta_s(q,\phi_p)$ is in estimating the critical point. Note also that 
$\phi_{c,c,\infty}$ and $\widetilde{P}_{c,\infty}$ differ significantly from
the values that would have been guessed by looking at Figs.~\ref{CP} and 
\ref{CP-press}, implying a significant crossover as $q$ increases beyond 3. 
These results differ from those of Ref.~\onlinecite{FT-08}, as their
expression for $\delta_s/R_c$ does not have the correct
behavior for $q$ large and $R_c \lesssim \xi$. \cite{DPP-depletion}
In Fig.~\ref{Binodals-Y} we report the large-$q$ binodals as a function of 
$Y$ and $\phi_c$. The curve shows a significant dependence on the interpolation
close to the critical point. However, differences decrease as one moves away
from the critical point towards the triple point.

\subsection{Comparison with previous work}

Let us now compare our results with those appearing in the literature.
We have first verified that our results for $z = 0$ (ideal polymers) are
in agreement with those obtained in similar studies (we have considered 
Ref.~\onlinecite{DBE-99}, which gives results for $q=0.6$ and 0.8). It is also
interesting to compare with the results for the AOV model. For $q=0.8$
simulations give\cite{VH-04,VH-04bis} 
$\phi_{p,\rm crit}^{(r)} = 0.766(2)$,
$\phi_{p,\rm crit} = 0.3562(6)$, and
$\phi_{c,\rm crit} = 0.1340(6)$ for the critical point, to be compared
with the free-volume theory predictions
$\phi_{p,\rm crit}^{(r)} \approx 0.602$,
$\phi_{p,\rm crit} = 0.214$,
$\phi_{c,\rm crit} = 0.141$. Analogously,
for $q = 1$, one finds \cite{DvR-02} $\phi_{p,\rm crit}^{(r)} \approx 0.7$ 
in the AOV model 
and  $\phi_{p,\rm crit}^{(r)} \approx 0.73$ by using free volume theory.
Also the triple-point position is quite accurately predicted: both theories
give $\phi_{p,t}^{(r)} \approx 6.0$ for $q=1$. 
Clearly, free volume theory provides a good description of the AOV model.

\begin{table}
\caption{Critical-point position and pressure 
($\widetilde{P} = \beta P V_c$) from full-monomer simulations 
of a lattice colloid-polymer system (They are obtained by 
using the results reported in the supplementary material of 
Ref.~\onlinecite{MLP-12}).
The polymer volume fraction has been determined using 
$R_g = 0.508 L^\nu$, where $L$ is the length of the chain. ``extr" gives 
the extrapolation $L\to \infty$, obtained as discussed in the text.
}
\label{FMdata}
\begin{tabular}{ccccc}
\hline\hline
$q$ &  $L$   &   $\phi_{c,\rm crit}$ & $\phi_{p,\rm crit}$ & 
    $\widetilde{P}_{\rm crit}$ \\
\hline
1   &  10    &  0.146 &  0.279  & 1.32 \\
    &  33    &  0.186 &  0.438  & 2.65 \\
    & 110    &  0.202 &  0.521  & 3.33 \\
    & extr   &  0.22  &  0.62   & 4.15 \\
    & GFVT   &  0.178 &  0.474  & 2.61  \\
2   &  33    &  0.120 &  0.537  & 0.64 \\
    & 110    &  0.162 &  0.770  & 1.27 \\
    & 350    &  0.176 &  0.905  & 1.55 \\
    & extr   &  0.19  &  1.076  & 1.88 \\
    & GFVT   &  0.115 &  1.205  & 1.34  \\
4   & 110    &  0.104 &  1.069  & 0.37 \\
    & 350    &  0.148 &  1.383  & 0.70 \\
    & extr   &  0.20  &  1.78   & 1.13  \\
\hline\hline
\end{tabular}
\end{table}

Let us now consider the good-solvent case, for which we can compare with 
full-monomer results.\cite{BML-03,CVPR-06,MLP-12,MIP-13} Given the 
computational complexity of these systems, the simulated chains are 
typically relatively short
and the results show significant corrections to scaling, which should be 
taken into account before comparing them with our GFVT results. Indeed, the
GFVT estimates have been obtained by using the universal asymptotic 
predictions for $K_p(\phi_p)$ and $\delta_s(q,\phi_p)$,
hence they apply to polymer systems in the scaling limit 
(large degree of polymerization). Ref.~\onlinecite{MLP-12} reports 
results for several chain lengths $L$ and several values of $q$. The critical
point position and pressure are reported in Table~\ref{FMdata}. To compute 
the polymer volume fraction we used $R_g = 0.508 L^\nu$.\cite{MLP-12} Note that 
scaling corrections are very large in all cases: as $L$ increases, there is a 
systematic drift of the critical parameters. We expect two types of scaling 
corrections. First, there are corrections that scale as $L^{-\Delta}$,
$\Delta = 0.528(12)$,\cite{Clisby-10} as in all polymer systems. 
In the presence of colloids,
a second type of corrections appear, related with the renormalization-group
operators associated with the colloid-polymer interactions.\cite{DPP-depletion}
They scale as $L^{-\nu}$, where $\nu \approx 0.5876$ is the Flory exponent.
The two exponents are very close, hence we have extrapolated 
the finite-length results as $a + b L^{-1/2}$. In Table~\ref{FMdata} 
lines labelled ``extr" report
the corresponding coefficient $a$. It is important to note that both types of 
corrections are not related to the lattice breaking of rotational invariance,
\cite{footnote-rotational} 
hence they are present both in lattice and continuum models.
Extrapolations for $q = 4$ should not be 
taken too seriously, given the very large difference between the data and the 
extrapolation results. In any case, they are roughly consistent
with the results of Ref.~\onlinecite{BML-03}.
They studied self-avoiding lattice walks with $L=2000$ monomers, finding 
$\phi_{c,\rm crit}\approx 0.24$ and 
$\phi_{p,\rm crit}\approx 1.9$ for $q=3.86$. No extrapolation 
can be done here, hence these results are affected by scaling corrections,
which are particularly large for SAWs in the semidilute
regime.\cite{Pelissetto-08} Still, they confirm that 
$\phi_{c,\rm crit}\gtrsim 0.20$ and $\phi_{p,\rm crit}\gtrsim 1.8$ for 
$q\approx 4$. Ref.~\onlinecite{BML-03} also presents results for 
$q = 5.58$ and 7.78. In all cases, they obtain $\phi_{c,\rm crit}\approx 0.25$.

For $q = 1$, the GFVT estimates of $\phi_{c,\rm crit}$
and $\phi_{p,\rm crit}$ both underestimate the extrapolated full-monomer 
values, differences being of the order of 20-25\%, while the pressure is 
largely underestimated. For $q = 2$, $\phi_{c,\rm crit}$ and 
$\widetilde{P}_{\rm crit}$ differ by a factor of 2 and 0.5, respectively,
from the numerical result,
while $\phi_{p,\rm crit}$ is in reasonable agreement. 
For larger values of $q$, differences are expected to increase. 
While numerical data indicate $\phi_{c,\rm crit}\gtrsim 0.20$, 
GFVT
predicts $\phi_{c,\rm crit}$ to converge to 0.02-0.03 as 
$q\to \infty$. These results show that GFVT looses predictivity in 
the protein limit. This
is not so surprising, since the theory describes polymers as spheres
which move in the free space left by the colloids. For $q>1$, this picture is 
clearly unrealistic.

Most of the other predictions for the good-solvent case are obtained by 
studying coarse-grained models in which each polymer is replaced 
by a monoatomic molecule, as in the AOV model. Ref.~\onlinecite{BLH-02} 
uses an exact coarse-graining procedure, which provides density-dependent
effective potentials that accurately reproduce pair correlation functions
as measured in full-monomer simulations. 
Monte Carlo studies of the resulting model provide 
accurate estimates of the critical and of the triple point. 
The estimates of the critical-point location 
$(\phi_{c,\rm crit},\phi_{p,\rm crit}) = 
(0.19,0.40)$ and $(0.18,0.51)$ for $q = 0.67$ and 1.05  are close to the 
GFVT predictions $(0.23,0.29)$ and $(0.17,0.50)$, respectively. 
The very good agreement for $q = 1.05$, however, is probably accidental
and should not be taken too seriously, given that full-monomer 
simulations, see Table~\ref{FMdata}, predict $(0.22,0.62)$ for $q\approx 1$.
The reservoir polymer
volume fraction at the triple point $\phi^{(r)}_{p,t}$ is also close to the 
GFVT estimate. The coarse-grained model gives $\phi^{(r)}_{p,t} = 0.90$,
1.62 (for $q=0.67,1.05$, respectively) 
to be compared with 0.97, 1.88 obtained using GFVT.
Again, discrepancies increase with $q$.

There are also results for several coarse-grained 
phenomenological models. However, 
in all these cases it is not obvious that they are really 
appropriate to describe polymers under good-solvent conditions interacting with 
hard-sphere colloids. Ref.~\onlinecite{AP-12} studied a simple model 
which reproduces the polymer thermodynamics for $\phi_c\to 0$ and provides
the correct second-virial combination $A_{2,cp}$. They obtained 
$\phi_{c,\rm crit} = 0.22$, $\phi_{p,\rm crit} = 0.93(2)$, and
$\phi_{p,\rm crit}^{(r)} = 1.321(4)$. While $\phi_{c,\rm crit}$ is in
reasonable agreement with the full-monomer results, the polymer volume 
fraction is too large: since $\phi_{p,\rm crit}$ increases with $q$, we would 
expect $\phi_{p,\rm crit}\lesssim 0.6$ for $q = 0.8$. Clearly, this
coarse-grained model is not so good as the one discussed in
Ref.~\onlinecite{BLH-02} which uses density-dependent accurate pair potentials.
A different model is presented in 
Refs.~\onlinecite{Dzubiella-etal-01,DLL-02,VJDL-05}. It reproduces 
\cite{MP-13} 
the second-virial combination $A_{2,pp}$ and also the combination 
$A_{2,cp}$ is in good agreement with the the full-monomer results of 
Ref.~\onlinecite{DPP-depletion}. For $q = 0.5$ and 1, the coarse-grained model
gives $A_{2,cp} = 109.4$, 29.3 to be compared with the full-monomer results 
107.4(3) and 27.54(6). However, the results for $q=0.56$ show\cite{VJDL-05}
the unexpected
feature that the good-solvent binodal is very close to the AOV binodal. 
This type of behavior is not in agreement with the results of 
Ref.~\onlinecite{BLH-02} and with the good-solvent 
GFVT predictions. This discussion 
shows that it is not enough to correctly reproduce the thermodynamic behavior 
of the systems in the dilute regime, a property that the models
of Ref.~\onlinecite{AP-12} and of
Refs.~\onlinecite{Dzubiella-etal-01,DLL-02,VJDL-05} both share.
An accurate parametrization of the effective potential appears to be a necessary
condition to obtain reasonably accurate predictions of the phase diagram.

Finally, let us discuss the results of Refs.~\onlinecite{SDB-03,ZVBHV-09},
which consider two different models of interacting polymers. 
If we compare the critical-point positions they obtain with the good-solvent 
GFVT predictions, we observe significant differences, much larger than those 
observed when comparing GFVT results with full-monomer data. As we now discuss, 
these differences are due to the fact that these models are not appropriate 
to describe good-solvent polymers, hence their results should not 
be compared with the corresponding good-solvent GFVT results. 
For instance, if we 
consider the model of Ref.~\onlinecite{SDB-03}---the interacting model 
with $\beta\epsilon=0.25$---we obtain for the polymer-polymer second virial
combination, $A_{2,pp} \approx 3.71$. Therefore, the model describes polymers 
in the thermal crossover region and the numerical results should rather 
be compared
with the GFVT ones for $z = z^{(3)}$. For $q = 1$, they obtain 
$\phi_{c,\rm crit} \approx 0.14$, 
$\phi_{p,\rm crit} \approx 0.2$,  which are not far from the GFVT
results 
$\phi_{c,\rm crit} \approx 0.139$, 
$\phi_{p,\rm crit} \approx 0.329$ for $z = z^{(3)}$. Discrepancies are
significantly reduced.
Also the model of
Ref.~\onlinecite{ZVBHV-09} does not appear to describe good-solvent polymers. 
As noted in Ref.~\onlinecite{AP-12}, $A_{2,pp} = 1.78$, hence the results 
should be better compared with those appropriate for $z = z^{(1)}$. 
With this identification, the Monte Carlo results are close to the GFVT ones.

\begin{figure}[t]
\begin{center}
\begin{tabular}{c}
\epsfig{file=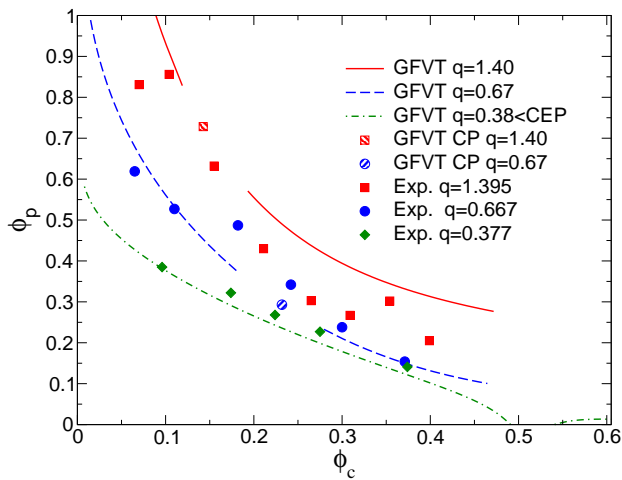,angle=0,width=8.2truecm} \hspace{0.5truecm} \\
\end{tabular}
\end{center}
\caption{Comparison of the experimental data of Ref.~\onlinecite{RFSZ-02}
with the good-solvent GFVT predictions. 
}
\label{Expt-comp}
\end{figure}

Finally, it is interesting to compare GFVT with experiments. An extensive discussion of
experiments is reported in Ref.~\onlinecite{FT-08}. Here we will only make 
a few comments
on the most recent ones. A solution of polystyrene in toluene at 
35$^\circ$C, which is a well-known example of good-solvent 
system,\cite{KK-83,FHLM-94}
mixed with silica particles was studied in Ref.~\onlinecite{RFSZ-02}. The experimental
results were compared (see their figure 5) with the original version of free-volume theory
(appropriate for $\theta$-point solutions), observing large discrepancies.
Here we perform the same comparison with GFVT, see Fig.~\ref{Expt-comp}. We observe a
relatively good agreement for $q = 0.337$ (which is below the critical endpoint)
and for $q = 0.667$. Discrepancies are observed instead for $q = 1.395$, confirming again that
GFVT is reasonably accurate in the colloid regime, but becomes unreliable as $q$ increases
beyond 1. In Ref.~\onlinecite{RFSZ-02} the experimental results are also compared with 
PRISM predictions.\cite{FS-00,FS-01} A comparison of their Fig. 8 with our
Fig.~\ref{Expt-comp} shows that PRISM is significantly less accurate than GFVT.

\begin{figure}
\begin{center}
\begin{tabular}{c}
\epsfig{file=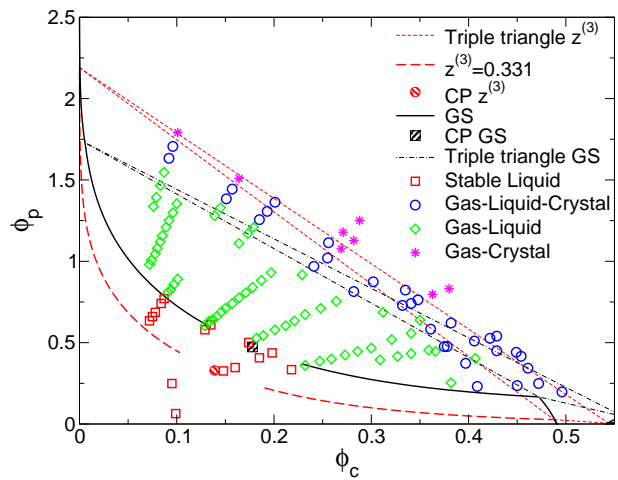,angle=0,width=8.2truecm} \hspace{0.5truecm} \\
\end{tabular}
\end{center}
\caption{Comparison of the experimental data of Ref.~\onlinecite{TSPEALF-08} 
for a solution of poly-methylmethacrylate colloids and linear
polystyrene in a mixture of cis-decalin and tetralin
with the GFVT predictions. We report the GFVT binodal, critical point,
and triple-point triangle 
for $z=z^{(3)}$ and $z = \infty$. We use $q = 0.95$. 
Symbols correspond to the experimental data.
}
\label{Expt-comp2}
\end{figure}

Finally, let us consider the experimental results of 
Ref.~\onlinecite{TSPEALF-08}. They 
consider poly-methylmethacrylate colloids and linear polystyrene in a 
mixture of 
cis-decalin and tetralin. Assuming that the addition of tetralin does 
not change the 
properties of the solution as argued by Tuinier {\em et al.},\cite{TSPEALF-08}
the results presented in the supplementary material for 
linear polystyrene
give $z = 0.30$-0.35 for a polymer of molar weight $15.4\cdot 10^6$ g/mol
and for $T-T_\theta = 3\, $K. Since $z^{(3)} \approx 0.32$,
the solution should not be considered as a good-solvent system, but rather as 
a system in the thermal crossover region. 
Using Eq.~(\ref{alpha-to-z}) we predict $\alpha_g = 1.12$
and $R_g = 123$ nm, so that $q = 0.95$ ($R_c \approx 130$ nm). 
Therefore, the experimental
data should be compared with the GFVT results for $z = z^{(3)}$, 
rather then with those
appropriate for good-solvent systems.  
In Fig.~\ref{Expt-comp2} we compare theory and
experiments. It is evident that GFVT for the appropriate value of 
$z$ underestimates the 
experimental binodal, as already observed when comparing theory and 
numerical data at $q \approx 1$. On the other hand, the triple-point triangle
obtained by GFVT for $z = z^{(3)}$ appears to be in better agremement with 
the experiment than that obtained by considering good-solvent conditions.

The large-$q$ phase diagram was investigated by Mutch {\em et al.},
\cite{MDEGH-08,MDEGH-09,MDEGH-10} considering water-in-oil microemulsion 
droplets mixed with polyisoprene in cyclohexane, which is an approximately
good-solvent system.\cite{footnote-MDEGH} For the colloid critical volume 
fraction, they quote \cite{MDEGH-09} 
$\phi_{c,\rm crit} = 0.21$ for $q=4,10$ and 
$\phi_{c,\rm crit} = 0.19$ for $q = 16$. It is quite clear that 
$\phi_{c,\rm crit}$ is much larger than the GFVT prediction, which is clearly 
not reliable for large $q$. On the other hand, they are close to the numerical
full-monomer estimates given in Table~\ref{FMdata}, 
provided that the extrapolation to the scaling limit
is performed. They are also reasonably consistent with the 
estimates of Ref.~\onlinecite{BML-03}. 
The expected large-$q$ scaling as a function of $\phi_c$ and 
$\phi_p q^{-1/\gamma}$ was also tested. Reasonable agreement was observed,
although small differences were observed for $\phi_c\gtrsim 0.2$. 
These differences are probably a consequence of the fact that 
polyisoprene is not exactly a good-solvent system for
the considered values of $M_w$. For instance, 
in the range of masses considered in
Ref.~\onlinecite{DFFHG-87}, the gyration radius scales as $M_w^{0.545}$ 
and not as $M_w^{0.5876}$. Note also that the best estimate of $A_{2,pp}$ 
for the sample considered in Ref.~\onlinecite{MDEGH-09} is \cite{footnote-MDEGH}
$A_{2,pp} \approx 4.80$, which is somewhat less than the good-solvent value
\cite{CMP-06} 5.50. A second possible reason for the discrepancy
is colloid polydispersity, which appears to increase with the size of the 
colloids.

\section{Generalization to charged-colloid systems} \label{sec5}

\begin{table}
\caption{Charged-colloid systems: Critical endpoint values $q_{CEP}$ for 
$z = 0$ (RW), $z = z^{(1)}$, $z^{(3)}$, and for good-solvent (GS)
polymers, as a function of $s$.}
\label{table:qCEP-generalized}
\begin{tabular}{ccccc}
\hline\hline 
$s$ & RW & $z^{(1)}$ & $z^{(3)}$ & GS \\
\hline
1.00 & 0.31 & 0.32 & 0.35 & 0.42 \\
1.01 & 0.32 & 0.33 & 0.36 & 0.44 \\
1.02 & 0.33 & 0.35 & 0.38 & 0.46 \\
1.03 & 0.34 & 0.36 & 0.40 & 0.49 \\
1.04 & 0.35 & 0.37 & 0.41 & 0.52 \\
1.05 & 0.37 & 0.39 & 0.43 & 0.55 \\
1.06 & 0.38 & 0.40 & 0.45 & 0.59 \\
1.07 & 0.40 & 0.42 & 0.48 & 0.63 \\
\hline\hline
\end{tabular}
\end{table}

\begin{figure*}
\begin{center}
\begin{tabular}{ccc}
\epsfig{file=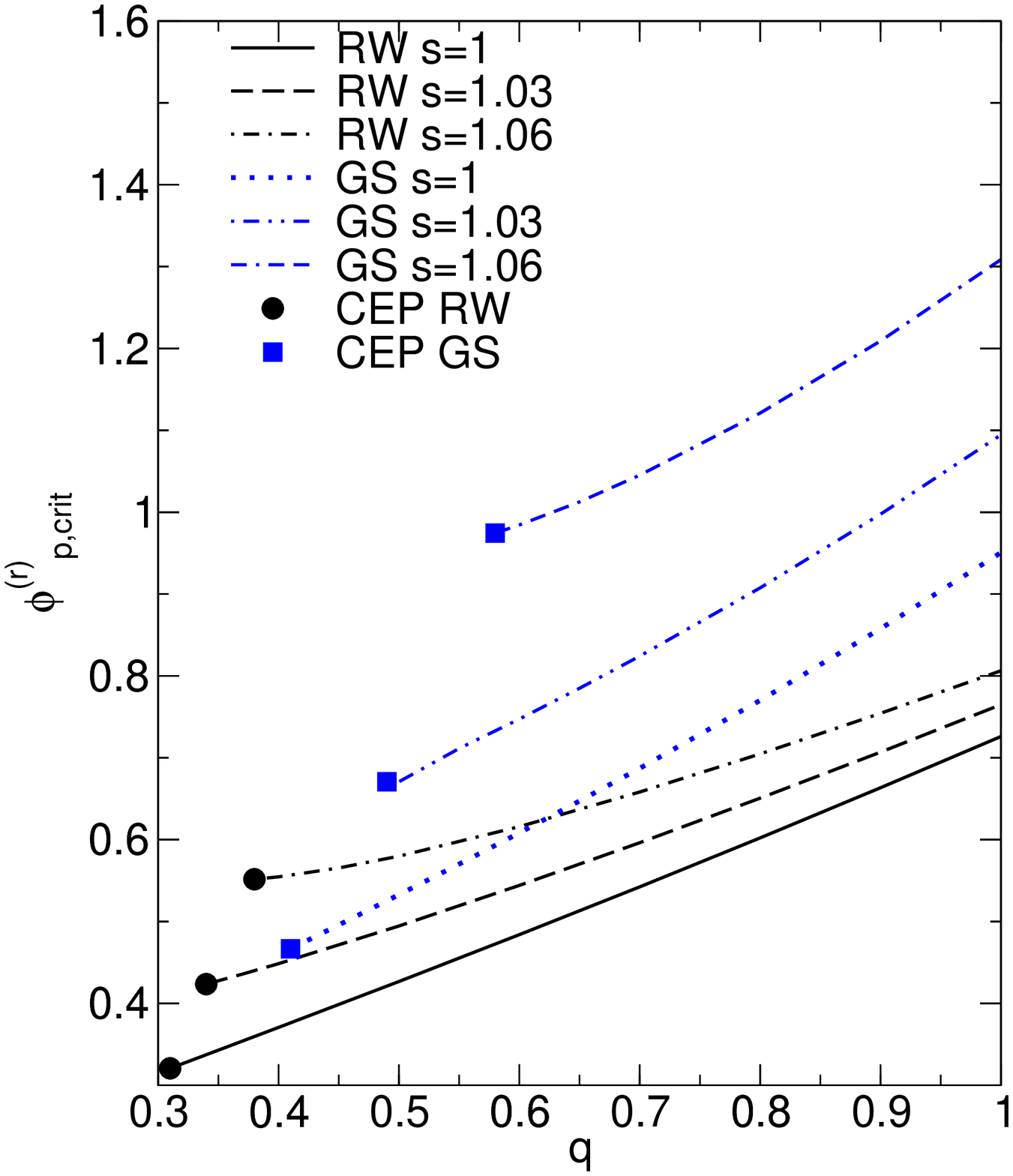,angle=0,width=5truecm} \hspace{0.5truecm}
\epsfig{file=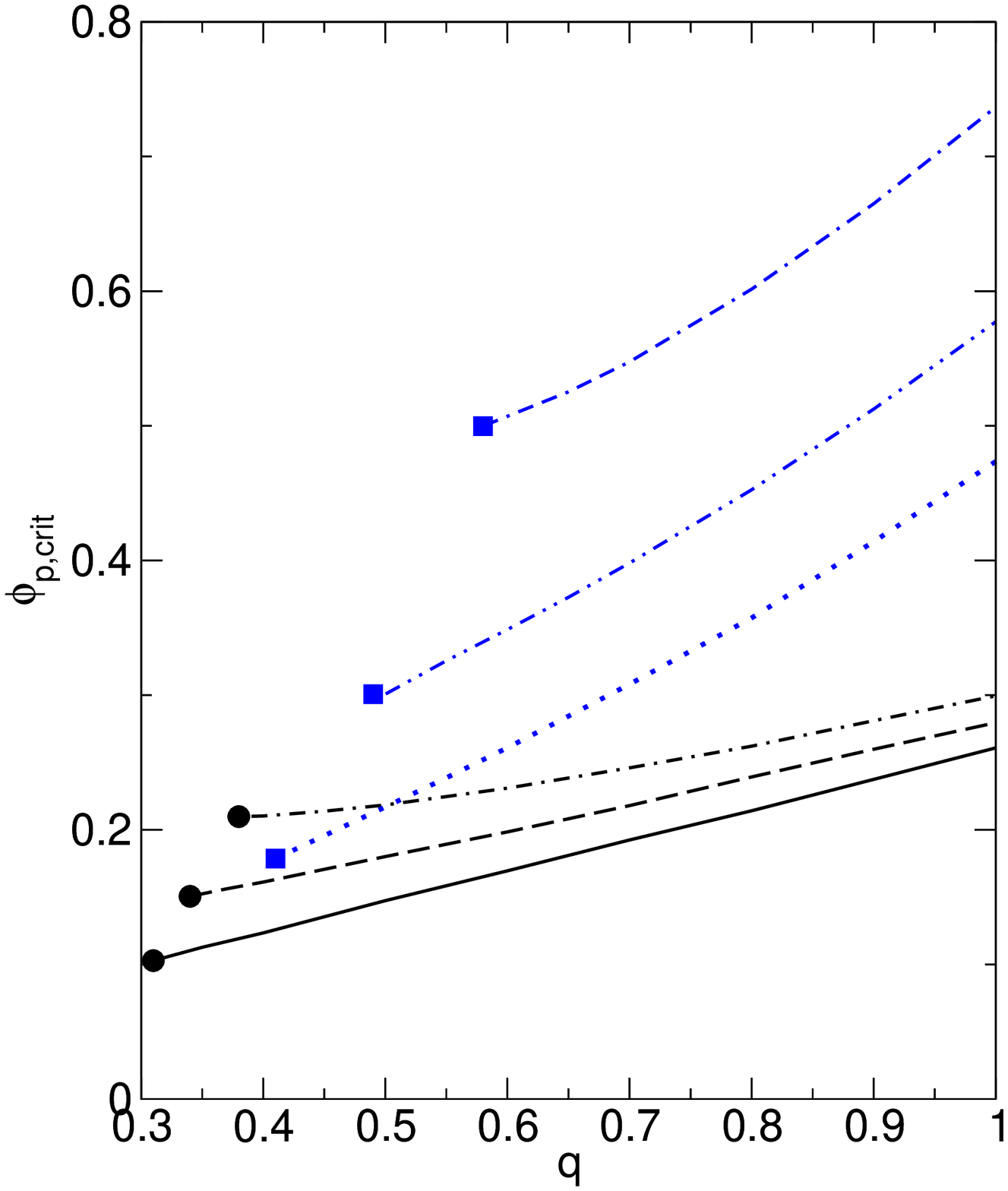,angle=0,width=5truecm} \hspace{0.5truecm}
\epsfig{file=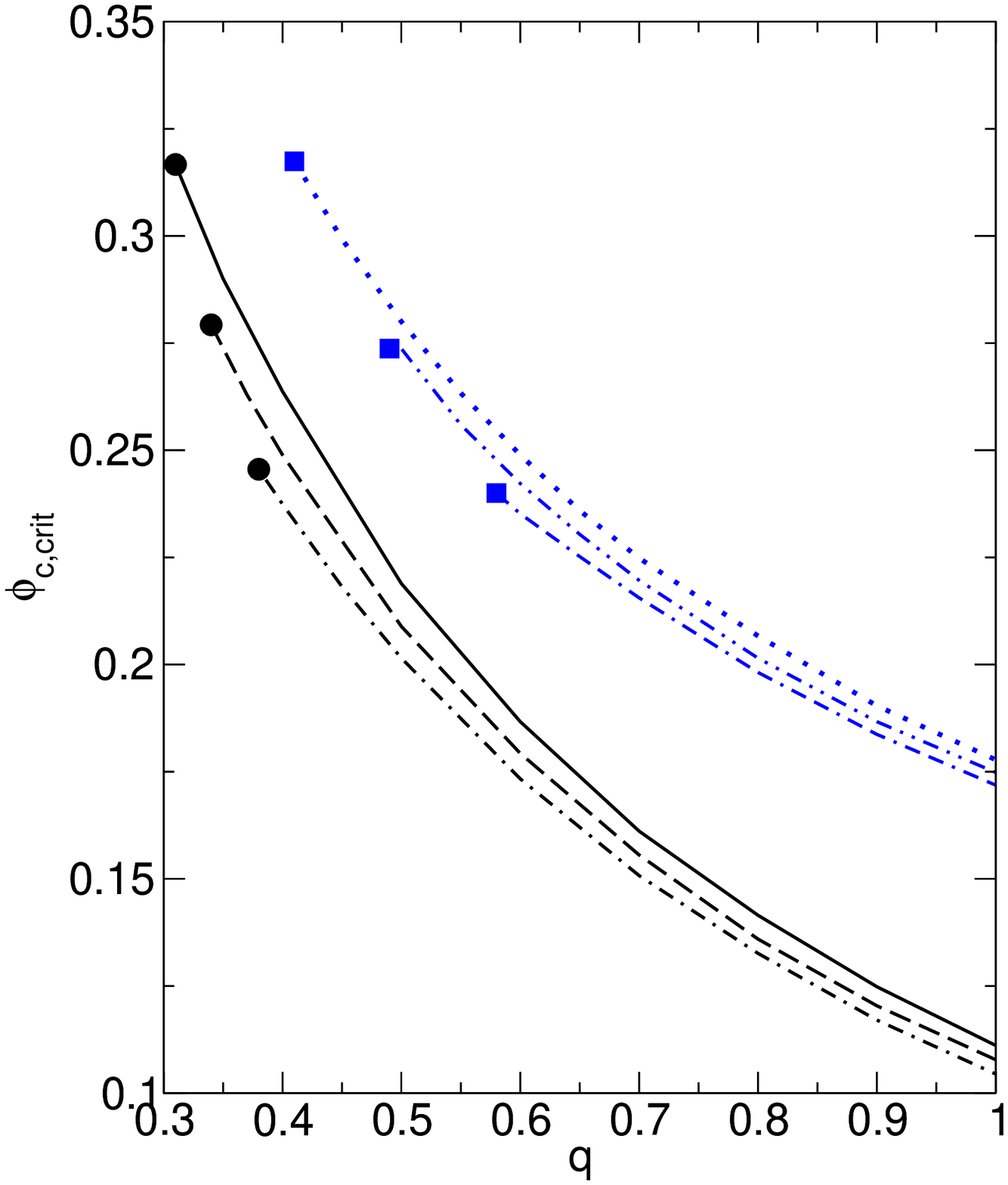,angle=0,width=5truecm} \hspace{0.5truecm}
\end{tabular}
\end{center}
\caption{Fluid-fluid critical-point volume fractions 
$\phi_{p,\rm crit}^{(r)}$ (left),
$\phi_{p,\rm crit}$  (middle), and
$\phi_{c,\rm crit}$ (right)
as a function of $q$ for different
values of the parameter $s$, for good-solvent (GS) and $\theta$ (RW) 
conditions. 
}
\label{phicrit-charged}
\end{figure*}

In the previous Sections we considered dispersions of hard-sphere colloids 
and nonadsorbing 
neutral polymer chains. However, in many practical applications colloids
cannot be modelled as hard spheres, since their interaction is characterized
by an additional repulsive tail. For instance, this is the case of charged
colloids in an aqueous salt solution, in which the effective colloid-colloid
potential  presents a Yukawa-like tail that depends on temperature,
on the dielectric constant, and on the salt concentration of the solvent. 
GFVT has been generalized to this class of systems
\cite{FDT-05,GT-08,vGTBS-13} and, more generally, to systems in which the
colloid-colloid interaction presents an additional repulsive tail.
As before, polymers are modelled as soft spheres of radius 
$\delta_s(q,\phi_p)$ interacting with colloids of radius $R_c$.
To keep into account the Yukawa tail in the colloid-colloid interaction,
the colloid potential is taken as 
\begin{eqnarray}
 & U = +\infty & \qquad |{\bf r}_{1,c} - {\bf r}_{2,c}| \le 2 s R_{c},  \\
 & U = 0 & \qquad |{\bf r}_{1,c} - {\bf r}_{2,c}| > 2 s R_{c}, \nonumber 
\end{eqnarray}
where $s$ is a parameter, which can be related to the original potential
by using the Barker-Henderson relation.\cite{BH-67} For $s=1$ we reobtain the 
hard-sphere potential, while the additional repulsive tail is mimicked by
taking $s > 1$. As before, we define $q = R_g/R_c$, $V_c = 4 \pi R_c^3/3$,
and $\phi_c = V_c \rho_c$, where $\rho_c$ is the colloid number density.
Note that, with these definitions, the colloid fluid phase ends at
$\phi_c \approx 0.49/s^3$ for $\phi_p\to 0$. 
To generalize the GFVT potential to this 
nonadditive system, two changes should be made.\cite{FDT-05} 
First, we must modify the colloid zero-polymer-density contribution to 
$\omega(\phi_c,\phi_p^{(r)})$. The function $f(\phi_c)$ given in
Eq.~(\ref{CS-f})
should be replaced by $f(s^3 \phi_c)/s^3$, to take into account the 
different interaction range of the potential. The same change should 
be made on the free energy in the solid phase, see Eq.~(\ref{solid-f}).
Second, we should modify the free-volume factor, which takes the form
(\ref{alpha-SPT}) with \cite{FDT-05}
\begin{eqnarray}
&& Q(\phi_c,d,s) = \\
&& \qquad
 (3 d + 3 d^2) y + 9 d^2/2 y^2 + d^3 y_s (1 + 3 y_s + 3 y_s^2),
\nonumber 
\end{eqnarray}
where $y = \phi_c/(1 - \phi_c)$ and $y_s = s^3 \phi_c/(1 - s^3 \phi_c)$.

Using these expressions, we have repeated the calculations presented in 
Refs.~\onlinecite{FDT-05,GT-08} for the ideal and the good-solvent case, 
and, moreover, we have extended the calculation to the thermal crossover
region. We consider values of $s$ in the range $1 \le s \lesssim 1.07$, 
corresponding to $1 \le m=s^3 \lesssim 1.2$, which is the expected range 
of validity of the approximation.\cite{vGTBS-13} We have first computed the 
values $q_{CEP}$ of the critical endpoints which determine the end of the 
fluid-fluid coexistence region. The results are reported in 
Table~\ref{table:qCEP-generalized}. The results for the ideal case ($z=0$)
should not agree with those reported in Table 1 of Ref.~\onlinecite{GT-08}
for the $\theta$ case. Indeed, Ref.~\onlinecite{GT-08} takes into account
the logarithmic corrections to the equation of state, that are present 
at the $\theta$ point. Such corrections are neglected here. The two approaches 
give similar results for neutral colloids. On the other hand, in the 
charged case, results differ, discrepancies increasing 
with increasing $s$. 
For instance, for $s = 1.07$ (corresponding to $m = s^3 = 1.225$) 
we obtain $q_{CEP} = 0.40$ to be compared with the result of 
Ref.~\onlinecite{GT-08}, that quotes $q_{CEP} = 0.49$. 
In any case, the qualitative conclusions of 
Ref.~\onlinecite{GT-08} are always in agreement with ours, and no 
appreciable differences can be observed by comparing the graphs 
reported in Refs.~\onlinecite{FDT-05,GT-08} with those obtained here. 
In the good-solvent case, in spite of the different expressions used for 
the depletion thickness and the polymer compressibility factor, our 
results for $q_{CEP}$ are fully consistent with those of
Ref.~\onlinecite{GT-08}. For instance, for $s = 1.07$ 
we obtain $q_{CEP} = 0.63$ to be compared with the result of 
Ref.~\onlinecite{GT-08}, quoting $q_{CEP} = 0.61$. It is interesting 
to compare also the results in the crossover region. For $z = z^{(1)}$ 
$q_{CEP}$ is very close to the value obtained in the ideal case. Apparently, 
in a large temperature interval around $T_\theta$, the phase diagram 
changes only slightly. On the other hand, $q_{CEP}$ varies significantly
for $z > z^{(3)}$, indicating that the phase diagram is quite sensitive 
to solvent quality close to the good-solvent regime, as already observed for 
neutral colloids.

\begin{figure*}
\begin{center}
\begin{tabular}{c}
\epsfig{file=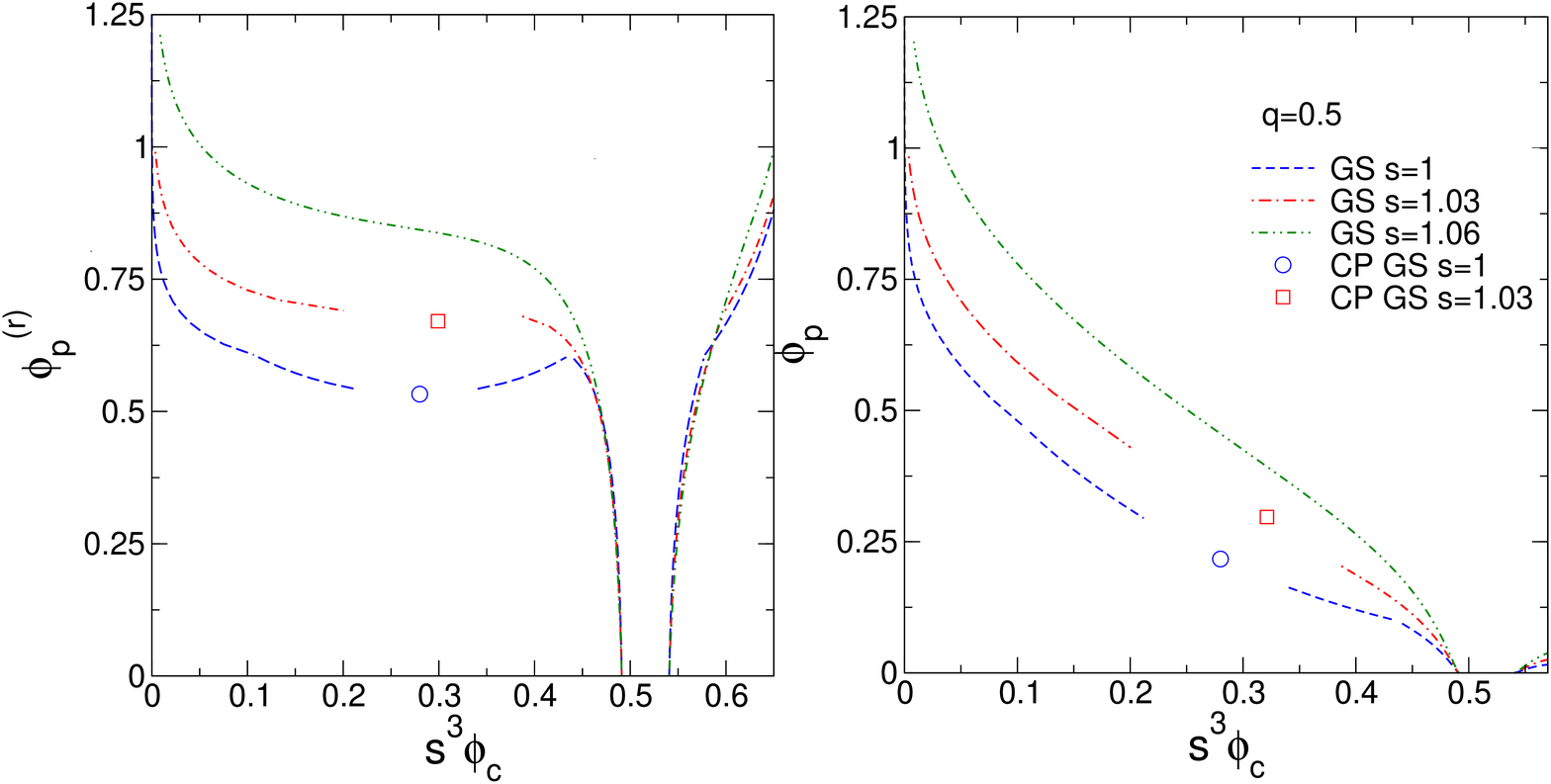,angle=0,width=14truecm} \hspace{0.5truecm} \\
\epsfig{file=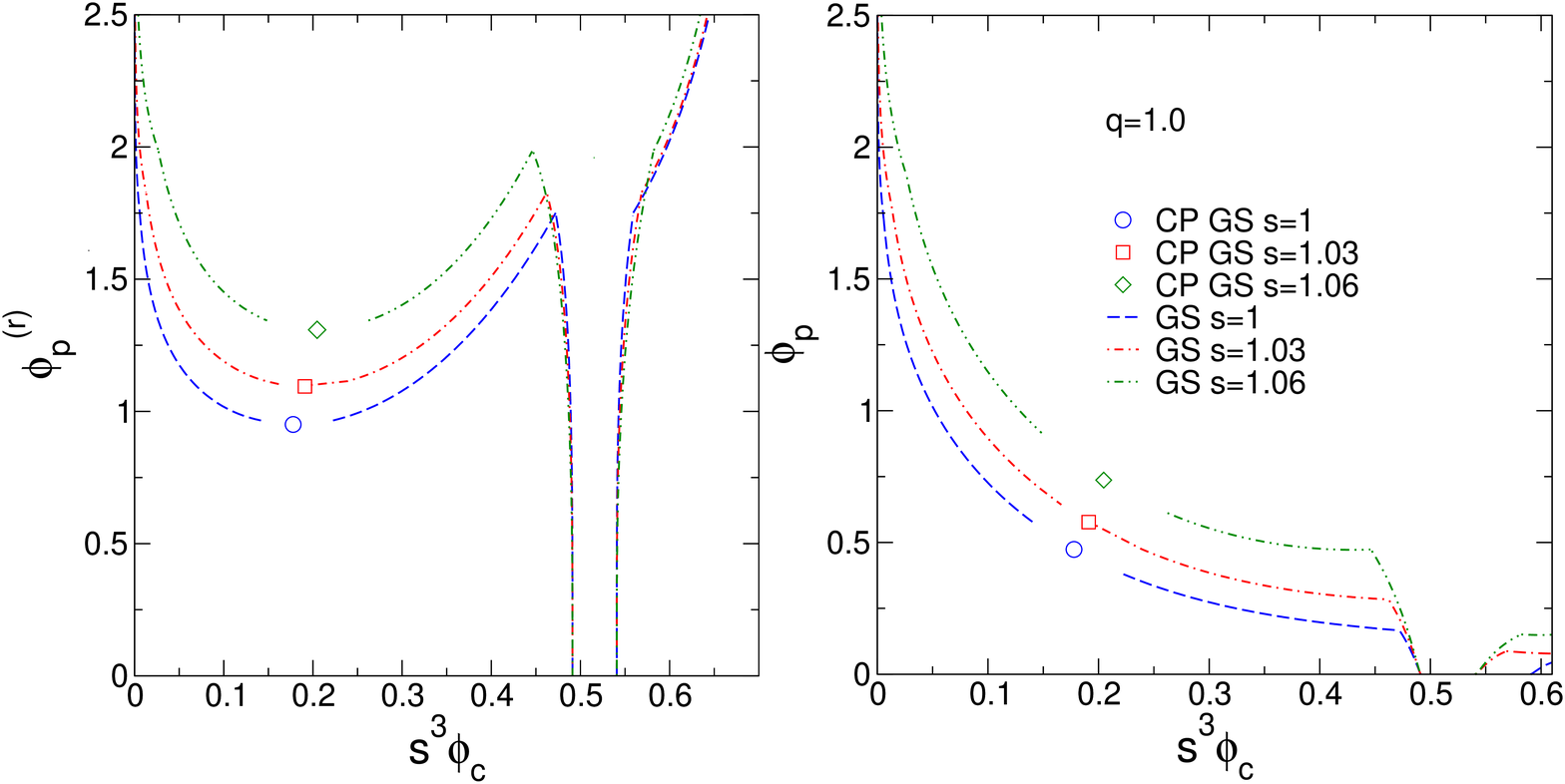,angle=0,width=14truecm} \hspace{0.5truecm} 
\end{tabular}
\end{center}
\caption{Binodals and critical points (CP) 
for good-solvent conditions for several values of $s$.
On the left we report $\phi_p^{(r)}$ and on the right $\phi_p$ as 
a function of $s^3 \phi_c$. The upper panels refer to $q = 0.5$, the lower ones 
to $q = 1$. No fluid-fluid transition occurs for $q=0.5$ and $s=1.06$.
}
\label{Phase-diag-charged-GS}
\end{figure*}

In Fig.~\ref{phicrit-charged} we report the volume fractions
$\phi_{c,\rm crit}$, $\phi_{p,\rm crit}^{(r)}$, 
and $\phi_{p,\rm crit}$ corresponding to the critical point
for good-solvent and $\theta$ conditions. The colloid $\phi_{c,\rm crit}$ 
shows a very tiny dependence on $s$. Solvent quality is here much more 
important that charge effects. 
In the $\theta$ region, also $\phi_{p,\rm crit}$
shows a relatively small dependence on $s$, 
which furthermore decreases as $q$ is increased. On the other hand, 
charge effects are important in the good-solvent regime. In particular, 
the difference between $\phi_{p,\rm crit}$ in the good-solvent and in the 
$\theta$ regime increases with $s$, indicating that a careful control 
of the quality of the solution is crucial to understand experimental results for 
charged systems. In the intermediate thermal crossover region, we observe
the same phenomenon discussed before. For $z = z^{(1)}$ and also, although
to a lesser extent, for $z = z^{(3)}$, $\phi_{p,\rm crit}$ shows a relatively
small dependence on $s$, indicating that charge effects play a relatively
small role in a large $z$ (i.e., temperature) interval around $\theta$
conditions. They become important only when one approaches the good-solvent 
region. 

\begin{figure}
\begin{center}
\begin{tabular}{c}
\epsfig{file=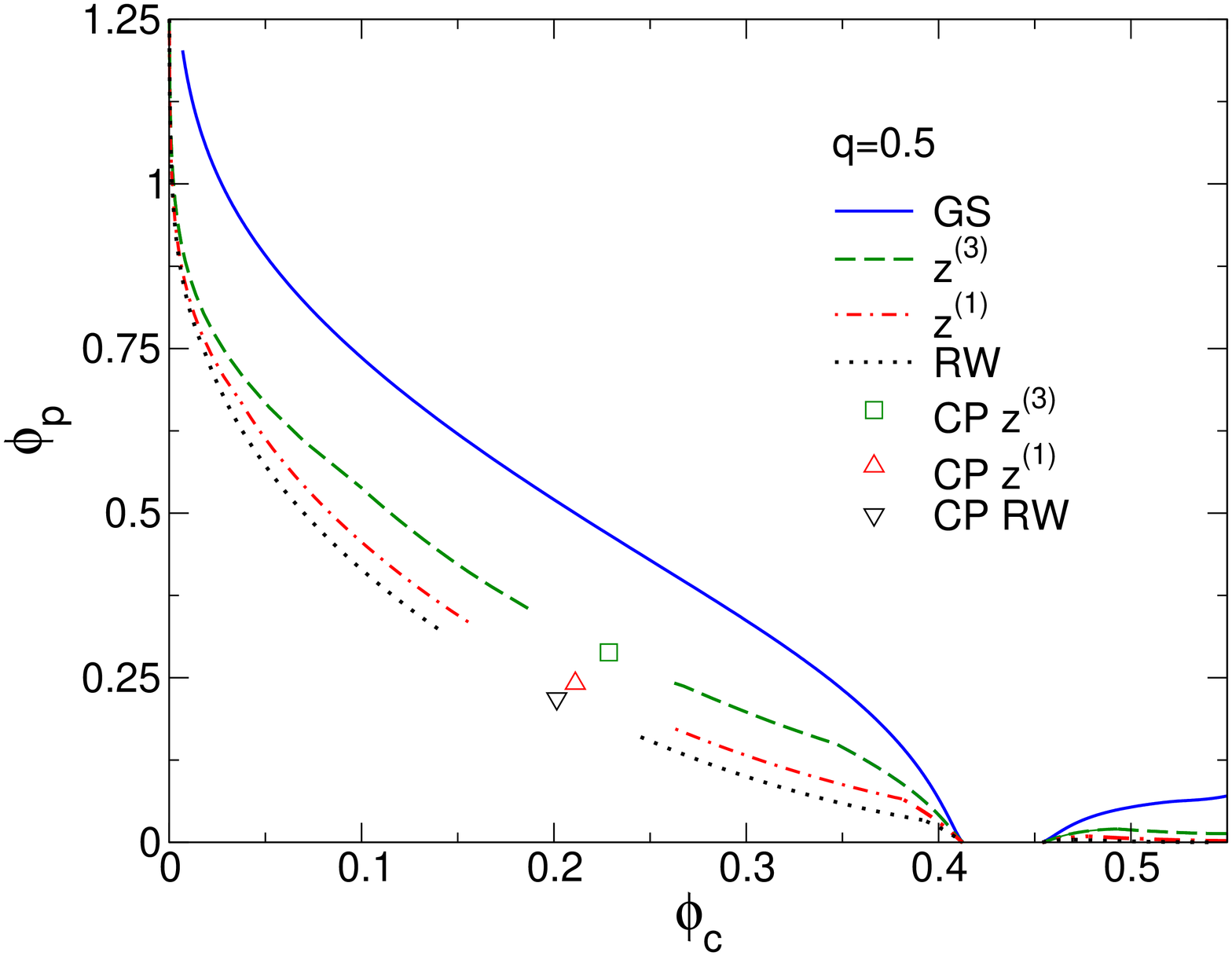,angle=0,width=8truecm} \hspace{0.5truecm} \\
\epsfig{file=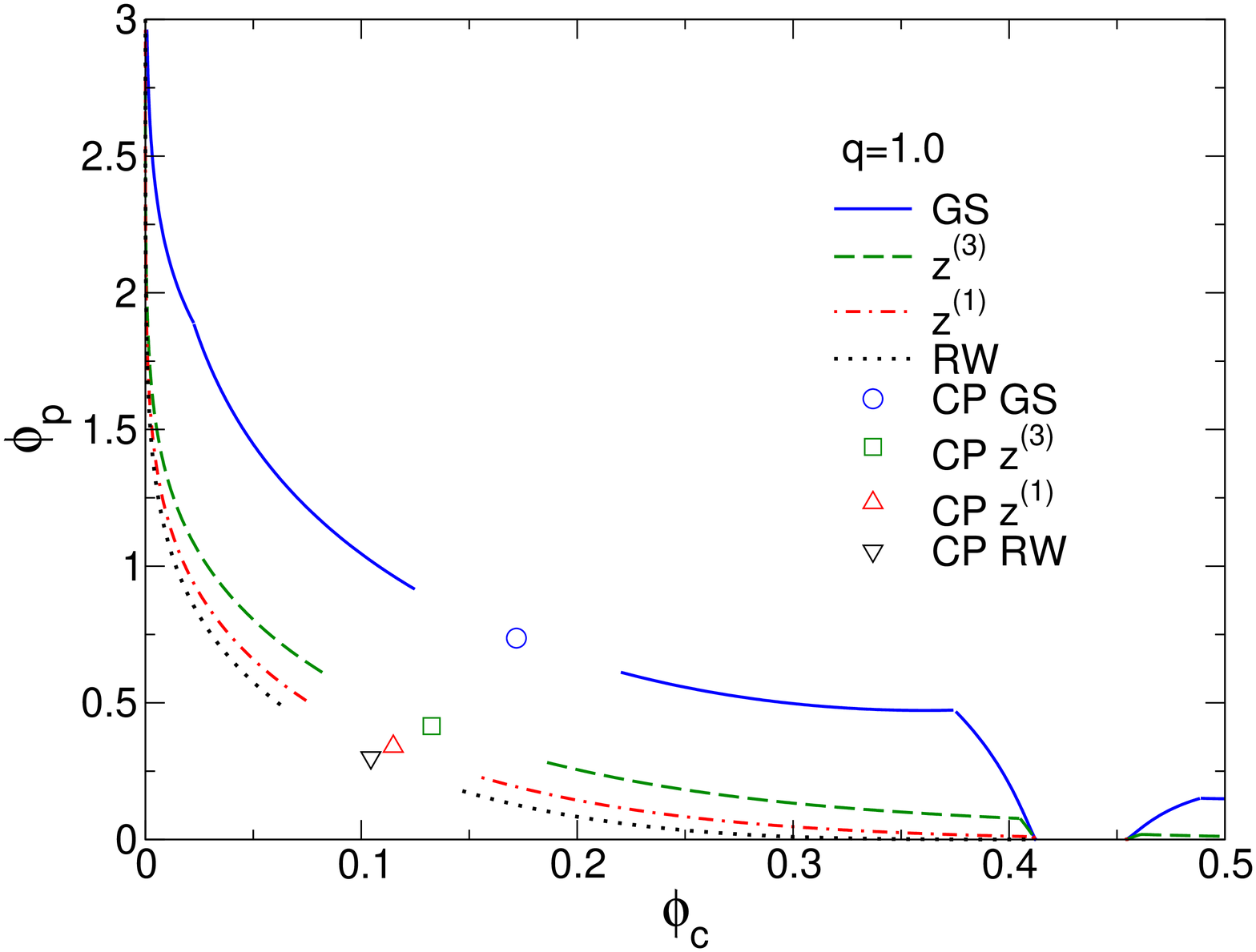,angle=0,width=8truecm} \hspace{0.5truecm} 
\end{tabular}
\end{center}
\caption{Phase diagram as a function of $\phi_c$ and $\phi_p$ for a 
system with $s = 1.06$, for several different solvent qualities: 
good-solvent (GS) conditions, $z = z^{(3)}$, $z = z^{(1)}$, and 
$\theta$ conditions (RW). In the upper panel we report results for 
$q = 0.5$, in the lower one results for $q = 1$.
We also report the critical points (CP). For $q=0.5$ and 
good-solvent conditions, no fluid-fluid transition occurs.
}
\label{Phase-diag-charged-s1p06}
\end{figure}

In Fig.~\ref{Phase-diag-charged-GS} we report the binodals for the good-solvent
case for $s=1$, 1.03, 1.06 as a function of $s^3 \phi_c$ (this guarantees 
that all binodals converge to the same values as $\phi_p\to 0$). 
As already discussed in Ref.~\onlinecite{GT-08}, charge effects are large,
moving the binodals towards larger values of $\phi_p$. Note that, for 
$q = 0.5$, liquid-gas coexistence disappears as $s$ is increased. 
In Fig.~\ref{Phase-diag-charged-s1p06} we report the binodals for 
$s=1.06$, for different solvent quality. As already discussed, 
the $\theta$ binodal and those for $z=z^{(1)}$ and $z=z^{(3)}$ are close, 
indicating that quality of the solution plays here a minor role, 
in spite of the fact that thermodynamic dilute properties vary significantly.
On the other hand, significant changes occur when the system is close to the 
good-solvent regime.

\section{Conclusions} \label{sec6}

In this paper we studied phase coexistence for dispersions of spherical 
colloids  and neutral polymers. We use an approximate scheme, known
as generalized free-volume theory (GFVT),\cite{ATL-02,FT-07,FT-08,LT-11}
that models polymers as soft spheres with a density-dependent radius
identified with the so-called depletion thickness.\cite{LT-11} This 
approach has already been applied\cite{FT-08} to good-solvent and 
$\theta$ polymers and is extended here to the thermal crossover region 
between these two regimes. 

First, we investigate the good-solvent behavior, using the accurate expressions
for the polymer equation of state and for the depletion thickness
of Refs.~\onlinecite{Pelissetto-08,DPP-thermal,DPP-depletion}. 
For $q\lesssim 1$, our results are in full agreement with those 
presented in Ref.~\onlinecite{FT-08}, in which slightly different expressions
for the polymer properties were used. On the other hand, our GFVT results 
differ significantly from those of Ref.~\onlinecite{FT-08} in the protein 
limit $q\gg 1$. This is due to the fact that their expression for the 
depletion thickness becomes inaccurate for $q\gtrsim 4$.\cite{DPP-depletion}
In any case, comparison with full-monomer Monte Carlo data \cite{BML-03,MLP-12}
and with experiments \cite{MDEGH-08,MDEGH-09,MDEGH-10} shows that GFVT 
is not predictive for $q\gtrsim 1$. This is not unexpected, since GFVT 
models polymers as soft spheres, an approximation that should be reasonably
accurate only for $q\lesssim 1$. Note that the comparison with 
full-monomer results is not straightforward, as the GFVT results refer 
to polymers in the scaling regime, while simulation are limited to relatively
short chains. Hence, a careful extrapolation of the Monte Carlo data is 
needed before making any comparison.

We also investigate the phase diagram in terms of the crossover variable $z$.
We find that phase behavior is not very sensitive to solvent quality 
close to the $\theta$ point. If we parametrize the thermal crossover in 
terms of the adimensional ratio $A_{2,pp}$ [or, equivalently, in terms
of the interpenetration ratio $\Psi = A_{2,pp}/(4 \pi^{3/2})$], we find that 
binodals and critical points do not change significantly as $A_{2,pp}$
changes from 0 to $0.2 A_{2,pp,GS}$, where $A_{2,pp,GS}$ is the 
value under good-solvent conditions. If $A_{2,pp}$ is increased further,
for instance by increasing temperature, we observe a systematic drift of the 
binodals towards larger values of $\phi_p$. Note, however, that in the 
range $A_{2,pp}/A_{2,pp,GS} \lesssim 0.6$ changes are rather small. 
Apparently, the phase diagram is very sensitive to solvent quality only
close to the good-solvent regime. 

Finally, we investigate the phase diagram for charged colloids in an 
aqueous salt solution. We use here a generalization of GFVT developed
in Refs.~\onlinecite{FDT-05,GT-08}. For $z \lesssim z^{(3)}$,
the additional repulsion between the charged colloids changes 
only slightly the phase diagram. On the other hand, close to the 
good-solvent region, phase behavior shows a strong dependence 
both on solvent quality and on the parameter $s$ that parametrizes 
the charge effects. Clearly, a meaningful comparison of experimental results 
with theory requires a careful determination of the solvent quality, 
i.e., the swelling ratio $\alpha$ or the second-virial combination $A_{2,pp}$,
as discussed in the supplementary material,\cite{suppl} as well as of 
the electric properties, i.e. Bjerrum and Debye screening length, 
colloid charge, etc., of the solution.

\section*{Acknowledgments}

We thank 
Profs. Rubinstein and Colby for useful correspondence and for providing the 
experimental data analyzed in the supplementary material.
C.P. is supported by the Italian Institute of Technology (IIT) under the
SEED project grant number 259 SIMBEDD – Advanced Computational Methods for
Biophysics, Drug Design and Energy Research.

\appendix 

\section{Third virial coefficients in the GFVT approach}

In this appendix we determine the second and third order virial coefficients 
predicted by the GFVT expression (\ref{GFVT-potential}) for the semigrand 
potential.

To begin with, we determine the low-density expansion of $\Omega$ in 
terms of the virial coefficients. If the
pressure has the expansion (\ref{virial-expansion}), the canonical free energy
can be expanded as
\begin{eqnarray}
&& {\beta F_{\rm can}\over V} \approx  \rho_c \ln (\rho_c\lambda^3_c) + 
      \rho_p \ln (\rho_p\lambda^3_p) - \rho_c - \rho_p \nonumber \\
&& \quad + B_{2,cc} \rho_c^2 + B_{2,pp} \rho_p^2 + B_{2,cp} \rho_c \rho_p 
    \\
&& \quad + {1\over 2} \left( 
   B_{3,ccc} \rho_c^3 + B_{3,ppp} \rho_p^3 + B_{3,ccp} \rho_c^2 \rho_p  + 
   B_{3,cpp} \rho_c \rho_p^2\right),
\nonumber 
\end{eqnarray}
where $\lambda_c$ and $\lambda_p$ are the colloid and polymer thermal lengths.
The expansion of the chemical potential follows immediately. If $z_p = e^{\beta
\mu_p}$, we have 
\begin{eqnarray}
&& z_p /(\lambda^3_p\rho_p)  \approx
    1 + B_{2,cp} \rho_c + 2 B_{2,pp} \rho_p \label{zp-to-rho} \\
&& \qquad + (B_{2,cp}^2 + B_{3,ccp}) \rho_c^2/2 +  
            (4 B_{2,pp}^2 + 3 B_{3,ppp}) \rho_p^2/2 
\nonumber \\
&& \qquad + 
           (2 B_{2,cp} B_{2,pp} + B_{3,cpp}) \rho_c\rho_p. 
\nonumber 
\end{eqnarray}
The fugacity $z_p$ can also be expressed in terms of the reservoir polymer
density $\rho_p^{(r)}$. Its expansion can be obtained by using
Eq.~(\ref{zp-to-rho}), replacing $\rho_p$ with $\rho_p^{(r)}$, and setting 
$\rho_c = 0$. These expressions can be used to obtain the expansion of 
$\rho_p$ in terms of $\rho_p^{(r)}$ and $\rho_c$. 
We obtain 
\begin{eqnarray}
&& \rho_p/\rho_p^{(r)} \approx 1 + 
   \rho_c \left[- B_{2,cp} + {1\over 2} (B_{2,cp}^2 - B_{3,ccp}) \rho_c 
\right. \nonumber \\
&& \qquad \left. 
    + (2 B_{2,cp} B_{2,pp} - B_{3,cpp}) \rho_p^{(r)} \right].
\end{eqnarray}
We are now in the position to compute 
\begin{equation}
{\Omega\over V} = {F_{\rm can}\over V} - \mu_p \rho_p.
\end{equation}
We obtain 
\begin{eqnarray}
&& {\beta \Omega\over V} \approx 
{\beta F_{\rm coll}(\rho_c)\over V} - \rho_p^{(r)}
\nonumber \\
&& \quad + \left ( B_{2,cp} \rho_c  - B_{2,pp} \rho_p^{(r)} \right)
\rho_p^{(r)} - B_{3,ppp} (\rho_p^{(r)})^3 
\label{semigrand-expansion} \\
&& \quad  + {1\over 2} \rho_c \rho_p^{(r)} 
   \left(- B_{2,cp}^2 \rho_c + B_{3,ccp} \rho_c + B_{3,cpp} \rho_p^{(r)}
    \right).
\nonumber
\end{eqnarray}
Let us now consider the GFVT case. The expressions we use for $K_p$ and 
$\delta_s$ have an expansion of the form \cite{DPP-depletion}
\begin{eqnarray}
&& K_p \approx 1 + 2 B_{2,pp}^{(e)} \rho_p + 3 B_{3,ppp}^{(e)} ,
\\
&& \delta_s/R_c \approx - 1 + C + {1\over 3} {\rho_p C \over B_{2,cp}^{(e)} } 
    (B_{3,cpp}^{(e)} - 2 B_{2,pp}^{(e)} B_{2,cp}^{(e)} ).
\nonumber 
\end{eqnarray}
Here, we set
\begin{eqnarray}
C = (B_{2,cp}^{(e)}/V_c)^{1/3},
\end{eqnarray}
and the superscript $(e)$ indicates that each quantity is our estimate 
of the corresponding virial coefficient. 
If we now expand Eq.~(\ref{GFVT-potential}), we obtain an expansion of the 
form (\ref{semigrand-expansion}) with $B_{2,cp} = B_{2,cp}^{(e)}$, 
$B_{2,pp} = B_{2,pp}^{(e)}$, $B_{3,cpp} = B_{3,cpp}^{(e)}$, 
$B_{3,ppp} = B_{3,ppp}^{(e)}$. The only nontrivial coefficient
is $B_{3,ccp}$, for which we obtain
\begin{equation}
B_{3,ccp}= 8 B_{2,cp}^{(e)} V_c + 2 V_c^2
 - 9 \left(B_{2,cp}^{(e)}/V_c\right)^{2/3} V_c^2.
\end{equation}
Eq.~(\ref{A3ccp-GFVT}) follows.

\clearpage

\section{Supplementary Material: Analysis of the experimental data}

\subsection{Two-parameter approach to thermal crossover}

In Section II we discussed the behavior of structural and 
thermodynamic properties
in the thermal crossover region. Here, we wish to compute 
the nonuniversal constants that allow one to relate theoretical results for 
Edward's two parameter model (TPM)
and experimental data. We will consider the experimental systems 
studied in Refs.~\onlinecite{PS-suppl,PCP-suppl,PIB-suppl},
which were already analyzed in the TPM framework. However,
the TPM expressions used in those analyses are not accurate and, 
in particular, differ significantly from the correct result close to the 
good-solvent regime. We have therefore decided to repeat the analyses,
matching the experimental results with the very precise TPM predictions for the 
crossover functions associated with the 
second virial coefficient and the swelling ratio reported in 
Refs.~\onlinecite{CMP-08-suppl,DPP-depletion-suppl}.

Before discussing the experimental results, 
let us first summarize the basic theoretical ideas,
which have already been presented in Sec.~II. 
General renormalization-group arguments
\cite{Duplantier-suppl,dCJ-book-suppl,Schaefer-99-suppl} indicate that,
in the thermal crossover region above $T_\theta$, global 
properties of the polymer solution satisfy a general scaling form:
\begin{equation}
\mathcal{O}(T,L,\rho)=\alpha_1 \mathcal{O}_G(L,\rho) 
   g_{\cal O}[L^{1/2} f_T(T) (\log L)^{-{4/11}}, \phi_p].
\label{scaling-suppl}
\end{equation}
Here $\mathcal{O}(T,L,\rho)$ 
is the value of the observable under consideration as 
a function of temperature $T$, the degree of polymerization $L$, and 
number density $\rho = N_p/V$, 
$\mathcal{O}_G(L,\rho) = \mathcal{O}(T_\theta,L,\rho)$ is 
the value of $\mathcal{O}$ at the $\theta$ temperature, 
$\phi_p = 4\pi \rho {R}_g^3/3$, where 
${R}_g$ is the radius of gyration 
of an isolated polymer of degree of polymerization $L$ at temperature $T$.
Chemical details are included in the amplitude $\alpha_1$ and in the function 
$f_T(T)$ of the temperature, which vanishes at the $\theta$ point:
$f_T(T_\theta) = 0$.
The function $g_{\cal O}$, which satisfies $g_{\cal O}(0,\phi_p) = 1$,
 is universal, i.e., independent of the microscopic chemical 
details, and can be identified with the TPM crossover function associated with 
${\cal O}$.\cite{Sokal-suppl} In Eq.~(\ref{scaling-suppl})
the relevant variable which parametrizes the crossover between $\theta$ and good-solvent
behavior is the combination $L^{1/2} f_T(T) (\log L)^{-{4/11}}$. However, 
the logarithmic dependence on $L$  is hardly measurable in experiments in which $L$ 
varies by no more than one order of magnitude, hence, 
as usual in most polymer literature, 
we will neglect the logarithmic dependence, 
rewriting Eq.~(\ref{scaling-suppl}) as
\begin{equation}
\mathcal{O}(T,L,\rho)=\alpha_1 \mathcal{O}_G(L,\rho) 
   g_{\cal O}[L^{1/2} f_T(T), \phi_p].
\label{scaling2-suppl}
\end{equation}
As a second comment, note that Eq.~(\ref{scaling-suppl}) is strictly
valid for $L\to \infty$.  Close to the $\theta$ point, theory
predicts\cite{Duplantier-suppl,dCJ-book-suppl,Schaefer-99-suppl} corrections 
that decrease as $1/\ln L$. However, as we discuss below, for the experimental
quantities 
we consider, such corrections are small compared to the typical experimental
uncertainties. Therefore, we will neglect these contributions and we will apply
Eq.~(\ref{scaling2-suppl}) to the analysis of the experimental systems without 
introducing any type of logarithmic correction. 

To define completely the arguments of the scaling function 
$g_{\cal O}$, 
we must specify the 
function $f_T(T)$, which represents what is called a {\em 
nonlinear scaling field} in the 
renormalization-group language.\cite{Wegner-76-suppl} It is an analytic function of
temperature which vanishes at the $\theta$ point. Therefore, it has 
a regular expansion around $T= T_\theta$ of the form
\begin{equation}
f_T(T) = \alpha_0 (T-T_\theta) [1 + \alpha_1 (T-T_\theta) + \alpha_2 (T-T_\theta)^2 + 
   \ldots].
\label{fT-suppl}
\end{equation}
The constants $\alpha_1$, $\alpha_2$, $\ldots$, are completely specified by the 
microscopic details of the model, while $\alpha_0$ is a normalization 
system-dependent constant  that has to be properly fixed to guarantee the 
universality of the scaling function $g_{\cal O}$.
Typically, experiments are not so precise to allow us 
to determine the $T$ dependence of the function $f_T(T)$. Therefore,
in the following we shall use two simple expressions that have the 
correct behavior for $T\to T_\theta$. We will consider
\begin{eqnarray}
&& f_T^{(1)}(T) = \alpha_0 {T-T_\theta\over T_\theta}, \\ 
&& f_T^{(2)}(T) = \alpha_0 {T-T_\theta\over T}. 
\label{fT12-suppl}
\end{eqnarray}
The first one corresponds to truncating Eq.~(\ref{fT-suppl}) at first order 
(with a simple redefinition of $\alpha_0$), while the second one
is the classical form that can be found in the 
experimental literature and in popular
textbooks of polymer physics (for instance, in Ref.~\onlinecite{DoiEdwards}).
The constant $\alpha_0$ is fixed by identifying $f_T(T) L^{1/2}$ with the 
Zimm-Fixman-Stockmayer variable\cite{ZSF-53-suppl} $z$. 
In other words, if $g_{{\cal O},\rm TPM}(z,\phi_p)$
is the scaling function associated with $\cal O$ computed in the TPM, 
we fix $\alpha_0$ so that 
\begin{equation}
g_{{\cal O},\rm TPM}(z,\phi_p) \approx g_{\cal O}(z=f_T(T) L^{1/2},\phi_p).
\label{TPM-expt-suppl}
\end{equation}
The use of the two functions $f_T^{(1)}(T)$ and $f_T^{(2)}(T)$ provides
slightly different estimates of the constant $\alpha_0$. Such a difference
gives us an indication of the uncertainty 
on the final result due to fact that 
we do not have a precise knowledge of the $T$ dependence of $f_T(T)$.
Note that, in the TPM framework, 
$\alpha_0$ does not depend on the observable $\cal O$. In other
words, if $\alpha_0$ is fixed by requiring the validity of 
Eq.~(\ref{TPM-expt-suppl}) for a given quantity $\cal O$, 
Eq.~(\ref{TPM-expt-suppl}) should also be satisfied by 
any other quantity.

\subsection{Experimental systems}

\begin{table}[tbp!]
\caption{Polymer solutions considered in the present analysis. 
For each sample
we report the weight-average molar mass $M_w$, the $\theta$ temperature,
the zero-density 
squared radius of gyration ${R}_{g,\theta}^2$ at the $\theta$ point, 
and the ratio  
$\ell_k = {R}_{g,\theta}/M_w^{1/2}$. We also report the 
polydispersity index $M_w/M_n$, where 
$M_n$ is the number-average molar mass. }
\label{tab:samplef}
\begin{center}
\begin{tabular}{ccccc}
\hline
\hline
\multicolumn{5}{c}{Polystyrene (PS) in decalin (Ref.~\onlinecite{PS-suppl})\footnote{
Solution of decalin with 61.7\% of the cis-isomer. 
Sample A-5 was also studied in a solution with $51.4\%$ of cis-decalin.} } \\
\multicolumn{5}{c}{$M_w/M_n\lesssim1.07$} \\
\hline
Sample & $M_w/10^6$ & $T_\theta(K) $ & 
     ${R}_{g,\theta}^2$ (nm$^2$) & $\ell_k$ \\
 & (g$\cdot$mol$^{-1}$) &&& [nm$\cdot$(mol/g)$^{1/2}$] \\
\hline 
A-30	  &	4.40	  &	288.2	& 4550 & 0.0321 \\
A-5	  &	1.56	  &	288.4	& 1320 & 0.0291	\\
A-16	  &	1.05	  &	288.2	&  820 & 0.0279	\\
A-13	  &	0.186     &	288.0	&  158 & 0.0291	\\
\hline
\hline
\multicolumn{5}{c}{ Polychloroprene (PCP) in trans-decalin 
(Ref.~\onlinecite{PCP-suppl})} \\
\multicolumn{5}{c}{$M_w/M_n\approx1.1$} \\
\hline
Sample & $M_w/10^6$ & $T_\theta(K) $ & 
     ${R}_{g,\theta}^2$ (nm$^2$) & $\ell_k$ \\
 & (g$\cdot$mol$^{-1}$) &&& [nm$\cdot$(mol/g)$^{1/2}$] \\
\hline
f-16B 	& 1.66	&    274.9& 1500 & 0.0300	\\	
f-14B 	& 0.865	&    274.9&  770 & 0.0298	\\	
f-12   	& 0.587	&    274.9&  519 & 0.0295	\\
\hline
\hline
\multicolumn{5}{c}{Polyisobutilene (PIB) in isoamyl-isovalerate 
   (Ref.~\onlinecite{PIB-suppl}) }
 \\
\multicolumn{5}{c}{$M_w/M_n\approx1.1$} \\
\hline
\hline
Sample & $M_w/10^6$ & $T_\theta(K) $ & 
     ${R}_{g,\theta}^2$ (nm$^2$) & $\ell_k$ \\
 & (g$\cdot$mol$^{-1}$) &&& [nm$\cdot$(mol/g)$^{1/2}$] \\
\hline
A-42 &	4.7	&    295.1	&4500 &	0.031	\\
A-32 &	3.1	&    295.1	&3200 & 0.032	\\	
A-73 &	1.4	&$\approx$295.2	&1350 & 0.031	\\	
A-54 &	0.81	&$\approx$295.3	& 730 & 0.030	\\
\hline
\hline
\end{tabular}
\end{center}
\end{table}

When comparing with the experimental data, it is more convenient to use the 
weight-average molar mass $M_w$ instead of $L$. Hence, we write $z$ as 
\begin{eqnarray}
z = a^{(1)} {T-T_\theta\over T_\theta} M_w^{1/2}, 
\label{z1-def-suppl} \\
z = a^{(2)} {T-T_\theta\over T} M_w^{1/2}.
\label{z2-def-suppl}
\end{eqnarray}
Here, we determine the nonuniversal constants $a^{(1)}$ and $a^{(2)}$ for three 
different polymer solutions that show an extensive 
thermal crossover close to room temperature:
polystyrene (PS) in a mixture of isomers of decalin,\cite{PS-suppl} 
polychloroprene (PCP) in trans-decalin,\cite{PCP-suppl} and
polyisobutilene (PIB) in isoamyl-isovalerate.\cite{PIB-suppl} 
We use the experimental results presented in 
Refs.~\onlinecite{PS-suppl,PIB-suppl,PCP-suppl}.
Details are reported in Table \ref{tab:samplef}. 
The samples are identified by using the same name abbreviations as in the 
original articles. In all cases 
the $\theta$ temperature is identified as the one at which 
the second virial coefficient vanishes. 
In Table~\ref{tab:samplef}  we  also report the ratio
$\ell_k$ defined by
\begin{equation}
\ell_k = {R}_g(T_\theta, M_w) M_w^{-1/2}
\end{equation}
(${R}_g(T_\theta, M_w)$ is the zero-density readius of gyration),
which is expected to be constant at the $\theta$ point. Data for PIB and PCP
are reasonably constant. Significant variations are instead observed for the 
PS data. In particular, $\ell_k$ for the sample with the largest molecular 
weight is significantly larger than that obtained for the other samples. 
Berry \cite{PS-suppl} explained this discrepancy as a polydispersity 
effect. However, if we assume that $\ell_k \approx 0.0287$ 
nm$\cdot$(mol/g)$^{1/2}$ for a monodisperse sample---this is the 
average of the results for samples A-5,A-16, and 
A-13---Eq.~(\ref{polydispRgtheta-suppl}), which relates 
${R}_g^2$  for a polydisperse sample with the 
corresponding monodisperse quantity,
gives $M_w/M_n \approx 1.33$. Therefore, the value 
$\ell_k \approx 0.0321$ nm$\cdot$(mol/g)$^{1/2}$ can be explained as 
a polydispersity effect only if the mass distribution is much broader than
that observed in fractionation studies of the same samples---they provide 
$1.01 \lesssim M_w/M_n \lesssim 1.07$. As we shall discuss below, such an 
interpretation is also inconsistent with the results for the 
second virial coefficient. More likely, the quoted value for ${R}_g^2$ 
is the result of an incorrect extrapolation to the zero-angle limit 
(note that in the experiment the angle varies between $18^\circ$ and 
$135^\circ$), which, as stated by Berry himself, becomes difficult 
when ${R}_g^2 \gtrsim 4\cdot 10^{-11}$ cm$^2$. For these reasons, we will
not use the measured value of ${R}_g^2$ for sample A-30. Instead, we will
use the scaling formula ${R}_g^2 = \ell_k^2 M_w$ with 
$\ell_k \approx 0.0287$ nm$\cdot$(mol/g)$^{1/2}$. This gives
\begin{equation}
{R}_{g,\theta}^2 = 3620 \,\hbox{nm}^2
\label{Rg2-A30}
\end{equation}
for sample A-30.

To determine the nonuniversal constants, we use two different quantities.
First, we consider the swelling ratio 
\begin{equation}
\alpha (T,M_w) = {{R}_g(T,M_w)\over {R}_g (T_\theta,M_w)}, 
\end{equation}
where ${R}_g(T,M_w)$ is the radius of gyration of the polymer in 
the infinite-dilution limit. For a monodisperse system, such a quantity 
should approach the TPM prediction \cite{CMP-08-suppl}
\begin{equation}
\alpha_{TPM}(z) = (1 + 10.9288 z + 35.1869 z^2 + 30.4463 z^3)^{0.0583867}
\label{alpha-to-z-suppl}
\end{equation}
for large values of $M_w$, with a proper choice of the normalization constant. 
Second, we consider the second virial coefficient 
$B_{2,{\rm expt}}$, which can be determined either by measuring 
the small density behavior of the osmotic pressure or from 
scattering experiments in dilute solutions (note that in the 
presence of polydispersity, these two methods provide different 
quantities, see Sec.~\ref{polydispersity-suppl} for the precise definitions).
The universal ratio $A_2$ (to lighten the notation, since we do not consider 
colloids here, we use $A_2$ instead of $A_{2,pp}$ as in the text)
can be expressed in terms of $B_{2,{\rm expt}}$ as 
\begin{equation}
A_2 (T,M_w) = {M^2_w B_{2,{\rm expt}}(T,M_w) \over N_A {R}_g(T,M_w)^3 },
\end{equation}
where $N_A$ is Avogadro's number. With a proper identification 
of the nonuniversal constant, $A_2 (T,M_w)$ should approach the TPM expression
\begin{eqnarray}
&& A_{2,TPM}(z) =
4 \pi^{3/2} z (1 + 19.1187 z
\label{A2-to-z-suppl}
\\
&& \qquad\quad
    + 126.783 z^2 + 331.99 z^3 + 268.96 z^4)^{-1/4}.
\nonumber
\end{eqnarray}
It is important to stress that the TPM expressions 
(\ref{alpha-to-z-suppl}) and (\ref{A2-to-z-suppl}) 
refer to a strictly monodisperse sample. 
In the presence of polydispersity, corrections should 
be considered. If the swelling ratio is defined in terms of the 
Z-averaged radius of gyration,\cite{Rubinstein-suppl} 
which is the quantity obtained in 
scattering experiments, we have 
\begin{equation}
\alpha_Z^2(z) = \alpha(z)^2 C(z),
\end{equation}
where $C(z)$ is the polydispersity correction, which depends on the
polymer mass distribution. The function $C(z)$ is computed in 
Sec.~\ref{polydispersity-suppl}. If the Schulz distribution is assumed,
\cite{Rubinstein-suppl} 
for systems which satisfy $M_w/M_n \lesssim 1.1$, $C(z)$ represents a small
correction, of the order of 2\% at most. Therefore, for the swelling ratio,
polydispersity effects are negligible, given the uncertainty on the data. 
On the other hand, polydispersity effect on $A_2$ are very large, at least
in the good-solvent region. For instance, in a system with Schulz parameter 
$s = 10$ (corresponding to $M_w/M_n  = 1.10$), 
$A_2$ is 40\% smaller than the value obtained for a monodisperse 
system,
see Sec.~\ref{polydispersity-suppl} and Ref.~\onlinecite{Polydisp-suppl}.
More precisely, the combination $A_{2,Z}$ which is determined in 
scattering experiments has a good-solvent value of 4.21 for $s = 10$,
while it takes the value\cite{CMP-06-suppl} 5.50 for a 
monodisperse solution. The results of Ref.~\onlinecite{Polydisp-suppl}
also allow us to exclude that sample A-30 is broadly polydispersed, 
with $M_w/M_n \approx 1.33$. Indeed, using the Schulz distribution,
we would obtain $s\approx 3$. Then, using the results reported in 
Ref.~\onlinecite{Polydisp-suppl}, we would estimate $A_{2,Z} \approx 2.7$ 
in the good-solvent regime, which is significantly lower than what is 
obtained experimentally. Indeed,
for $T = 363.2$ K and 377.8 K, which are probably close to the 
good-solvent regime ($T_\theta \approx 288$ K), the experimental results 
for sample A-30 give $A_2 = 3.94$ and $A_2 = 4.01$, if the measured value of 
${R}_g$ is used. Both values are significantly larger than 
$A_{2,Z} \approx 2.7$, excluding that polydispersity is the origin 
of the quite large value for the radius of gyration at the $\theta$ point.
Note that, if we use instead 
${R}_{g,\theta}^2 = 3620$ nm$^2$, 
for $T = 363.2$ K and 377.8 K we obtain 
$A_2 = 5.34$ and 5.44, which are close to the good-solvent 
value 5.50 appropriate for monodisperse solutions.\cite{CMP-06-suppl} 

\begin{figure}[tbp]
\begin{center}
\epsfig{file=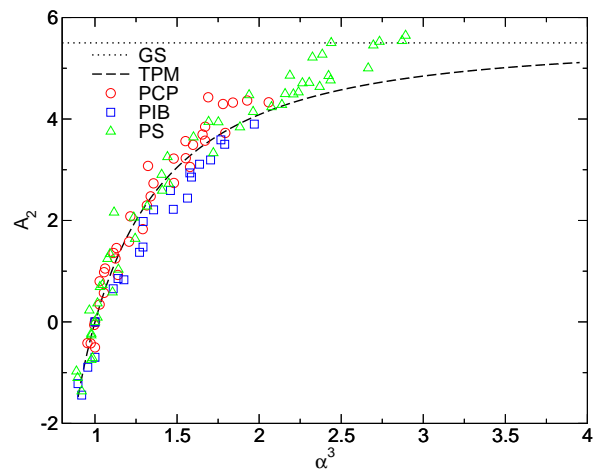,width=8cm,angle=0}
\end{center}
\caption{Combination $A_2(T,M_w)$ versus $\alpha(T,M_w)^3$ for three 
different experimental systems. The dashed line is the TPM prediction.}
\label{fig:generalmap}
\end{figure}

Before computing $a^{(1)}$ and $a^{(2)}$, we verify that the logarithmic 
scaling corrections related the three-body interactions can be neglected 
even at the $\theta$ point for the experimental quantities we consider. 
This is of course a necessary requirement 
to apply TPM results to the data. For this purpose,
in Fig.~\ref{fig:generalmap}  we plot $A_2(T,M_w)$ 
versus $[\alpha(T,M_w)]^3$. We also report the theoretical TPM curve,
obtained by expressing $A_{2,TPM}(z)$ in terms of $[\alpha_{TPM}(z)]^3$. This 
comparison does not require us to fix the normalization of $z$, hence there 
are no free parameters. Within the accuracy of the data, 
the experimental results fall on top of the theoretical curve. 
Again, note that it is crucial to correct the results for PS sample A-30.
Otherwise, no agreement would be observed for large $\alpha^3$.
Clearly, the 
logarithmic three-body corrections are not detectable, at least 
with the experimental resolution of 
Refs.~\onlinecite{PCP-suppl,PIB-suppl,PS-suppl}. A second clear indication that 
logarithmic corrections are small for the quantities at hand is the fact that 
the $\theta$ temperature (which is defined as the temperature at which 
the second virial coefficient vanishes) is essentially independent of $M_w$.
This conclusion is related to the considered quantities and to 
the precision of the data: for 
polystyrene in cyclohexane, tricritical effects were observed in 
Ref.~\onlinecite{DJdC-86-suppl} very close to the $\theta$ point. 
Moreover, tricritical
corrections are essential to describe the behavior of higher-order virial
coefficients. For instance, the experimental results for PS
in cyclohexane \cite{NNT-91-suppl} and for PIB in  
isoamyl-isovalerate,\cite{ANNT-94-suppl} as well as numerical results 
for lattice self-avoiding walks,\cite{PH-05-suppl} 
show that the third virial coefficient does not vanish at the 
$\theta$ point and that it 
increases as $T$ is lowered below $T_\theta$. This behavior 
cannot be explained by the TPM: in the TPM the third virial coefficient
behaves as $z^3$ close to the $\theta$ point (in particular, it is 
negative below the $\theta$ point). Analogously, tricritical effects 
cannot be neglected in the equation of state in the semidilute regime. 
Hence, the results presented here 
apply in general only away from the $\theta$ point, in the region in which 
tricritical effects are negligible. The combination
$A_2$ and the swelling ratio $\alpha^2$ appear to be exceptions, 
since the TPM apparently provides a good description of the data even at the 
$\theta$ point.

\subsection{Direct determination of the nonuniversal constants}
\label{secB-suppl}

\begin{table}
\caption{Constants $a^{(1)}$ and $a^{(2)}$ (in units $(\hbox{mol/g})^{1/2}$), 
obtained by using different
methods: from the analysis of the swelling ratio ($\alpha^2$), 
from the analysis of $A_2$ ($A_2$), and by using the recursive method of Berry
\cite{PS-suppl} ($\alpha^2$ B and $A_2$ B). For the PS estimates
obtained by using the recursive method, we also report an error related 
to the uncertainty with which $s_B$ and $s_S$ are estimated.
In the last line we report the results of Ref.~\onlinecite{PS-suppl}
for PS, \onlinecite{PIB-suppl} for PIB, and 
\onlinecite{PCP-suppl} for PCP.}
\label{tab:results-aconstants}
\begin{tabular}{ccccccc}
\hline\hline
& \multicolumn{2}{c}{PS}  
& \multicolumn{2}{c}{PIB}  
& \multicolumn{2}{c}{PCP}  \\
Method&  
  $a^{(1)}$ & $a^{(2)}$ & $a^{(1)}$ & $a^{(2)}$ & $a^{(1)}$ & $a^{(2)}$ \\
\hline
$\alpha^2$ & 0.0047 & 0.0057  & 0.0031 & 0.0035 & 0.0046 & 0.0052 \\
$A_2$      & 0.0072 & 0.0084  & 0.0027 & 0.0028 & 0.0056 & 0.0060 \\
$\alpha^2$ B & 
   0.0069(4) & 0.0066(4) & 0.0031 & 0.0036 & 0.0049 & 0.0046 \\
$A_2$ B & 0.0070(5) & 0.0074(5) & 0.0027 & 0.0030 & 0.0051 & 0.0059 \\
liter.   & & \multicolumn{1}{c}{0.00975} &
& \multicolumn{1}{c}{0.003} &
& \multicolumn{1}{c}{0.0057} \\
\hline\hline
\end{tabular}
\end{table}

\begin{figure}
\begin{tabular}{c}
\epsfig{file=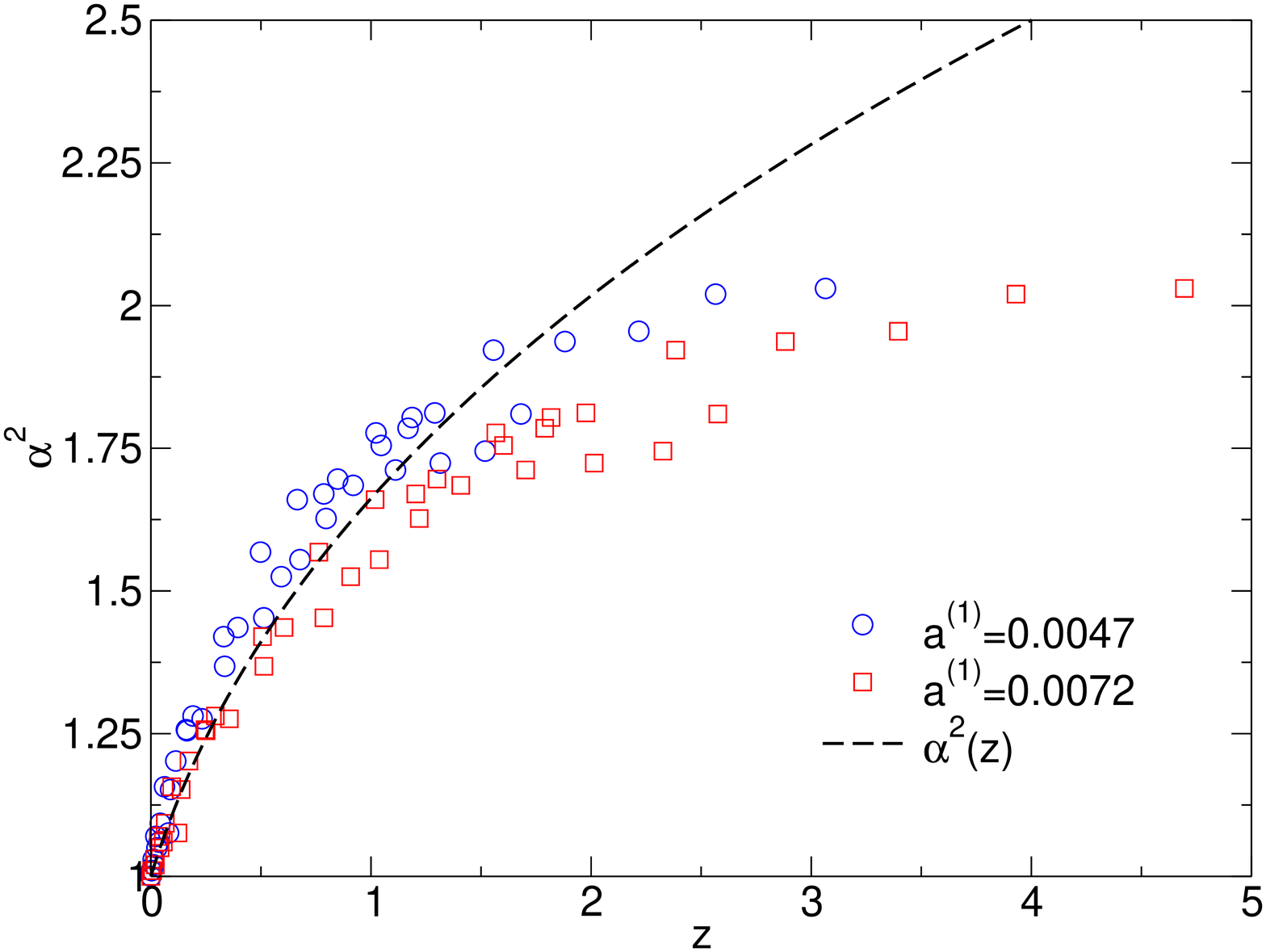,angle=0,width=9truecm} \hspace{0.5truecm} \\
\epsfig{file=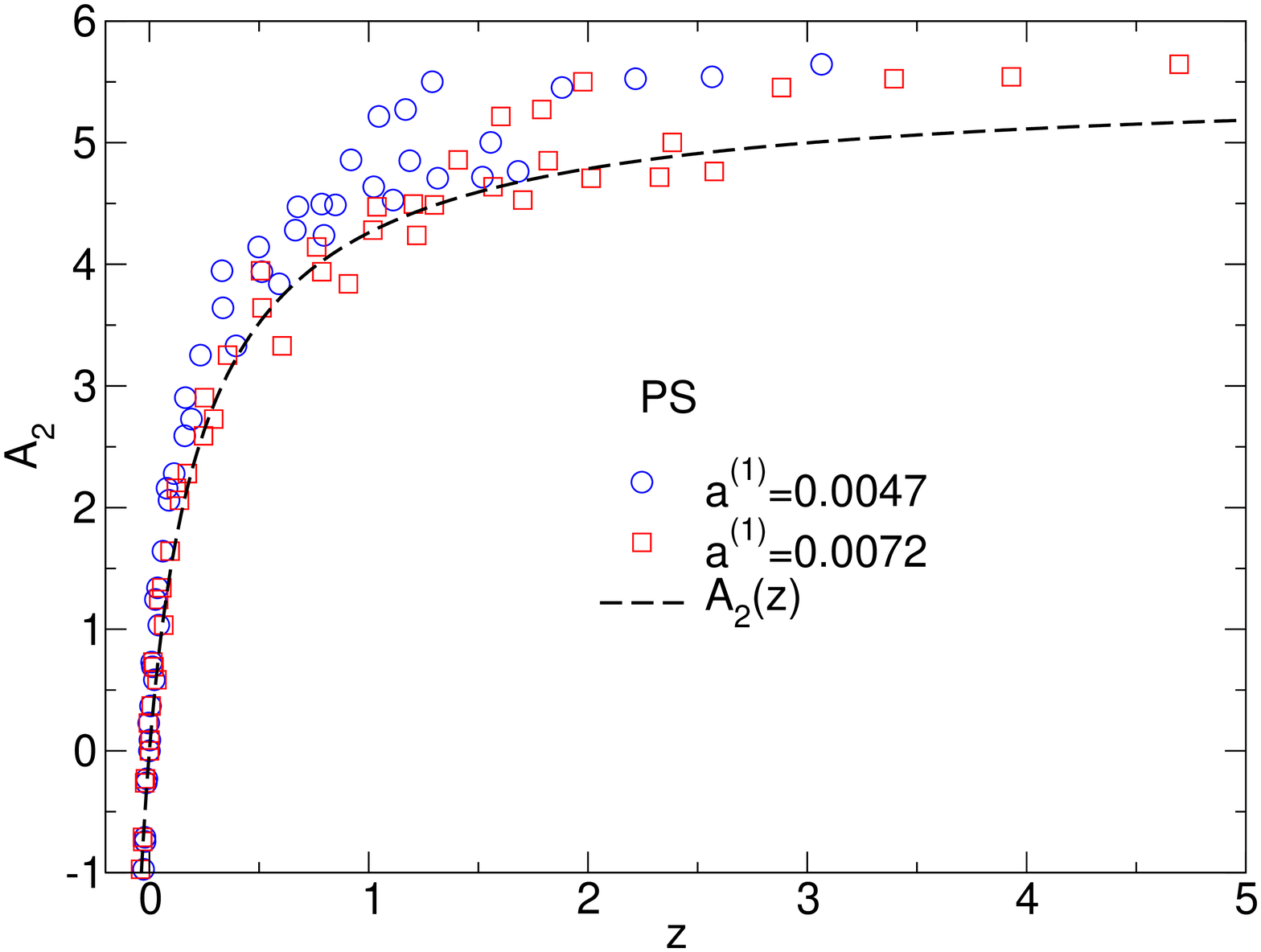,angle=0,width=9truecm} \hspace{0.5truecm} 
\end{tabular}
\caption{
Comparison of the experimental data for PS (Ref.~\onlinecite{PS-suppl})
and the TPM prediction. We define $z$ as in Eq.~(\ref{z1-def-suppl})
and use $a^{(1)} = 0.0047$ $(\hbox{mol/g})^{1/2}$ and 
$a^{(1)} = 0.0072$ $(\hbox{mol/g})^{1/2}$. We report  $\alpha^2$ (top) and 
$A_2$ (bottom). }
\label{fig-comparison-PS-suppl}
\end{figure}

The nonuniversal constants can be determined directly, by requiring the 
experimental data to match the TPM expression. 
For instance, if we consider the data for the second virial coefficient, 
we define
\begin{equation}
R(a^{(i)}) = \sum_a [A_a(T,M_w) - A_{TPM}(z_a) ]^2,
\end{equation}
where the sum is over all experimental data and $z_a$ is computed by 
using the appropriate function $f_T(T)$, with nonuniversal constant $a^{(i)}$.
The function $R(a^{(i)})$ is the usual goodness of the fit, in the absence of 
any knowledge of the error affecting the experimental data. 
The value of $a^{(i)}$ is then determined by minimizing $R$. 
The results are reported in Table ~\ref{tab:results-aconstants} 
(line $A_2$). The same procedure can be applied to $\alpha^2$. The
corresponding results are reported in Table ~\ref{tab:results-aconstants}
(line $\alpha^2$). 

\begin{figure*}[tbp]
\begin{center}
\begin{tabular}{cc}
\epsfig{file=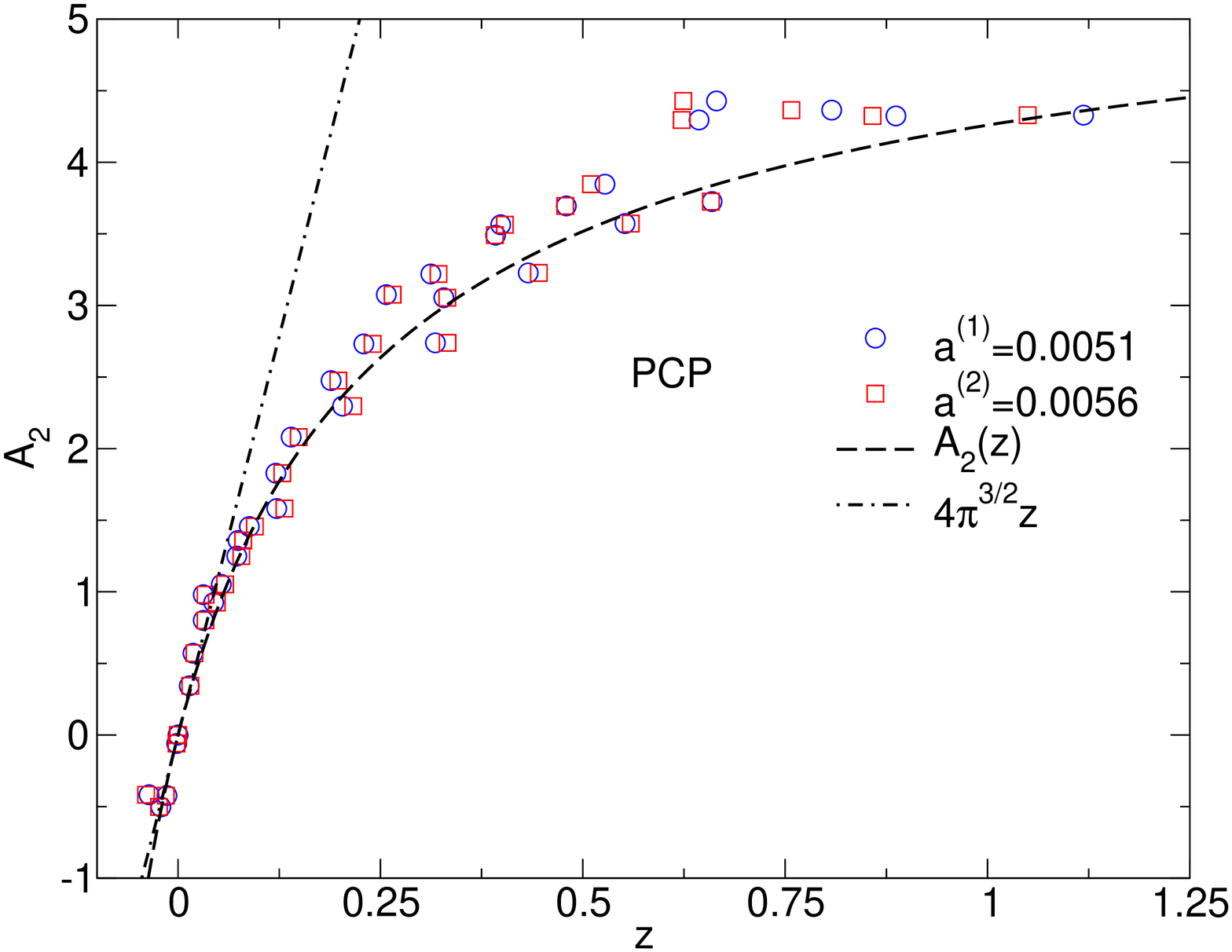,width=8cm,angle=0} & 
\epsfig{file=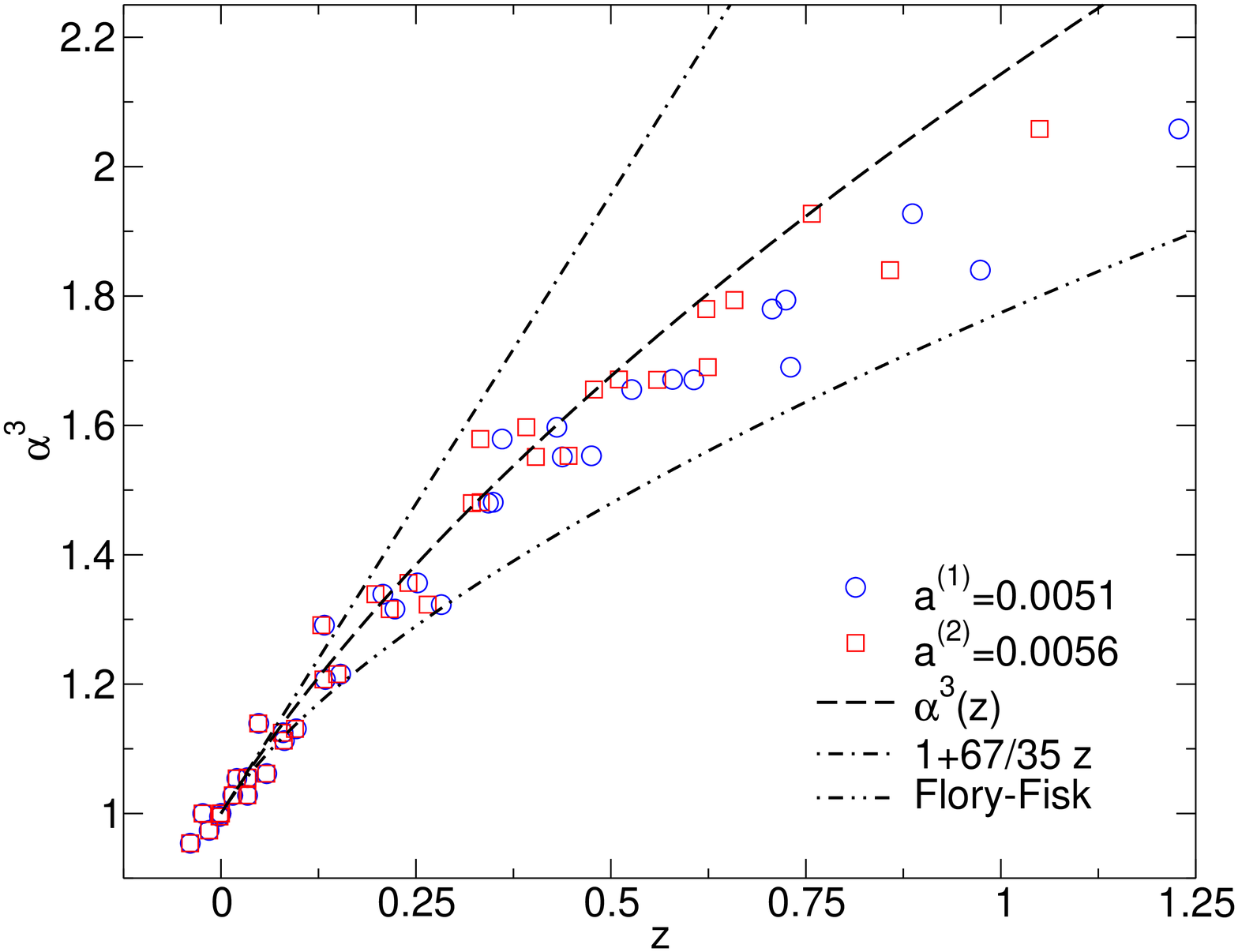,width=8cm,angle=0} \\
\epsfig{file=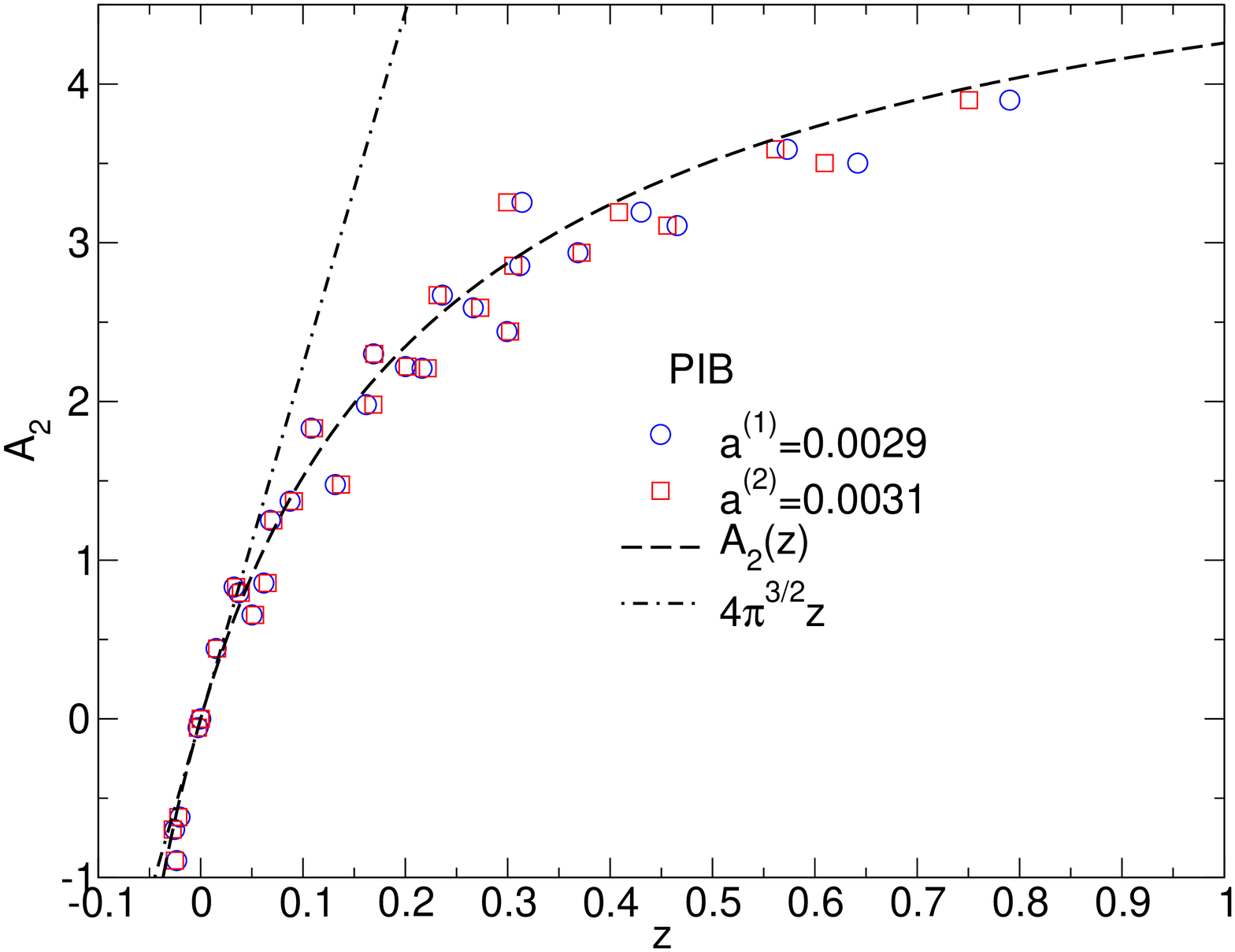,width=8cm,angle=0} & 
\epsfig{file=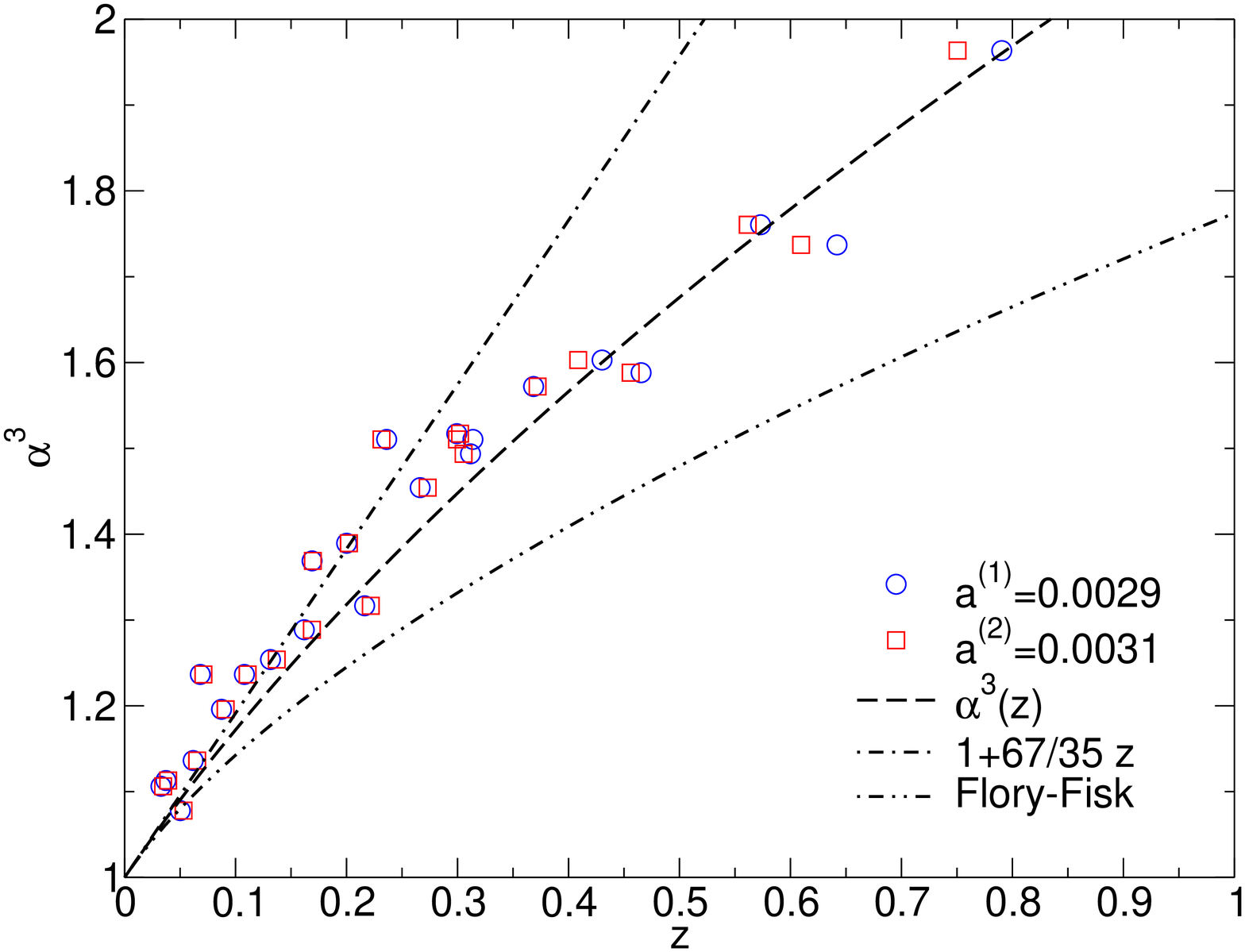,width=8cm,angle=0} \\
\epsfig{file=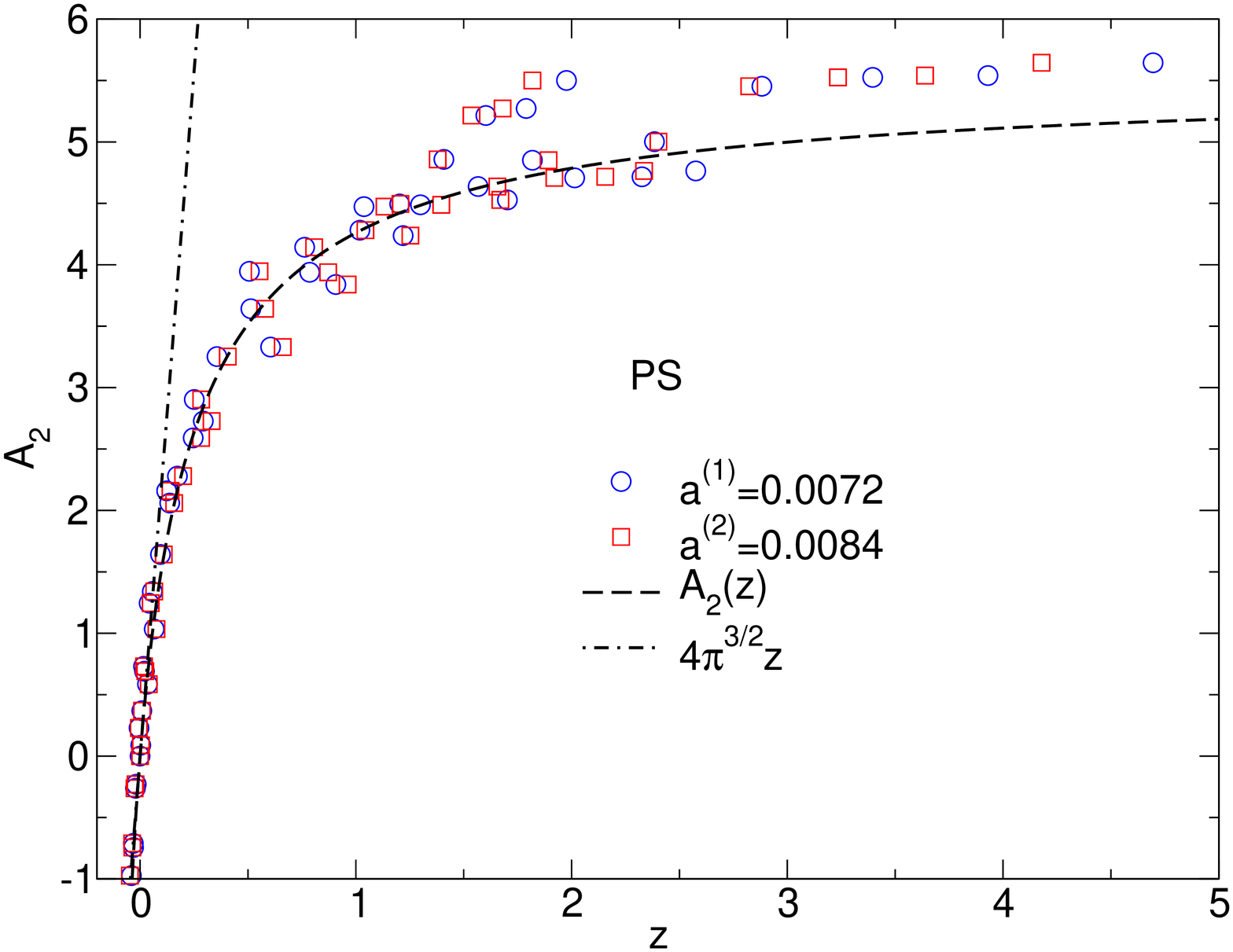,width=8cm,angle=0} & 
\epsfig{file=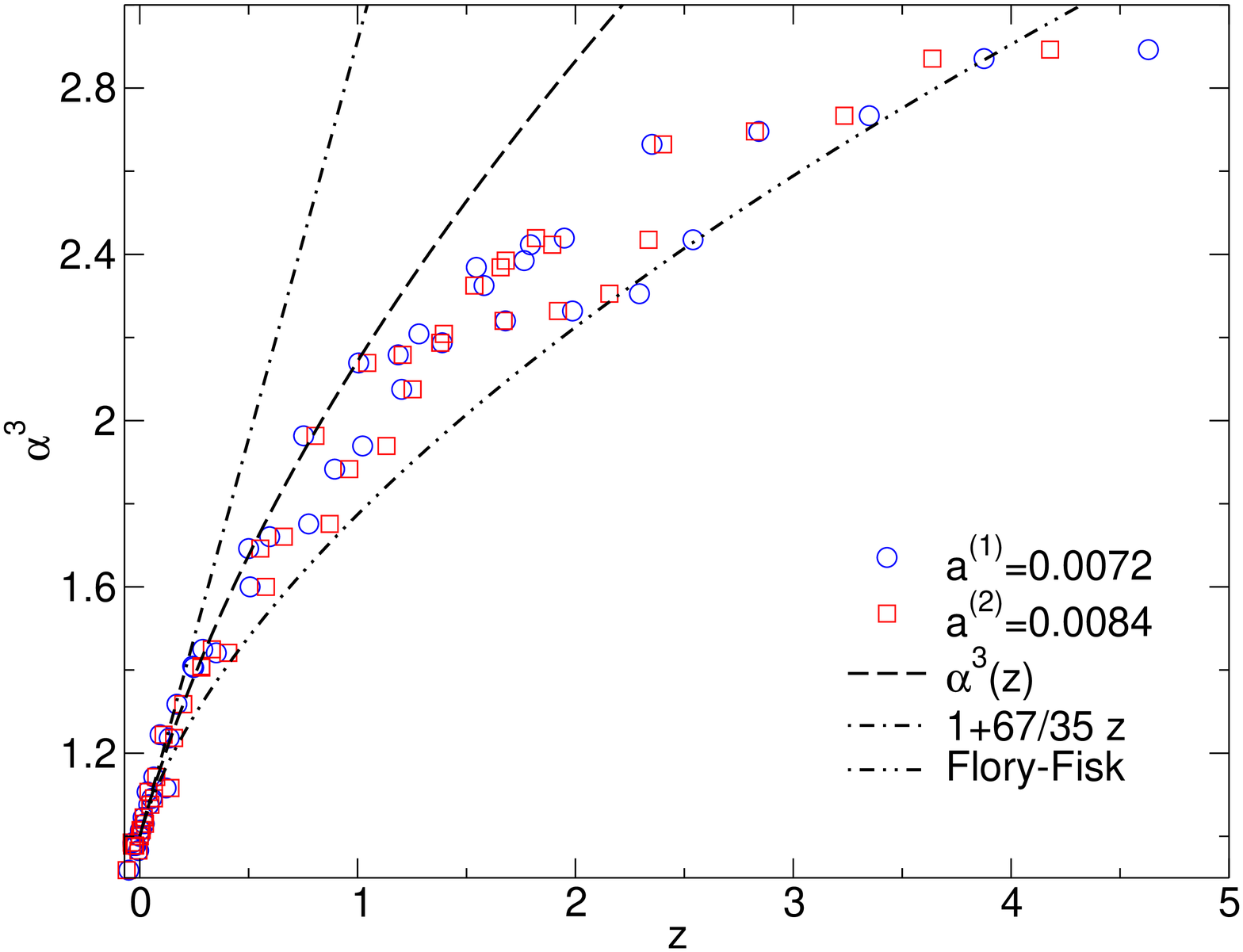,width=8cm,angle=0} \\
\end{tabular}
\end{center}
\caption{Comparison of experiments with the TPM predictions. We  fix $a^{(1)}$ 
and $a^{(2)}$ using Eqs.~(\ref{PCP-const-suppl}), 
(\ref{PIB-const-suppl}), and (\ref{PS-const-suppl}).
}
\label{fig:results-direct-suppl}
\end{figure*}

The results for PIB and PCP show little dependence on the choice of the 
function $f_T(T)$. This is probably related to the fact that the experimental
data belong to a small temperature region ($z\lesssim 1$), so that
the linear approximation works well. On the other hand, for PS 
differences are larger, but in this case data extend up to $z\approx 4$, i.e.,
one observes the full crossover from $\theta$ to good-solvent behavior. 
The analysis of $\alpha^2$ and $A_2$ should in principle provide the 
same result for the constants $a^{(i)}$. For PIB and PCP, differences are 
small. They are probably a consequence of experimental errors, 
polydispersity effects,
and scaling corrections. As a final result we quote the average of the
estimates obtained by using $A_2$ and $\alpha^2$:
\begin{eqnarray}
&& a^{(1)}(\hbox{PIB}) = 0.0029\, (\hbox{mol/g})^{1/2},\nonumber \\
&& a^{(2)}(\hbox{PIB}) = 0.0031\, (\hbox{mol/g})^{1/2},
\label{PIB-const-suppl} \\
&& a^{(1)}(\hbox{PCP}) = 0.0051\, (\hbox{mol/g})^{1/2},\nonumber \\
&& a^{(2)}(\hbox{PCP}) = 0.0056\, (\hbox{mol/g})^{1/2} .
\label{PCP-const-suppl} 
\end{eqnarray}
For PS, instead, the results of the two analyses differ quite significantly. 
One may suspect that the systematic 
deviations are due the inclusion of sample A-30, for which the 
radius of gyration was estimated by using the expected scaling at the 
$\theta$ point, see Eq.~(\ref{Rg2-A30}). 
To understand the role of data A-30 in the analysis, 
we have repeated the fitting procedure using only samples A-5 and A-16. 
From the analysis of $\alpha^2$ we obtain 
\begin{equation}
a^{(1)} = 0.0051\,(\hbox{mol/g})^{1/2} \qquad\qquad 
a^{(2)} = 0.0061\,(\hbox{mol/g})^{1/2}, 
\end{equation}
while from the analysis of $A_2$ we obtain 
\begin{equation}
a^{(1)} = 0.0080\,(\hbox{mol/g})^{1/2} \qquad\qquad 
a^{(2)} = 0.0086\,(\hbox{mol/g})^{1/2}.
\end{equation}
Discrepancies are not reduced, allowing us to conclude that the 
problem does not lie with sample A-30. To understand better the origin 
of the observed differences, in Fig.~\ref{fig-comparison-PS-suppl} 
we compare the data for $\alpha^2$ and $A_2$ with the TPM predictions.
We define $z$ as in Eq.~(\ref{z1-def-suppl}) [similar plots would 
be obtained by using Eq.~(\ref{z2-def-suppl})] and use the estimates of 
$a^{(1)}$ reported in Table~\ref{tab:results-aconstants}.
For $a^{(1)} = 0.0072$ $(\hbox{mol/g})^{1/2}$, 
the estimate obtained from the analysis of 
$A_2$, the second-virial coefficient data are quite
well described by the TPM curve. The results for $\alpha^2$ are also 
quite well described by the TPM curve up to $z\approx 1$. The data 
for $z\gtrsim 1$ lie instead well below the TPM curve. 
If we instead use $a^{(1)} = 0.0047$ $(\hbox{mol/g})^{1/2}$, 
even the data for $\alpha^2$ are 
poorly fitted. Indeed, the best fit is obtained by letting the data close
the $\theta$ point lie significantly above the TPM curve, while those 
corresponding to $z\gtrsim 1$ lie somewhat below the TPM curve. 
Clearly, the value $a^{(1)} = 0.0047$ $(\hbox{mol/g})^{1/2}$
cannot be trusted. For these reasons,
for PS, we only consider the results obtained by using $A_2$. Hence, we quote
\begin{eqnarray}
&& a^{(1)}(\hbox{PS}) = 0.0072\,(\hbox{mol/g})^{1/2}, \nonumber \\
&&a^{(2)}(\hbox{PS}) = 0.0084\,(\hbox{mol/g})^{1/2}.
\label{PS-const-suppl} 
\end{eqnarray}
The experimental data are compared with the TPM predictions in
Fig.~\ref{fig:results-direct-suppl}. The $A_2$ data are always reasonably well
described by the TPM expression (small deviations are observed for PS 
close to the good-solvent regime). On the other hand, for the swelling ratio,
deviations occur as soon as $z\gtrsim 1$. These deviations are probably due to
scaling corrections (polydispersity effects should not be 
important as we discuss below), which increase as $z$
increases.\cite{CMP-06-suppl}
We can compare our results with those 
appearing in the literature. While our estimates for PCP and PIB are 
in good agreement with those of Refs.~\onlinecite{PCP-suppl,PIB-suppl},
for PS our estimate of $a^{(2)}$  is slightly smaller than the 
one proposed in Ref.~\onlinecite{PS-suppl}. 
We suspect two possible reasons for this discrepancy. 
First, in this analysis we only used a subset of the data, 
samples A-30, A-16, A-5, that have the same percentage of cis-isomer
and correspond to polymers with the largest molar mass.
Second, in Ref.~\onlinecite{PS-suppl} the crossover function  
is approximated by its expansion to order $z^2$,
while here we use a precise expression that is valid for all values of $z$.

Finally, we note that Berry,\cite{PS-suppl} 
by using his estimate 
of the nonuniversal constant, found an
excellent agreement between the experimental data for  $\alpha^3$ 
and the corresponding
crossover function of Flory and Fisk (Ref.~\onlinecite{FF-66-suppl}),
defined implicitly by
\begin{equation}
\alpha^5(z)-\alpha^3(z)=0.648 z [1+0.969(1+10z/\alpha^3)^{-2/3}].
\end{equation}
Unfortunately, this expression is not consistent with general scaling 
arguments for large values of $z$. Indeed, for $z$ large, it predicts 
$\alpha^5\propto 0.648 z$, while theory\cite{CMP-08-suppl} gives 
$\alpha^5\propto 2.711 z^{5(2\nu-1)} \approx 2.71 z^{0.87597}$.
Moreover, it is significantly different from the accurate 
expression of 
Ref.~\onlinecite{CMP-08-suppl}. Hence, the agreement appears to be 
fortuitous, as it is also confirmed by the fact that Flory-Fisk 
expression does not agree
with experiments, once we use the estimates of the nonuniversal constants 
obtained by using the second-virial coefficient data.

\subsection{Determination of the nonuniversal constants using Berry's method}

As we have discussed, for PS a direct estimate of the nonuniversal
constants gives results that are somewhat different from those reported in 
Ref.~\onlinecite{PS-suppl}. We wish now to understand the origin of the 
discrepancy, whether it is due to the expressions of the TPM functions used,
or rather to the method used to fix the constants. We thus adopt the method 
originally
proposed by Berry in Ref.~\onlinecite{PS-suppl} and apply it in 
combination with the precise estimates of the TPM functions of 
Ref.~\onlinecite{CMP-08-suppl}.

To analyze the data for the second virial coefficient,
we define a universal ratio that involves 
only the radius of gyration at the $\theta$ point:
\begin{equation}
A_2' (T,M_w) = {M^2_w B_{2,{\rm expt}}(T,M_w) \over N_A 
{R}_g(T_\theta,M_w)^3 }.
\end{equation}
It converges to $4 \pi^{3/2} z F'(z)$, where 
\begin{eqnarray}
\label{eq:correction}
&&F'(z)=\frac{A_2(z) \alpha^3(z)}{4\pi^{3/2} z}\\ \nonumber
&=&\frac{\left(30.4463 z^3+35.1869 z^2+10.9288
   z+1\right)^{0.17516}}{\left(268.96 z^4+331.99 z^3+126.783 z^2+19.1187
   z+1\right)^{0.25}}, 
\end{eqnarray}
where in the last line we have used the accurate expressions 
for $A_2(z)$ and $\alpha(z)$ of Ref.~\onlinecite{CMP-08-suppl}.
For $z$ small, $A_2'(z) \approx 4 \pi^{3/2} z = 4 \pi^{3/2} f_T(T) M_w^{1/2}$. 
A first estimate of $a^{(1)}$ or $a^{(2)}$ can be obtained by 
determining the dependence of $B_{2,\rm expt}(T,M_w)$ close to the 
$\theta$ temperature. We write 
\begin{equation}
 B_{2,\rm expt}(T,M_w) = s_B {T-T_\theta\over T_\theta} + \ldots,
\end{equation}
where the coefficient $s_B$ is independent of $M_w$ in the 
TPM limit. Then,  we obtain
\begin{equation}
a^{(i)} = \frac{s_{B} }{4\pi^{3/2} N_A \ell_k^3},
\end{equation}
where $\ell_k^2 = {R}_g(T_\theta,M_w)^2/M_w$ (see Table \ref{tab:samplef}).
Once a first estimate is available, the procedure is iterated, by setting
\begin{equation}
a^{(i)}=\frac{s_{B} }{4\pi^{3/2} N_A \ell_k^3 F'(z)},
\end{equation}
where, at each step, we use the previous determination of $a^{(i)}$ to 
determine $z$ in the right-hand side. The procedure converges in a few steps
and provides the estimates reported in Table
\ref{tab:results-aconstants}. They are slightly different from those 
obtained in the analysis of the previous section, essentially because here 
more emphasis is given in reproducing correctly the small-$z$ behavior. 
In the analysis of Sec.~\ref{secB-suppl}, instead, 
one aims at obtaining the best approximation
in the whole range of values of $z$ considered.

We can use a similar method to analyze the swelling ratio $\alpha$. We define
the ratio
\begin{equation}
\label{eq:Berryalpha3}
S=\frac{R_g^3(T,M_w)-R_g^3(T_\theta,M_w)}{M_w^2},
\end{equation}
which should converge to the TPM expression 
\begin{equation}
S\approx \frac{67}{35} \frac{\ell_k^3}{M_w^{1/2}} z G'(z),
\end{equation}
where $G'(z)$ is defined as
\begin{eqnarray}
&&G'(z)=\frac{35}{67}\frac{\alpha^3(z)-1}{z} = \frac{35}{67 z} \\
&&\times 
\left[\left(1+10.9288 z+35.1869 z^2+30.4463 z^3\right)^{0.17516}-1\right].
\nonumber 
\end{eqnarray}
For $z = 0$, we have $G'(0) = 1$ (note that $10.9288\times 0.17516 \approx
67/35$). Hence, to compute the nonuniversal constants, we first determine
the behavior of $S$ close to the $\theta$ point:
\begin{equation}
S(T,M_w) \approx s_S {T-T_\theta\over T_\theta} + \ldots
\end{equation}
A first guess for the constants is obtained by setting
\begin{equation}
a^{(1)} = a^{(2)}  = 
\frac{35 s_{S} }{67 \ell_k^3}.
\end{equation}
The procedure is then iterated by setting at each step
\beq
a^{(i)} =\frac{35 s_{S} }{67\ell_k^3  G'(z)}
\eeq
where $z$ in the right-hand side is computed by using the 
value of the constants at the previous step.
The results are reported in Table 
\ref{tab:results-aconstants} (line $\alpha^2$ Berry).

\subsection{Polydispersity effects} \label{polydispersity-suppl}

In order to compare with experimental results, it is crucial to take into
account polydispersity effects. For the ratio $A_2$, an accurate analysis 
in the good-solvent regime was presented in Ref.~\onlinecite{Polydisp-suppl}.
In the polydisperse case, two different virial coefficients can be defined.
\cite{Rubinstein-suppl}
The weight-averaged $B_{2,{\rm expt},w}$  is obtained by measuring 
the small-density behavior of the (osmotic) pressure:
\begin{equation}
{P\over c R T} = {1\over M_n} + B_{2,{\rm expt},w} c + O(c^2),
\end{equation}
where $c$ is the weight concentration. The Z-averaged $B_{2,{\rm expt},Z}$
is instead obtained from the the small-concentration behavior 
of the Rayleigh ratio $R_\theta$ measured in scattering experiments:
\begin{equation}
{K c\over R_\theta} = {1\over M_w} + 2 B_{2,{\rm expt},Z} c + O(c^2),
\end{equation}
where $K$ is a constant. As for the radius of gyration, scattering 
experiments provide the $Z$-averaged square radius of gyration 
$R^2_{g,Z}$.  \cite{Rubinstein-suppl} 
We can thus define two different ratios corresponding to $A_2$:
\begin{eqnarray}
A_{2,w} &=& {B_{2,{\rm expt},w} M^2_w\over N_A {R}^3_{g,Z} }, \\
A_{2,Z} &=& {B_{2,{\rm expt},Z} M^2_w\over N_A {R}^3_{g,Z} }, 
\end{eqnarray}
where $N_A$ is the Avogadro number. The analysis of the behavior of
these two quantities as a function of polydispersity was performed in 
Ref.~\onlinecite{Polydisp-suppl}. The Schulz distribution 
\begin{equation}
 P(m) = {s\over m_n \Gamma(s)} \left({s m\over m_n}\right)^{s-1}
    \exp\left(-{s m\over m_n}\right),
\end{equation}
is usually considered. Here $P(m)$ is the probability of having a 
polymer of mass $m$ in the solution and $m_n$ is the number-average
mass of the polymer (in terms of the number average molar mass $M_n = m_n
N_A$), i.e., 
\begin{equation}
m_n = \int_0^\infty m P(m) dm.
\end{equation}
For the Schulz distribution we have
\begin{equation}
{M_w\over M_n} = 1 + 1/s.
\end{equation}
Therefore, since all samples considered in the previous paragraph have 
$M_w/M_n\lesssim 1.1$, we can conclude that $s\gtrsim 10$. Even though 
$s$ is quite large so that the distribution is peaked around $m\approx m_n$, 
the second-virial combinations differ significantly from the monodisperse 
value $A_{2,{\rm mono}} \approx 5.50$.\cite{CMP-06-suppl} 
Indeed, \cite{Polydisp-suppl} for 
$s= 10$ we have $A_{2,w} = 3.91$ and $A_{2,Z} = 4.21$. Polydispersity effects
decrease very slowly, as it can also be seen from the expressions
\cite{Polydisp-suppl}
\begin{equation}
A_{2,w} = 5.50 - 20.5/s, \qquad A_{2,Z} = 5.50 - 16.3/s
\end{equation}
valid for $s \gtrsim 30$. The combinations $A_{2,w}$ and $A_{2,Z}$ 
are also discussed in Ref.~\onlinecite{Schaefer-99-suppl} by using 
perturbative methods. Also field theory predicts large polydispersity
corrections, although not as large as determined numerically. At 
one-loop order, field theory predicts $A_{2,Z} \approx 4.78$ for $s=10$, 
which, although it  differs significantly from the good-solvent value $5.50$, 
is nonetheless larger than the accurate numerical estimate $A_{2,Z} =
4.21$.\cite{Polydisp-suppl}

\begin{figure}
\epsfig{file=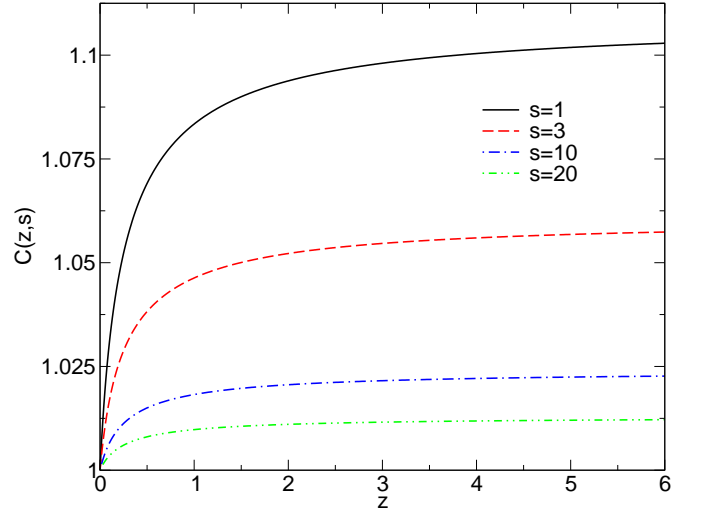,width=9cm,angle=0} 
\caption{Polydispersity function $C(z,s)$ for $s = 1,3,10,20$ and $0\le z\le
6$. }
\label{polydisp-C-suppl}
\end{figure}

Let us now consider the effect of polydispersity on the swelling ratio
$\alpha(z)$. For a polymer of mass $m$ the zero-density 
radius of gyration
behaves as 
\begin{equation}
R^2_g(T,m) = {\ell_k}^2 m f[(m N_A)^{1/2} f_T(T)],
\end{equation}
where $f(z) = \alpha(z)^2$ is the 
swelling factor computed for a monodisperse system [hence, we can use 
Eq.~(\ref{alpha-to-z-suppl})]. 
It follows that the Z-averaged radius of gyration is 
given by
\begin{equation}
R^2_{g,Z}(T) = 
         {\ell_k^2} 
     {\int_0^\infty m^3 P(m) f[(m N_A)^{1/2} f_T(T)] dm\over 
      \int_0^\infty m^2 P(m) dm}.
\end{equation}
Now, the Schulz distribution can be rewritten as $P(m) = p(m/m_n)/m_n$. 
Redefining $m = x m_n$ and setting $z = M_w^{1/2} f_T(T)$, we obtain 
\begin{equation}
R^2_{g,Z}(T) = 
    {\ell_k^2 m_n} 
     {\int_0^\infty x^3 p(x) f[(x M_n/M_w)^{1/2} z] dx\over 
     \int_0^\infty x^2 p(x) dx}.
\end{equation}
We can use this expression to
compare $R^2_{g,Z}(T_\theta)$ for the polydisperse system of 
weight-averaged mass $M_w$ and 
$R^2_{g}(T_\theta,M_w)$ for a monodisperse system of polymers of the 
same molar mass $M_w$. For the Schulz distribution we obtain
\begin{equation}
{R^2_{g,Z}(T_\theta) \over R^2_{g}(T_\theta,M_w)} = 
   {s+2\over s+1} = 2 - {M_n\over M_w}.
\label{polydispRgtheta-suppl}
\end{equation}
For a polydisperse system, we can define an effective swelling ratio as 
\begin{equation}
\alpha_Z(z) = {R_{g,Z}(T) \over R_{g,Z}(T_\theta)},
\end{equation}
obtaining 
\begin{equation}
\alpha_Z(z)^2 = {\int_0^\infty x^3 p(x) f[(x M_n/M_w)^{1/2} z] dx\over 
               \int_0^\infty x^3 p(x) dx}.
\end{equation}
For the Schulz distribution, we obtain explicitly
\begin{equation}
\alpha_Z(z)^2 = {s^2\over (1+s)(2+s)} 
                \int_0^\infty x^3 p(x) f[(sx/(1+s))^{1/2} z] dx.
\end{equation}
We define a polydispersity function $C(z,s)$ as 
\begin{equation}
\alpha_Z(z)^2 = C(z,s) \alpha(z)^2,
\end{equation}
where $\alpha(z)$ is the swelling ratio for a monodisperse system. 
The function $C(z,s)$ is an increasing function of $z$ which 
approaches the constant
\begin{equation}
C(z,s) \approx C_\infty(s) = {(1+s)^{1-2\nu} \Gamma(s + 2 + 2\nu) \over 
        \Gamma(s+3)}
\label{Cs-largez}
\end{equation}
for large $z$.
Such a quantity is close 
to one for any $s\gtrsim 10$. For instance, 
$C_\infty(s) = 1.108, 1.024$ for $s = 1,10$, showing that for 
$s\gtrsim 10$ (i.e., $M_w/M_n\lesssim 1.10$) corrections are at most of 
2\%. The curve $C(z,s)$ is plotted versus $z$ for a few values of $s$ in 
Fig.~\ref{polydisp-C-suppl}.

\section{Supplementary material: Polymer properties}

\subsection{Depletion thickness} 

We report here the results for the depletion thickness we used in the paper. 
They are taken from Ref.~\onlinecite{DPP-depletion-suppl}.

The zero-density depletion thickness $\delta_s(q,0)$
can be expressed in terms of the colloid-polymer second virial 
combination $A_{2,cp}$, as\cite{DPP-depletion-suppl}
\begin{equation}
{\delta_s(q,0)\over R_c} = - 1 + \left({3 q^3 A_{2,cp} \over 4\pi}\right)^{1/3}.
\end{equation}
The quantity $A_{2,cp}$ has been accurately determined by Monte Carlo
simulations. The results can be parametrized as 
\begin{eqnarray}
A_{2,cp}&=& \frac{4\pi }{3q^3}
  \left[ \frac{1+a_1 q + a_2 q^2+a_3 q^3}{1+a_4 q}
  \right]^{1/(2\nu)},
\label{A2cp-param-suppl}
\end{eqnarray}
where $\nu = 0.5876$ for the good-solvent case and
$\nu=1/2$ for $z=z^{(1)}$ and $z^{(3)}$. The coefficients $a_i$ are 
reported in Table~\ref{tab:VirialPar-suppl}. 

\begin{table}
\caption{Coefficients parametrizing $A_{2,cp}$:
The interpolation is accurate in the range $0\le q \le q_{\rm max}$.}
\label{tab:VirialPar-suppl}
\begin{center}
\begin{tabular}{ccccccccc}
\hline\hline
& $z$ & $a_1$ & $a_2$ & $a_3$ & $a_4$ & $\beta$ & $q_{\rm max}$\\
\hline
\hline
$A_{2,cp}$ & $\infty$ & 4.1329 & 5.4906 &  2.12578 & 0.3942 &  & 50 \\
& $z_3$ & 3.4774 & 3.37453 & 0.39752 & 0.15763 &  & 3\\
& $z_1$ & 3.4378 & 3.18934 & 0.20253 & 0.071526 & & 30\\
\hline\hline
\end{tabular}
\end{center}
\end{table}

\begin{table}[tb!]
\caption{Coefficients parametrizing the depletion-thickness interpolations
(\ref{deltas-fit-suppl})
as function of density. The parametrizations should hold for
$\phi_p\le \phi_{\rm max}$.}
\label{tab:depletion-fits-suppl}
\begin{center}
\begin{tabular}{ccccccccc}
\hline\hline
$z$ & $q$ & $n$ & $a_1$ & $a_2$ & $a_3$ & $\eta$ & $\phi_{\rm max}$\\
\hline
$\infty$ & 0 & 3 & 3.9467 & 4.3305 & 5.8889 & 0.770 & $4$ \\
& 0.5   &  3 & 4.0909   & 6.8272   & 2.6728 & 0.770 & $4$ \\
& 1.0   &  3 & 4.0987   &  4.4818 & 4.92968 & 0.770 & $4$ \\
& 2.0   &  3 & 4.0753   & 6.12348   & 1.92624 & 0.770 & $4$ \\
\hline
$z^{(3)}$ & 0.5 & 2 & 1.8641 & 1.0753 & 0 &0.5579 & $2$ \\
& 1.0 & 2 & 1.8747  & 1.0279 & 0          &0.5579 & $4$ \\
& 2.0  & 2 & 2.0279 & 1.16915 & 0         &0.5128 & $4$ \\
\hline
$z^{(1)}$ & 0.5/1.0/2.0 & 3 & 1.7682 & 1.8151 & 0.6591 & 0.2845 & $4$ \\
\hline\hline
\end{tabular}
\end{center}
\end{table}

For $q\lesssim 2$, which are the values of $q$ investigated in 
Ref.~\onlinecite{DPP-depletion-suppl}, the ratio 
$\delta_s(q,\phi_p)/\delta_s(q,0)$ 
shows a tiny dependence on $q$. The results for $q=0$ (good-solvent case only)
and $q=0.5,1,2$ can be parametrized as 
\begin{equation}
{\delta_s(q,\phi_p)\over \delta_s(q,0)} =
    \left(1 + \sum_{k=1}^n a_k \phi_p^k\right)^{-\eta/n}.
\label{deltas-fit-suppl}
\end{equation}
The corresponding coefficients are reported in
Table~\ref{tab:depletion-fits-suppl}.
For values of $q$ that differ from those considered in 
Ref.~\onlinecite{DPP-depletion-suppl}, we use a linear interpolation 
if possible. 
For $2 \le q\le 3$, we simply set ${\delta_s(q,\phi_p)/\delta_s(q,0)} = 
{\delta_s(2,\phi_p)/\delta_s(2,0)}$ and analogously, for $q\le 0.5$ 
and $z\not=\infty$, we set
${\delta_s(q,\phi_p)/\delta_s(q,0)} = 
{\delta_s(0.5,\phi_p)/\delta_s(0.5,0)}$.

In the paper we also consider the expression for the depletion thickness
proposed in Refs.~\onlinecite{FST-07-suppl,LT-11-suppl} 
for colloids in good-solvent
polymers:
\begin{equation}
\delta_s/R_c = 0.865 q^{0.88} (1 + 3.95 \phi_p^{1.54})^{-0.44}.
\end{equation}
As already discussed in Ref.~\onlinecite{DPP-depletion-suppl}, this expression
is in agreement with Monte Carlo full-monomer results and with 
renormalization-group predictions only for $0.2\lesssim q \lesssim 4$.

\subsection{Full-monomer equation of state}

One of the quantities needed in the GFVT calculation is the pressure derivative
$K(\phi_p)$ for the polymer system in the absence of colloids. 
Such a quantity has been derived from the explicit expressions for the 
compressibility factor $Z = \beta P/\rho$ reported in 
Ref.~\onlinecite{Pelissetto-08-suppl} (good-solvent case) and in 
Ref.~\onlinecite{DPP-thermal-suppl} ($z=z^{(1)}$ and $z = z^{(3)}$).

\begin{table}[t!]
\caption{Coefficients for the interpolations of the
compressibility factor $Z = \beta P/\rho_p$ 
obtained from the full-monomer results
for $K$.}
\label{ZFM-from-interp-suppl}
\begin{center}
\begin{tabular}{cccccc}
\hline
\hline
$z$ & $a_1$&$a_2$&$a_3$&$a_4$ &  $b$ \\
\hline
$z^{(1)}$ &  0.455489 & 0.057089 &  $-$0.00117019 & $-$0.0184279 & ---\\
$z^{(3)}$ &  2.69479  & 3.07853 &  1.79532 & 0.614611 & 1.04964 \\
$\infty$  &  2.50342  & 2.54540 &  1.02975 & 0.500318 & 0.6555 \\
\hline\hline
\end{tabular}
\end{center}
\end{table}

\noindent
For $z = z^{(1)}$ 
we parametrize the compressibility factor as
\begin{equation}
Z(z^{(1)},\phi_p)=
    \frac{\sqrt{1+a_1 \phi_p+a_2 \phi_p^2+a_3\phi_p^3}}{\sqrt{1+a_4\phi_p}},
\label{Z-interp-FM-suppl}
\end{equation}
which behaves linearly for large $\phi_p$. 
For $z=z^{(3)}$ and $\phi_p\lesssim 30$, we use
\begin{equation}
Z(z^{(3)},\phi_p)=(1+a_1\phi_p+a_2\phi_p^2+a_3\phi_p^3+a_4\phi_p^4)^{b/4},
\end{equation}
For the good-solvent case we use the parametrization
\begin{equation}
Z(z=\infty,\phi_p)={(1+a_1 \phi_p+a_2 \phi_p^2+a_3\phi_p^3)^{b} \over 
                 (1+a_4\phi_p)^{b}}.
\end{equation}
The coefficients are reported in Table~\ref{ZFM-from-interp-suppl}.
These parametrizations are found to be accurate with deviations of less
than 1\%.

In the paper we will also consider the expression for $K$
proposed in Refs.~\onlinecite{FST-07-suppl,LT-11-suppl} 
for polymers under good-solvent conditions:
\begin{equation}
K(\phi_p) = 1 + 3.73 \phi_p^{1.31}.
\end{equation}
This expression agrees (small deviations only occur in the dilute regime)
with numerical simulations \cite{Pelissetto-08-suppl} and 
field-theory predictions.\cite{Schaefer-99-suppl}

\section{Supplementary matherial: GFVT results}

We report here some numerical GFVT results. In Tables~
\ref{GFVTres-crit-suppl-GS}, \ref{GFVTres-crit-suppl-z3}, and 
\ref{GFVTres-crit-suppl-z1}, we give the critical-point position
for different values of $q$ (up to $q = 3$) and for the three different types of
solvent quality we consider in the paper.  In Table \ref{GFVTres-crit-suppl-z0}
we report the analogous quantity for noninteracting polymers.
We report the polymer and colloid volume fractions at the critical point,
the corresponding reservoir polymer volume fraction,
the reduced critical pressure $\widetilde{P} = \beta P V_c$, 
and the excess chemical potentials.

In Tables \ref{Triple-points-GS-suppl}, \ref{Triple-points-z3-suppl}, 
\ref{Triple-points-z1-suppl}, and \ref{Triple-points-z0-suppl}
we report the triple-point position. We give the reduced 
triple-point pressure $\widetilde{P} = \beta P V_c$, 
and the volume fractions of the 
three coexisting phases.

\onecolumngrid

\begin{table}
\caption{Critical-point position, critical pressure 
($\widetilde P = \beta P V_c$), 
and excess chemical potentials for the good-solvent case.}
\label{GFVTres-crit-suppl-GS}
\begin{spacing}{0.65}
\begin{tabular}{cccccccc}
\hline\hline
$z$  & $q$ & $\phi_{p,\rm crit}^{(r)}$ & $\phi_{c,\rm crit}$ & 
             $\phi_{p,\rm crit}$ & 
    $\tilde{P}_{\rm crit}$ & $\beta \mu_{c,\rm crit}^{(exc)}$  & 
    $\beta \mu_{p,\rm crit}^{(exc)}$ \\
\hline
$\infty$ &  0.45 &  0.496 &  0.299 &  0.196 &  9.85 &  22.8 &  2.42 \\  
 &  0.50 &  0.533 &  0.280 &  0.217 &  8.02 &  19.6 &  2.51 \\  
 &  0.55 &  0.570 &  0.263 &  0.239 &  6.70 &  17.1 &  2.61 \\  
 &  0.60 &  0.608 &  0.249 &  0.261 &  5.72 &  15.3 &  2.72 \\  
 &  0.65 &  0.647 &  0.236 &  0.284 &  4.97 &  13.8 &  2.83 \\  
 &  0.70 &  0.687 &  0.225 &  0.308 &  4.39 &  12.7 &  2.95 \\  
 &  0.75 &  0.728 &  0.215 &  0.332 &  3.92 &  11.7 &  3.08 \\  
 &  0.80 &  0.770 &  0.207 &  0.357 &  3.55 &  10.9 &  3.22 \\  
 &  0.85 &  0.813 &  0.198 &  0.386 &  3.25 &  10.3 &  3.36 \\  
 &  0.90 &  0.858 &  0.190 &  0.414 &  3.00 &  9.71 &  3.50 \\  
 &  0.95 &  0.903 &  0.184 &  0.442 &  2.79 &  9.24 &  3.66 \\  
 &  1.00 &  0.951 &  0.178 &  0.474 &  2.61 &  8.85 &  3.83 \\  
 &  1.05 &  0.996 &  0.172 &  0.502 &  2.45 &  8.47 &  3.99 \\  
 &  1.10 &  1.042 &  0.167 &  0.533 &  2.31 &  8.15 &  4.15 \\  
 &  1.15 &  1.088 &  0.162 &  0.563 &  2.19 &  7.86 &  4.33 \\  
 &  1.20 &  1.136 &  0.157 &  0.595 &  2.08 &  7.61 &  4.50 \\  
 &  1.25 &  1.184 &  0.153 &  0.627 &  1.99 &  7.38 &  4.69 \\  
 &  1.30 &  1.233 &  0.150 &  0.660 &  1.90 &  7.18 &  4.88 \\  
 &  1.35 &  1.283 &  0.146 &  0.694 &  1.83 &  7.01 &  5.08 \\  
 &  1.40 &  1.334 &  0.143 &  0.728 &  1.77 &  6.85 &  5.28 \\  
 &  1.45 &  1.386 &  0.139 &  0.765 &  1.71 &  6.70 &  5.49 \\  
 &  1.50 &  1.438 &  0.136 &  0.801 &  1.65 &  6.57 &  5.70 \\  
 &  1.55 &  1.492 &  0.134 &  0.837 &  1.61 &  6.46 &  5.92 \\  
 &  1.60 &  1.546 &  0.131 &  0.875 &  1.57 &  6.35 &  6.15 \\  
 &  1.65 &  1.601 &  0.129 &  0.914 &  1.53 &  6.26 &  6.38 \\  
 &  1.70 &  1.657 &  0.126 &  0.954 &  1.49 &  6.17 &  6.62 \\  
 &  1.75 &  1.713 &  0.124 &  0.994 &  1.46 &  6.09 &  6.87 \\  
 &  1.80 &  1.771 &  0.122 &  1.034 &  1.43 &  6.02 &  7.12 \\  
 &  1.85 &  1.829 &  0.120 &  1.077 &  1.41 &  5.96 &  7.38 \\  
 &  1.90 &  1.888 &  0.119 &  1.118 &  1.38 &  5.90 &  7.65 \\  
 &  1.95 &  1.948 &  0.117 &  1.162 &  1.36 &  5.85 &  7.92 \\  
 &  2.00 &  2.009 &  0.115 &  1.205 &  1.34 &  5.80 &  8.20 \\  
 &  2.05 &  2.082 &  0.114 &  1.259 &  1.34 &  5.79 &  8.53 \\  
 &  2.10 &  2.156 &  0.113 &  1.314 &  1.34 &  5.79 &  8.88 \\  
 &  2.15 &  2.232 &  0.111 &  1.370 &  1.34 &  5.79 &  9.24 \\  
 &  2.20 &  2.311 &  0.110 &  1.429 &  1.34 &  5.80 &  9.61 \\  
 &  2.25 &  2.391 &  0.110 &  1.488 &  1.34 &  5.82 &  10.0 \\  
 &  2.30 &  2.475 &  0.109 &  1.549 &  1.35 &  5.83 &  10.4 \\  
 &  2.35 &  2.560 &  0.108 &  1.613 &  1.36 &  5.86 &  10.8 \\  
 &  2.40 &  2.648 &  0.107 &  1.679 &  1.37 &  5.89 &  11.3 \\  
 &  2.45 &  2.738 &  0.107 &  1.748 &  1.38 &  5.92 &  11.7 \\  
 &  2.50 &  2.831 &  0.106 &  1.816 &  1.39 &  5.96 &  12.2 \\  
 &  2.55 &  2.926 &  0.106 &  1.887 &  1.40 &  6.00 &  12.7 \\  
 &  2.60 &  3.025 &  0.105 &  1.963 &  1.42 &  6.04 &  13.2 \\  
 &  2.65 &  3.126 &  0.105 &  2.036 &  1.44 &  6.09 &  13.7 \\  
 &  2.70 &  3.229 &  0.105 &  2.114 &  1.45 &  6.14 &  14.3 \\  
 &  2.75 &  3.336 &  0.104 &  2.196 &  1.48 &  6.20 &  14.9 \\  
 &  2.80 &  3.446 &  0.104 &  2.276 &  1.50 &  6.26 &  15.5 \\  
 &  2.85 &  3.558 &  0.104 &  2.359 &  1.52 &  6.32 &  16.1 \\  
 &  2.90 &  3.674 &  0.104 &  2.449 &  1.55 &  6.39 &  16.7 \\  
 &  2.95 &  3.793 &  0.104 &  2.538 &  1.57 &  6.46 &  17.4 \\  
 &  3.00 &  3.915 &  0.104 &  2.630 &  1.60 &  6.54 &  18.1 \\  
\hline\hline
\end{tabular}
\end{spacing}
\end{table}

\begin{table}
\caption{Critical-point position, critical pressure 
$\widetilde{P} = \beta P V_c$, 
and excess chemical potentials for $z=z^{(3)}$.}
\label{GFVTres-crit-suppl-z3}
\begin{spacing}{0.65}
\begin{tabular}{cccccccc}
\hline\hline
$z$  & $q$ & $\phi_{p,\rm crit}^{(r)}$ & $\phi_{c,\rm crit}$ & 
    $\phi_{p,\rm crit}$ & 
    $\tilde{P}_{\rm crit}$ & $\beta \mu_{c,\rm crit}^{(exc)}$  & 
    $\beta \mu_{p,\rm crit}^{(exc)}$ \\
\hline
$z^{(3)}$ &  0.40 &  0.391 &  0.288 &  0.141 &  8.07 &  19.7 &  1.59 \\  
 &  0.45 &  0.420 &  0.265 &  0.155 &  6.21 &  16.3 &  1.61 \\  
 &  0.50 &  0.449 &  0.246 &  0.170 &  4.94 &  13.8 &  1.64 \\  
 &  0.55 &  0.479 &  0.228 &  0.185 &  4.04 &  12.0 &  1.66 \\  
 &  0.60 &  0.509 &  0.213 &  0.200 &  3.37 &  10.6 &  1.69 \\  
 &  0.65 &  0.539 &  0.200 &  0.216 &  2.86 &  9.46 &  1.72 \\  
 &  0.70 &  0.569 &  0.188 &  0.231 &  2.47 &  8.56 &  1.75 \\  
 &  0.75 &  0.600 &  0.178 &  0.247 &  2.15 &  7.82 &  1.79 \\  
 &  0.80 &  0.631 &  0.169 &  0.262 &  1.90 &  7.21 &  1.82 \\  
 &  0.85 &  0.662 &  0.160 &  0.278 &  1.70 &  6.69 &  1.86 \\  
 &  0.90 &  0.694 &  0.153 &  0.295 &  1.53 &  6.25 &  1.90 \\  
 &  0.95 &  0.726 &  0.146 &  0.311 &  1.38 &  5.87 &  1.94 \\  
 &  1.00 &  0.758 &  0.139 &  0.329 &  1.26 &  5.54 &  1.98 \\  
 &  1.05 &  0.791 &  0.133 &  0.345 &  1.16 &  5.25 &  2.03 \\  
 &  1.10 &  0.824 &  0.128 &  0.362 &  1.07 &  4.99 &  2.08 \\  
 &  1.15 &  0.857 &  0.123 &  0.381 &  0.989 & 4.77 &  2.12 \\  
 &  1.20 &  0.890 &  0.119 &  0.398 &  0.921 & 4.56 &  2.17 \\  
 &  1.25 &  0.924 &  0.114 &  0.417 &  0.861 & 4.39 &  2.21 \\  
 &  1.30 &  0.958 &  0.110 &  0.435 &  0.807 & 4.22 &  2.26 \\  
 &  1.35 &  0.992 &  0.106 &  0.453 &  0.759 & 4.07 &  2.31 \\  
 &  1.40 &  1.026 &  0.103 &  0.473 &  0.717 & 3.94 &  2.36 \\  
 &  1.45 &  1.061 &  0.100 &  0.491 &  0.678 & 3.81 &  2.41 \\  
 &  1.50 &  1.096 &  0.096 &  0.511 &  0.645 & 3.70 &  2.46 \\  
 &  1.55 &  1.131 &  0.093 &  0.531 &  0.612 & 3.60 &  2.51 \\  
 &  1.60 &  1.166 &  0.091 &  0.550 &  0.583 & 3.50 &  2.57 \\  
 &  1.65 &  1.202 &  0.088 &  0.570 &  0.557 & 3.41 &  2.62 \\  
 &  1.70 &  1.237 &  0.086 &  0.591 &  0.533 & 3.33 &  2.67 \\  
 &  1.75 &  1.273 &  0.083 &  0.611 &  0.511 & 3.26 &  2.73 \\  
 &  1.80 &  1.309 &  0.081 &  0.632 &  0.491 & 3.19 &  2.78 \\  
 &  1.85 &  1.345 &  0.079 &  0.654 &  0.472 & 3.12 &  2.84 \\  
 &  1.90 &  1.382 &  0.077 &  0.674 &  0.455 & 3.06 &  2.90 \\  
 &  1.95 &  1.418 &  0.075 &  0.695 &  0.438 & 3.00 &  2.95 \\  
 &  2.00 &  1.455 &  0.073 &  0.716 &  0.423 & 2.95 &  3.01 \\  
 &  2.05 &  1.493 &  0.071 &  0.741 &  0.410 & 2.90 &  3.07 \\  
 &  2.10 &  1.532 &  0.070 &  0.763 &  0.397 & 2.86 &  3.13 \\  
 &  2.15 &  1.571 &  0.069 &  0.787 &  0.386 & 2.82 &  3.20 \\  
 &  2.20 &  1.611 &  0.067 &  0.811 &  0.375 & 2.78 &  3.26 \\  
 &  2.25 &  1.651 &  0.066 &  0.832 &  0.365 & 2.74 &  3.32 \\  
 &  2.30 &  1.691 &  0.064 &  0.858 &  0.355 & 2.70 &  3.39 \\  
 &  2.35 &  1.731 &  0.063 &  0.882 &  0.346 & 2.67 &  3.45 \\  
 &  2.40 &  1.772 &  0.062 &  0.910 &  0.338 & 2.64 &  3.52 \\  
 &  2.45 &  1.813 &  0.061 &  0.934 &  0.330 & 2.61 &  3.58 \\  
 &  2.50 &  1.854 &  0.060 &  0.960 &  0.322 & 2.59 &  3.65 \\  
 &  2.55 &  1.896 &  0.059 &  0.987 &  0.315 & 2.56 &  3.72 \\  
 &  2.60 &  1.938 &  0.057 &  1.014 &  0.308 & 2.54 &  3.79 \\  
 &  2.65 &  1.980 &  0.057 &  1.038 &  0.302 & 2.51 &  3.86 \\  
 &  2.70 &  2.023 &  0.056 &  1.067 &  0.296 & 2.49 &  3.93 \\  
 &  2.75 &  2.066 &  0.055 &  1.092 &  0.290 & 2.47 &  4.00 \\  
 &  2.80 &  2.109 &  0.054 &  1.123 &  0.285 & 2.45 &  4.07 \\  
 &  2.85 &  2.153 &  0.053 &  1.149 &  0.280 & 2.43 &  4.15 \\  
 &  2.90 &  2.197 &  0.052 &  1.176 &  0.275 & 2.41 &  4.22 \\  
 &  2.95 &  2.241 &  0.051 &  1.204 &  0.270 & 2.40 &  4.30 \\  
 &  3.00 &  2.286 &  0.051 &  1.232 &  0.266 & 2.38 &  4.37 \\  
\hline\hline
\end{tabular}
\end{spacing}
\end{table}

\begin{table}
\caption{Critical-point position, critical pressure 
($\widetilde{P} = \beta P V_c$, 
and excess chemical potentials for $z=z^{(1)}$.}
\label{GFVTres-crit-suppl-z1}
\begin{spacing}{0.65}
\begin{tabular}{cccccccc}
\hline\hline
$z$  & $q$ & $\phi_{p,\rm crit}^{(r)}$ & $\phi_{c,\rm crit}$ & 
             $\phi_{p,\rm crit}$ & 
    $\tilde{P}_{\rm crit}$ & $\beta \mu_{c,\rm crit}^{(exc)}$  & 
    $\beta \mu_{p,\rm crit}^{(exc)}$ \\
\hline
$z^{(1)}$ &  0.35 &  0.348 &  0.300 &  0.116 &  8.99 &  21.4 &  1.26 \\  
 &  0.40 &  0.376 &  0.271 &  0.130 &  6.57 &  17.0 &  1.24 \\  
 &  0.45 &  0.405 &  0.248 &  0.143 &  5.01 &  14.0 &  1.23 \\  
 &  0.50 &  0.433 &  0.228 &  0.155 &  3.95 &  11.8 &  1.23 \\  
 &  0.55 &  0.462 &  0.210 &  0.168 &  3.19 &  10.2 &  1.23 \\  
 &  0.60 &  0.491 &  0.195 &  0.181 &  2.64 &  8.96 &  1.23 \\  
 &  0.65 &  0.520 &  0.183 &  0.193 &  2.22 &  7.98 &  1.24 \\  
 &  0.70 &  0.550 &  0.170 &  0.207 &  1.89 &  7.19 &  1.24 \\  
 &  0.75 &  0.579 &  0.161 &  0.218 &  1.64 &  6.54 &  1.25 \\  
 &  0.80 &  0.610 &  0.151 &  0.232 &  1.43 &  6.00 &  1.26 \\  
 &  0.85 &  0.640 &  0.143 &  0.245 &  1.26 &  5.55 &  1.27 \\  
 &  0.90 &  0.671 &  0.135 &  0.259 &  1.13 &  5.16 &  1.27 \\  
 &  0.95 &  0.703 &  0.127 &  0.274 &  1.01 &  4.83 &  1.28 \\  
 &  1.00 &  0.734 &  0.121 &  0.287 &  0.913 & 4.54 &  1.29 \\  
 &  1.05 &  0.766 &  0.115 &  0.301 &  0.830 & 4.28 &  1.30 \\  
 &  1.10 &  0.799 &  0.110 &  0.316 &  0.758 & 4.06 &  1.31 \\  
 &  1.15 &  0.831 &  0.105 &  0.323 &  0.696 & 3.86 &  1.32 \\  
 &  1.20 &  0.864 &  0.100 &  0.346 &  0.642 & 3.69 &  1.33 \\  
 &  1.25 &  0.897 &  0.096 &  0.360 &  0.595 & 3.53 &  1.34 \\  
 &  1.30 &  0.931 &  0.092 &  0.374 &  0.553 & 3.38 &  1.36 \\  
 &  1.35 &  0.965 &  0.088 &  0.391 &  0.516 & 3.25 &  1.37 \\  
 &  1.40 &  0.999 &  0.085 &  0.406 &  0.483 & 3.14 &  1.38 \\  
 &  1.45 &  1.033 &  0.081 &  0.422 &  0.453 & 3.03 &  1.39 \\  
 &  1.50 &  1.068 &  0.079 &  0.438 &  0.426 & 2.93 &  1.40 \\  
 &  1.55 &  1.103 &  0.076 &  0.455 &  0.402 & 2.84 &  1.42 \\  
 &  1.60 &  1.138 &  0.073 &  0.471 &  0.380 & 2.76 &  1.43 \\  
 &  1.65 &  1.174 &  0.070 &  0.488 &  0.360 & 2.68 &  1.44 \\  
 &  1.70 &  1.210 &  0.068 &  0.504 &  0.342 & 2.61 &  1.46 \\  
 &  1.75 &  1.246 &  0.066 &  0.521 &  0.325 & 2.54 &  1.47\\  
 &  1.80 &  1.282 &  0.064 &  0.539 &  0.310 & 2.49 &  1.48 \\  
 &  1.85 &  1.319 &  0.062 &  0.555 &  0.296 & 2.43 &  1.50 \\  
 &  1.90 &  1.355 &  0.060 &  0.572 &  0.283 & 2.37 &  1.51 \\  
 &  1.95 &  1.392 &  0.058 &  0.594 &  0.271 & 2.33 &  1.52 \\  
 &  2.00 &  1.430 &  0.056 &  0.610 &  0.259 & 2.28 &  1.54 \\  
 &  2.05 &  1.467 &  0.054 &  0.631 &  0.249 & 2.24 &  1.55 \\  
 &  2.10 &  1.505 &  0.053 &  0.649 &  0.239 & 2.20 &  1.57 \\  
 &  2.15 &  1.542 &  0.051 &  0.668 &  0.230 & 2.16 &  1.58 \\  
 &  2.20 &  1.581 &  0.050 &  0.683 &  0.222 & 2.12 &  1.60 \\  
 &  2.25 &  1.619 &  0.049 &  0.704 &  0.214 & 2.09 &  1.61 \\  
 &  2.30 &  1.657 &  0.047 &  0.726 &  0.207 & 2.06 &  1.63 \\  
 &  2.35 &  1.696 &  0.046 &  0.744 &  0.200 & 2.03 &  1.64 \\  
 &  2.40 &  1.735 &  0.045 &  0.762 &  0.193 & 2.00 &  1.66 \\  
 &  2.45 &  1.774 &  0.044 &  0.782 &  0.187 & 1.97 &  1.68 \\  
 &  2.50 &  1.813 &  0.043 &  0.803 &  0.181 & 1.95 &  1.69 \\  
 &  2.55 &  1.852 &  0.042 &  0.824 &  0.175 & 1.92 &  1.71 \\  
 &  2.60 &  1.892 &  0.041 &  0.840 &  0.170 & 1.90 &  1.73 \\  
 &  2.65 &  1.931 &  0.040 &  0.864 &  0.165 & 1.88 &  1.74 \\  
 &  2.70 &  1.971 &  0.039 &  0.881 &  0.161 & 1.85 &  1.76 \\  
 &  2.75 &  2.011 &  0.038 &  0.906 &  0.156 & 1.84 &  1.77 \\  
 &  2.80 &  2.052 &  0.037 &  0.925 &  0.152 & 1.82 &  1.79 \\  
 &  2.85 &  2.092 &  0.037 &  0.945 &  0.148 & 1.80 &  1.81 \\  
 &  2.90 &  2.132 &  0.036 &  0.965 &  0.144 & 1.78 &  1.83 \\  
 &  2.95 &  2.173 &  0.035 &  0.986 &  0.141 & 1.76 &  1.84 \\  
 &  3.00 &  2.214 &  0.034 &  1.008 &  0.137 & 1.75 &  1.86 \\  
\hline\hline
\end{tabular}
\end{spacing}
\end{table}

\begin{table}
\caption{Critical-point position, 
critical pressure ($\widetilde{P} = \beta P V_c$), and excess chemical 
potentials for $z=0$.}
\label{GFVTres-crit-suppl-z0}
\begin{spacing}{0.65}
\begin{tabular}{cccccccc}
\hline\hline
$z$  & $q$ & $\phi_{p,\rm crit}^{(r)}$ & $\phi_{c,\rm crit}$ & 
    $\phi_{p, \rm crit}$ & 
    $\tilde{P}_{\rm crit}$ & $\beta \mu_{c,\rm crit}^{(exc)}$  & 
    $\beta \mu_{p,\rm crit}^{(exc)}$ \\
\hline
0 &  0.35 &  0.343 &  0.290 &  0.113 &  8.19 &  20.0 &  0 \\  
 &  0.40 &  0.371 &  0.264 &  0.123 &  5.95 &  15.8 &  0 \\  
 &  0.45 &  0.399 &  0.239 &  0.136 &  4.51 &  13.0 &  0 \\  
 &  0.50 &  0.427 &  0.219 &  0.147 &  3.53 &  10.9 &  0 \\  
 &  0.55 &  0.455 &  0.201 &  0.159 &  2.84 &  9.42 &  0 \\  
 &  0.60 &  0.484 &  0.187 &  0.170 &  2.33 &  8.25 &  0 \\  
 &  0.65 &  0.513 &  0.174 &  0.180 &  1.95 &  7.33 &  0 \\  
 &  0.70 &  0.542 &  0.161 &  0.192 &  1.65 &  6.59 &  0 \\  
 &  0.75 &  0.572 &  0.150 &  0.204 &  1.42 &  5.98 &  0 \\  
 &  0.80 &  0.602 &  0.141 &  0.214 &  1.24 &  5.47 &  0 \\  
 &  0.85 &  0.633 &  0.133 &  0.226 &  1.09 &  5.05 &  0 \\  
 &  0.90 &  0.663 &  0.125 &  0.237 &  0.96 &  4.68 &  0 \\  
 &  0.95 &  0.695 &  0.118 &  0.249 &  0.86 &  4.37 &  0 \\  
 &  1.00 &  0.726 &  0.111 &  0.261 &  0.770 & 4.10 &  0 \\  
 &  1.05 &  0.758 &  0.106 &  0.271 &  0.696 & 3.86 &  0 \\  
 &  1.10 &  0.790 &  0.100 &  0.283 &  0.632 & 3.65 &  0 \\  
 &  1.15 &  0.823 &  0.095 &  0.295 &  0.577 & 3.46 &  0 \\  
 &  1.20 &  0.855 &  0.090 &  0.308 &  0.529 & 3.30 &  0 \\  
 &  1.25 &  0.889 &  0.086 &  0.319 &  0.486 & 3.15 &  0 \\  
 &  1.30 &  0.922 &  0.082 &  0.331 &  0.449 & 3.01 &  0 \\  
 &  1.35 &  0.956 &  0.078 &  0.343 &  0.417 & 2.89 &  0 \\  
 &  1.40 &  0.990 &  0.075 &  0.355 &  0.387 & 2.79 &  0 \\  
 &  1.45 &  1.024 &  0.071 &  0.367 &  0.361 & 2.68 &  0 \\  
 &  1.50 &  1.059 &  0.069 &  0.379 &  0.338 & 2.59 &  0 \\  
 &  1.55 &  1.094 &  0.066 &  0.392 &  0.316 & 2.50 &  0 \\  
 &  1.60 &  1.129 &  0.063 &  0.404 &  0.297 & 2.42 &  0 \\  
 &  1.65 &  1.164 &  0.060 &  0.417 &  0.280 & 2.35 &  0 \\  
 &  1.70 &  1.200 &  0.058 &  0.432 &  0.264 & 2.29 &  0 \\  
 &  1.75 &  1.235 &  0.056 &  0.445 &  0.249 & 2.23 &  0 \\  
 &  1.80 &  1.271 &  0.054 &  0.455 &  0.236 & 2.17 &  0 \\  
 &  1.85 &  1.308 &  0.052 &  0.471 &  0.224 & 2.12 &  0 \\  
 &  1.90 &  1.344 &  0.050 &  0.483 &  0.212 & 2.07 &  0 \\  
 &  1.95 &  1.381 &  0.048 &  0.498 &  0.202 & 2.03 &  0 \\  
 &  2.00 &  1.418 &  0.046 &  0.508 &  0.192 & 1.98 &  0 \\  
 &  2.05 &  1.455 &  0.045 &  0.520 &  0.183 & 1.94 &  0 \\  
 &  2.10 &  1.492 &  0.043 &  0.538 &  0.175 & 1.91 &  0 \\  
 &  2.15 &  1.530 &  0.041 &  0.552 &  0.167 & 1.87 &  0 \\  
 &  2.20 &  1.567 &  0.040 &  0.561 &  0.160 & 1.83 &  0 \\  
 &  2.25 &  1.605 &  0.039 &  0.578 &  0.153 & 1.80 &  0 \\  
 &  2.30 &  1.643 &  0.038 &  0.589 &  0.147 & 1.77 &  0 \\  
 &  2.35 &  1.681 &  0.037 &  0.601 &  0.141 & 1.74 &  0 \\  
 &  2.40 &  1.720 &  0.035 &  0.622 &  0.136 & 1.72 &  0 \\  
 &  2.45 &  1.758 &  0.034 &  0.629 &  0.130 & 1.69 &  0 \\  
 &  2.50 &  1.797 &  0.033 &  0.644 &  0.125 & 1.67 &  0 \\  
 &  2.55 &  1.836 &  0.032 &  0.661 &  0.121 & 1.65 &  0 \\  
 &  2.60 &  1.875 &  0.031 &  0.679 &  0.116 & 1.63 &  0 \\  
 &  2.65 &  1.914 &  0.030 &  0.689 &  0.112 & 1.61 &  0 \\  
 &  2.70 &  1.953 &  0.030 &  0.700 &  0.108 & 1.58 &  0 \\  
 &  2.75 &  1.992 &  0.029 &  0.721 &  0.105 & 1.57 &  0 \\  
 &  2.80 &  2.032 &  0.028 &  0.734 &  0.101 & 1.55 &  0 \\  
 &  2.85 &  2.071 &  0.027 &  0.748 &  0.098 & 1.54 &  0 \\  
 &  2.90 &  2.111 &  0.026 &  0.763 &  0.095 & 1.52 &  0 \\  
 &  2.95 &  2.156 &  0.024 &  0.831 &  0.092 & 1.57 &  0 \\  
 &  3.00 &  2.202 &  0.023 &  0.866 &  0.089 & 1.58 &  0 \\  
\hline\hline
\end{tabular}
\end{spacing}
\end{table}

\begin{table}
\caption{Triple points: $(\phi_{cg}, \phi_{pg})$, $(\phi_{cl}, \phi_{pl})$
$(\phi_{cs}, \phi_{ps})$ are the volume fractions of the three coexisting
phases, $\widetilde{P}_{tp} = \beta P_{tp} V_c$ is
the corresponding reduced pressure. Good-solvent case.}
\label{Triple-points-GS-suppl}
\begin{tabular}{cccccccccc}
\hline\hline
$z$ & $q$ & $\phi^{(r)}_{tp}$ &
    $\phi_{cg}$ & $\phi_{pg}$ & $\phi_{cl}$ & $\phi_{pl}$ &
    $\phi_{cs}$ & $\phi_{ps}$ & $\widetilde{P}_{tp}$ \\
\hline
 &  0.5 &  0.607 &  0.107 &  0.468 &  0.436 &  0.100 &  0.578 &  
  0.017 &  9.70 \\  
 &  0.6 &  0.797 &  0.043 &  0.719 &  0.458 &  0.096 &  0.568 &  
  0.021 &  8.69 \\  
 &  0.7 &  1.01 &  0.021 &  0.961 &  0.466 &  0.104 &  0.563 &  
  0.026 &  8.22 \\  
 &  0.8 &  1.24 &  0.012 &  1.21 &  0.470 &  0.120 &  0.561 &  
  0.031 &  8.01 \\  
 &  0.9 &  1.49 &  0.0076 &  1.46 & 0.471 &  0.141 &  0.560 &  
  0.038 &  7.93 \\  
 &  1.0 &  1.75 &  0.0053 &  1.73 & 0.472 &  0.167 &  0.559 &  
  0.046 &  7.92 \\  
 &  1.1 &  2.02 &  0.0040 &  2.00 & 0.473 &  0.191 &  0.559 &  
  0.054 &  7.89 \\  
 &  1.2 &  2.29 &  0.0033 &  2.28 & 0.473 &  0.217 &  0.559 &  
  0.062 &  7.88 \\  
 &  1.3 &  2.58 &  0.0028 &  2.56 & 0.473 &  0.246 &  0.559 &  
  0.071 &  7.89 \\  
 &  1.4 &  2.89 &  0.0024 &  2.86 & 0.473 &  0.276 &  0.559 &  
  0.080 &  7.91 \\  
 &  1.5 &  3.16 &  0.0021 &  3.15 & 0.473 &  0.309 &  0.559 &  
  0.090 &  7.93 \\  
 &  1.6 &  3.47 &  0.0019 &  3.46 & 0.473 &  0.343 &  0.559 &  
  0.100 &  7.95 \\  
 &  1.7 &  3.78 &  0.0018 &  3.77 & 0.473 &  0.379 &  0.559 &  
  0.111 &  7.99 \\  
 &  1.8 &  4.10 &  0.0017 &  4.09 & 0.473 &  0.417 &  0.559 &  
  0.123 &  8.02 \\  
 &  1.9 &  4.42 &  0.0016 &  4.42 & 0.472 &  0.456 &  0.560 &  
  0.135 &  8.05 \\  
 &  2.0 &  4.75 &  0.0015 &  4.75 & 0.472 &  0.498 &  0.560 &  
  0.148 &  8.08 \\  
\hline\hline
\end{tabular}
\end{table}

\begin{table}
\caption{Triple points: $(\phi_{cg}, \phi_{pg})$, $(\phi_{cl}, \phi_{pl})$
$(\phi_{cs}, \phi_{ps})$ are the volume fractions of the three coexisting
phases, $\widetilde{P}_{tp}$ is the corresponding reduced pressure. Results for 
$z = z^{(3)}$.}
\label{Triple-points-z3-suppl}
\begin{tabular}{cccccccccc}
\hline\hline
$z$ & $q$ & $\phi^{(r)}_{tp}$ &
    $\phi_{cg}$ & $\phi_{pg}$ & $\phi_{cl}$ & $\phi_{pl}$ &
    $\phi_{cs}$ & $\phi_{ps}$ & $\widetilde{P}_{tp}$ \\
\hline
 &  0.4 &  0.439 &  0.114 &  0.323 &  0.437 &  0.056 &  0.576 &  
  0.0070 &  9.21 \\  
 &  0.5 &  0.632 &  0.024 &  0.593 &  0.470 &  0.037 &  0.559 &  
  0.0067 &  7.52 \\  
 &  0.6 &  0.877 &  $5\cdot 10^{-3}$ &  0.865 & 0.480 &  0.030 &  0.551 &  
  0.0061 &  6.84 \\  
 &  0.7 &  1.16  &  $1\cdot 10^{-3}$ &  1.16 &  0.484 &  0.028 &  0.547 &  
  0.0057 &  6.54 \\  
 &  0.8 &  1.48 &   $4\cdot 10^{-4}$ &  1.48 &  0.486 &  0.027 &  0.546 &  
  0.0053 &  6.39 \\  
 &  0.9 &  1.82 &   $1\cdot 10^{-4}$ &  1.82 &  0.487 &  0.027 &  0.545 &  
  0.0051 &  6.30 \\  
 &  1.0 &  2.19 &   $4\cdot 10^{-5}$ &  2.19 &  0.488 &  0.027 &  0.544 &  
  0.0049 &  6.24 \\  
 &  1.1 &  2.57 &   $1\cdot 10^{-5}$ &  2.57 &  0.488 &  0.027 &  0.543 &  
  0.0048 &  6.21 \\  
 &  1.2 &  2.97 &   $8\cdot 10^{-6}$ &  2.97 &  0.488 &  0.027 &  0.543 &  
  0.0046 &  6.17 \\  
 &  1.3 &  3.39 &   $4\cdot 10^{-6}$ &  3.39 &  0.489 &  0.027 &  0.543 &  
  0.0044 &  6.15 \\  
 &  1.4 &  3.82 &   $2\cdot 10^{-6}$ &  3.82 &  0.489 &  0.026 &  0.543 &  
  0.0042 &  6.13 \\  
 &  1.5 &  4.26 &   $1\cdot 10^{-6}$ &  4.26 &  0.489 &  0.026 &  0.542 &  
  0.0040 &  6.12 \\  
 &  1.6 &  4.72 &   $6\cdot 10^{-7}$ &  4.72 &  0.489 &  0.026 &  0.542 &  
  0.0038 &  6.11 \\  
 &  1.7 &  5.19 &   $3\cdot 10^{-7}$ &  5.19 &  0.490 &  0.025 &  0.542 &  
  0.0036 &  6.09 \\  
 &  1.8 &  5.69 &   $2\cdot 10^{-7}$ &  5.69 &  0.490 &  0.024 &  0.542 &  
  0.0033 &  6.09 \\  
 &  1.9 &  6.17 &   $1\cdot 10^{-7}$ &  6.17 &  0.490 &  0.024 &  0.542 &  
  0.0031 &  6.08 \\  
 &  2.0 &  6.68 &   $7\cdot 10^{-8}$ &  6.68 &  0.490 &  0.023 &  0.542 &  
  0.0029 &  6.07 \\  
\hline\hline
\end{tabular}
\end{table}

\begin{table}
\caption{Triple points: $(\phi_{cg}, \phi_{pg})$, $(\phi_{cl}, \phi_{pl})$
$(\phi_{cs}, \phi_{ps})$ are the volume fractions of the three coexisting
phases, $\widetilde{P}_{tp}$ is the corresponding reduced pressure. Results for 
$z = z^{(1)}$.}
\label{Triple-points-z1-suppl}
\begin{tabular}{cccccccccc}
\hline\hline
$z$ & $q$ & $\phi^{(r)}_{tp}$ &
    $\phi_{cg}$ & $\phi_{pg}$ & $\phi_{cl}$ & $\phi_{pl}$ &
    $\phi_{cs}$ & $\phi_{ps}$ & $\widetilde{P}_{tp}$ \\
\hline
 &  0.4 &  0.467 &  $5.7\cdot 10^{-2}$  &  0.397 &  0.458 &  
  $3.2\cdot 10^{-2}$  &  0.566 &  $4.1\cdot 10^{-3}$  &  8.16 \\  
 &  0.5 &  0.722 &  $6.4\cdot 10^{-3}$  &  0.708 &  0.479 &  
  $1.7\cdot 10^{-2}$  &  0.551 &  $2.7\cdot 10^{-3}$  &  6.78 \\  
 &  0.6 &  1.08 &   $5.0\cdot 10^{-4}$  &  1.08  &  0.486 &  
  $1.0\cdot 10^{-2}$  &  0.545 &  $1.6\cdot 10^{-3}$  &  6.31 \\  
 &  0.7 &  1.54 &   $5.0\cdot 10^{-5}$  &  1.53  &  0.489 &  
  $6.8\cdot 10^{-3}$  &  0.543 &  $8.8\cdot 10^{-4}$  &  6.13 \\  
 &  0.8 &  2.07 &   $5.2\cdot 10^{-6}$  &  2.07  &  0.490 &  
  $4.7\cdot 10^{-3}$  &  0.542 &  $4.9\cdot 10^{-4}$  &  6.07 \\  
 &  0.9 &  2.66 &   $5.5\cdot 10^{-7}$  &  2.67  &  0.490 &  
  $3.3\cdot 10^{-3}$  &  0.541 &  $2.8\cdot 10^{-4}$  &  6.04 \\  
 &  1.0 &  3.32 &   $7\cdot 10^{-8}$  &  3.32 &  0.491 &  
  $2.3\cdot 10^{-3}$  &  0.541 &  $1.6\cdot 10^{-4}$  &  6.03 \\  
 &  1.1 &  4.02 &   $9\cdot 10^{-9}$  &  4.02 &  0.491 &  
  $1.6\cdot 10^{-3}$  &  0.541 &  $9\cdot 10^{-5}$  &  6.02 \\  
 &  1.2 &  4.77 &   $1\cdot 10^{-9}$  &  4.77 &  0.491 &  
  $1.1\cdot 10^{-3}$  &  0.541 &  $5\cdot 10^{-5}$  &  6.02 \\  
 &  1.3 &  5.55 &  0 &  5.55 &  0.491 &
   $8\cdot 10^{-4}$  &  0.541 &  $3\cdot 10^{-5}$ & 6.01 \\
 &  1.4 &  6.37 &  0 &  6.37 &  0.491 &
   $6\cdot 10^{-4}$  &  0.541 &  $2\cdot 10^{-5}$ & 6.01 \\
 &  1.5 &  7.23 &  0 &  7.23 &  0.491 &  
   $4\cdot 10^{-4}$  &  0.541 &  $8\cdot 10^{-6}$ & 6.01 \\
 &  1.6 &  8.12 &  0 &  8.12 &  0.491 &  
   $3\cdot 10^{-4}$  &  0.541 &  $5\cdot 10^{-6}$ & 6.01 \\
 &  1.7 &  9.03 &  0 &  9.03 &  0.491 &  
   $2\cdot 10^{-4}$  &  0.541 &  $3\cdot 10^{-6}$ & 6.01 \\
 &  1.8 &  10.0 &  0 &  10.0 &  0.491 &  
   $1\cdot 10^{-4}$  &  0.541 &  $1\cdot 10^{-6}$ & 6.01 \\
 &  1.9 &  11.0 &  0 &  11.0 &  0.491 &  
   $9\cdot 10^{-5}$  &  0.541 &  $8\cdot 10^{-7}$ & 6.01 \\
 &  2.0 &  12.0 &  0 &  12.0 &  0.491 &  
   $6\cdot 10^{-5}$  &  0.541 &  $4\cdot 10^{-7}$ & 6.01 \\
\hline\hline
\end{tabular}
\end{table}

\begin{table}
\caption{Triple points: $(\phi_{cg}, \phi_{pg})$, $(\phi_{cl}, \phi_{pl})$
$(\phi_{cs}, \phi_{ps})$ are the volume fractions of the three coexisting
phases, $\widetilde{P}_{tp}$ is the corresponding reduced pressure. Results for 
$z = 0$.}
\label{Triple-points-z0-suppl}
\begin{tabular}{cccccccccc}
\hline\hline
$z$ & $q$ & $\phi^{(r)}_{tp}$ &
    $\phi_{cg}$ & $\phi_{pg}$ & $\phi_{cl}$ & $\phi_{pl}$ &
    $\phi_{cs}$ & $\phi_{ps}$ & $\tilde{P}_{tp}$ \\
\hline
 &  0.5 & 0.811 &  $1.8\cdot 10^{-3}$ &  0.806 & 0.484 &  
   $7.8\cdot 10^{-3}$ &  0.547 &  $9.6\cdot 10^{-4}$ & 6.48 \\  
 &  0.6 &  1.32 &  $6.6\cdot 10^{-5}$ &  1.32  & 0.489 &
   $2.4\cdot 10^{-3}$ &  0.542 &  $2.0\cdot 10^{-4}$ & 6.12 \\
 &  0.7 &  2.07 &  $9.9\cdot 10^{-7}$ &  2.07  & 0.491 &
   $5.8\cdot 10^{-4}$ &  0.541 &  $2.6\cdot 10^{-5}$ & 6.03 \\
 &  0.8 &  3.08 &  $9\cdot 10^{-9}$ &  3.08 &  0.491 &  
   $1.1\cdot 10^{-4}$ &  0.541 &  $1.9\cdot 10^{-6}$ & 6.01 \\  
 &  0.9 &  4.38 &  $6\cdot 10^{-11}$ & 4.38 &  0.491 &  
   $1.4\cdot 10^{-5}$ &  0.541 &  $9\cdot 10^{-8}$   & 6.01 \\  
 &  1.0 &  6.01 &  0 &  6.01 &  0.491 &  
   $1.4\cdot 10^{-6}$ &  0.541 &  $3\cdot 10^{-9}$   & 6.01 \\
 &  1.1 &  8.00 &  0 &  8.00 &  0.491 &
   $1.0\cdot 10^{-7}$ &  0.541 &  $5\cdot 10^{-11}$  & 6.01 \\
 &  1.2 &  10.4 &  0 &  10.4 &  0.491 &
   $5\cdot 10^{-9}$   &  0.541 &  $5\cdot 10^{-13}$  & 6.01 \\
 &  1.3 &  13.2 &  0 &  13.2 &  0.491 &
   $2\cdot 10^{-10}$  &  0.541 &  0 & 6.01 \\
 &  1.4 &  16.5 &  0 &  16.5 &  0.491 &  
   $5\cdot 10^{-12}$  &  0.541 &  0 & 6.01 \\
 &  1.5 &  20.3 &  0 &  20.3 &  0.491 &  
   $1\cdot 10^{-13}$  &  0.541 &  0 & 6.01 \\
 &  1.6 &  24.6 &  0 &  24.6 &  0.491 &  0 &  0.541 &  0 &  6.01 \\  
 &  1.7 &  29.5 &  0 &  29.5 &  0.491 &  0 &  0.541 &  0 &  6.01 \\  
 &  1.8 &  35.1 &  0 &  35.1 &  0.491 &  0 &  0.541 &  0 &  6.01 \\
 &  1.9 &  41.2 &  0 &  41.2 &  0.491 &  0 &  0.541 &  0 &  6.01 \\
 &  2.0 &  48.1 &  0 &  48.1 &  0.491 &  0 &  0.541 &  0 &  6.01 \\
\hline\hline
\end{tabular}
\end{table}
\clearpage

\medskip

{\bf References for the supplementary material}


\begin{thebibliography}{99}
\bibitem{footnote-err}
Note that the second equation appearing in Aarts {\em et al.} 
(Ref.~\onlinecite{ATL-02}) is incorrect. The term $2 (\Delta/R_c)^3$ should 
be replaced by $3 (\Delta/R_c)^3$.
\bibitem{footnote-MCSL} We consider 
a bidisperse solution of hard spheres of radius $R_c$ and $R_p$, $R_p/R_c = d$, 
at volume fractions $\phi_c$ and $\phi_p$ and define, 
see Eq.~(\ref{phip-GFVT}), 
$\alpha = \lim_{\phi_p\to 0} \phi_p/\phi_p^{(r)}$. In this limit we have
$\rho_p^{(r)} \lambda^3_p  = e^{\beta \mu_p}$. Then, if we define
the excess chemical potential $\mu_p^{(exc)}$ as 
$e^{\beta \mu_p} = \rho_p\lambda^3_p e^{\beta \mu_p^{(exc)}}$, we end up
with $\alpha = \lim_{\phi_p\to 0} e^{-\beta \mu_p^{(exc)}}$. Such a quantity
can be easily determined from the equation of state of the bidisperse system.
\bibitem{foot-extrapola}
To extrapolate $A(L,V) = \beta R_c^3/\kappa_T$, we assume
$A(L,V) = A_{\rm scal} + b L^{-\nu} + c L^{s\nu}/M^s$, where $M$ is 
the size of the box (its volume is $M^3$) and $L$ the chain length. 
The number $s$ depends on the choice of the finite-volume definition 
(see Ref.~\onlinecite{DPP-depletion}, Sec.~5.1), while $\nu$ 
is equal to 0.5876 for good-solvent conditions and $\nu = 1/2$ in the 
thermal crossover region. Note the
peculiar $L$ dependence of the volume correction. It is justified in
detail in Ref.~\onlinecite{DPP-thermal}, Sec.~IV.B.
\end{thebibliography}
\end{document}